\documentclass[9pt]{article}
\usepackage{amsmath}
\usepackage{amssymb}
\usepackage{graphicx}
\usepackage{color}
\usepackage[table,xcdraw]{xcolor}
\usepackage{orcidlink}
\usepackage{hyperref}
\hypersetup{colorlinks=true, linkcolor=blue, citecolor=blue, urlcolor=blue}
\usepackage{array}
\usepackage{longtable}
\usepackage{booktabs}
\usepackage{colortbl}
\usepackage{subcaption}
\usepackage{geometry}
\renewcommand{\arraystretch}{1.8}
\RequirePackage[numbers,sort&compress]{natbib}
\begin{document}
\baselineskip=14pt

\begin{center}
{\LARGE Probing the Charged Hayward Black Hole in Dark Matter and String Cloud \\[6pt] Environments through Shadow, Geodesics, and Quasinormal Spectrum}
\end{center}

\vspace{0.1cm}

\begin{center}
{\bf Faizuddin Ahmed}\orcidlink{0000-0003-2196-9622}\\Department of Physics, The Assam Royal Global University, Guwahati, 781035, Assam, India\\
e-mail: faizuddinahmed15@gmail.com\\
\vspace{0.1cm}
{\bf Ahmad Al-Badawi}\orcidlink{0000-0002-3127-3453}\\
Department of Physics, Al-Hussein Bin Talal University, 71111,
Ma'an, Jordan. \\
e-mail: ahmadbadawi@ahu.edu.jo\\
\vspace{0.1cm}
{\bf \.{I}zzet Sakall{\i}}\orcidlink{0000-0001-7827-9476}\\
Physics Department, Eastern Mediterranean University, Famagusta 99628, North Cyprus via Mersin 10, Turkey\\
e-mail: izzet.sakalli@emu.edu.tr (Corresponding author)
\end{center}
\vspace{0.1cm}

\begin{abstract}
We investigate the spacetime geometry of a charged Hayward black hole surrounded by perfect fluid dark matter (PFDM) and a cloud of strings (CoS). The metric function incorporates the Hayward regularization parameter $g$, electric charge $Q$, PFDM $\beta$, and CoS parameter $\alpha$. We derive the horizon structure and identify non-extremal, extremal, single-horizon, and naked singularity configurations. For null geodesics, we compute the effective potential, photon sphere radius, and shadow radius, showing how $\alpha$ and $\beta$ modify the gravitational barrier. The dynamics of neutral particles are analyzed through the specific energy, angular momentum, and innermost stable circular orbit conditions. Quasiperiodic oscillations are studied using the relativistic precession model, yielding orbital, radial, vertical, and periastron frequencies. We examine scalar perturbations governed by the Klein-Gordon equation, compute greybody factors using semi-analytical bounds, and establish the connection between quasinormal modes and the BH shadow in the eikonal limit. Our results show that increasing $\alpha$ and $\beta$ suppresses effective potential peaks for photons and scalar waves, while producing opposite effects on the specific energy of orbiting particles. These findings potentially provide testable predictions for Event Horizon Telescope and X-ray timing observations.
\\
\\
{\bf Keywords}: Hayward black hole; perfect fluid dark matter; quasiperiodic oscillations; photon sphere; greybody factors
\end{abstract}

\tableofcontents

\section{Introduction} \label{isec1}

BHs represent one of the most fascinating predictions of general relativity (GR), serving as natural laboratories for testing gravitational physics in the strong-field regime. The detection of gravitational waves from binary BH mergers by LIGO and Virgo \cite{isz01,isz02} and the imaging of supermassive BH shadows by the Event Horizon Telescope (EHT) collaboration \cite{EHTL1,EHTL4,EHTL6,EHTL12,EHTL14,EHTL15,EHTL16,EHTL17} have opened new observational windows into BH physics. These groundbreaking observations have stimulated theoretical investigations into BH solutions beyond the classical Schwarzschild and Kerr metrics, particularly those incorporating additional matter fields and modified gravity effects \cite{isz06,isz07}.

A fundamental issue in classical GR is the presence of singularities at the centers of BHs, where curvature invariants diverge and predictability breaks down. Regular BHs, which are singularity-free solutions to the field equations, have attracted considerable attention as potential resolutions to this problem. Bardeen \cite{isz08} introduced the first regular BH model, later interpreted as a gravitational field of a nonlinear magnetic monopole \cite{isz09}. Subsequently, Hayward \cite{isz10} proposed an alternative regular BH solution characterized by a different mass function $m(r) = Mr^3/(r^3 + g^3)$, where $M$ is the total mass and $g$ is the Hayward regularization parameter. This mass function ensures de Sitter behavior at the core while recovering Schwarzschild geometry at large distances. The Hayward metric has been extensively studied in various contexts including thermodynamics \cite{isz11,isz12}, geodesic motion \cite{isz13}, quasinormal modes (QNMs) \cite{isz14}, and gravitational lensing \cite{isz15,isz16}. Charged generalizations of the Hayward BH have also been investigated, incorporating electric or magnetic charges that further modify the spacetime geometry \cite{isz17,isz18}.

Astrophysical BHs are expected to be surrounded by various forms of matter, including dark matter halos that dominate the mass budget of galaxies. PFDM provides a phenomenological framework for incorporating dark matter effects around compact objects \cite{isz19,isz20}. The PFDM model, derived from an anisotropic fluid energy-momentum tensor, introduces a logarithmic correction to the metric function of the form $(\beta/r)\ln(r/|\beta|)$, where $\beta$ is the PFDM parameter. This logarithmic dependence produces distinct observational signatures compared to other dark matter profiles such as the Navarro-Frenk-White \cite{isz21} or pseudo-isothermal models \cite{isz22}. Recent studies have examined how PFDM affects BH shadows \cite{isz23,isz24}, geodesics \cite{isz25}, and QNMs \cite{isz26}, demonstrating that PFDM can produce measurable deviations from vacuum BH predictions.

Another astrophysically motivated modification involves the CoS, introduced by Letelier \cite{isz27} as a representation of one-dimensional extended objects in the spacetime. The CoS arises naturally in string theory contexts and produces a solid angle deficit characterized by the string parameter $\alpha$ satisfying $0 \leq \alpha < 1$. The presence of CoS modifies the asymptotic structure of the spacetime such that $\lim_{r\to\infty} g_{tt} = -(1-\alpha) \neq -1$, leading to a conical geometry at spatial infinity. BHs coupled to CoS have been studied in the context of Schwarzschild \cite{isz28}, Reissner-Nordstr\"{o}m (RN) \cite{isz29}, and various modified gravity theories \cite{isz30,isz31}. The combination of CoS with other matter fields provides a framework for investigating how multiple environmental factors simultaneously influence BH properties.

The study of geodesic motion around BHs is central to understanding their observational signatures. Null geodesics determine the paths of photons and govern phenomena such as gravitational lensing, BH shadows, and photon sphere properties \cite{isz32,isz33}. The photon sphere radius $r_s$, defined by the condition $2f(r_s) - r_s f'(r_s) = 0$ for a spherically symmetric metric with $g_{tt} = -f(r)$, corresponds to the location of unstable circular photon orbits. The shadow radius observed by a distant observer is directly related to the photon sphere properties and provides constraints on BH parameters through EHT observations \cite{isz34,isz35}. Timelike geodesics describe the motion of massive particles and determine the structure of accretion disks through quantities such as the specific energy, specific angular momentum, and innermost stable circular orbit (ISCO) radius \cite{isz36}.

Quasiperiodic oscillations (QPOs) observed in the X-ray emission from accreting BHs offer another powerful probe of strong-field gravity \cite{isz37,isz38}. High-frequency QPOs appear in pairs with frequencies in commensurable ratios, suggesting a connection to the fundamental frequencies of geodesic motion. The relativistic precession model relates the observed QPO frequencies to the orbital, radial epicyclic, and vertical epicyclic frequencies of test particles near the BH \cite{isz39}. The periastron precession frequency, defined as the difference between the orbital and radial frequencies, characterizes the precession of slightly eccentric orbits. These epicyclic frequencies depend sensitively on the spacetime geometry and can distinguish between different BH models when compared with observational data from sources such as GRS 1915+105 and XTE J1550-564 \cite{isz40}. Circular orbits, in particular the innermost stable circular orbits (ISCOs), play a fundamental role in understanding the dynamics of test particles in the vicinity of black holes. Observations of accretion disks offer valuable constraints on black hole parameters and their physical properties \cite{JFS2011,LG2014,JEM2015}. A large body of work has investigated particle dynamics around black holes in a variety of configurations within general relativity and modified theories of gravity. These studies demonstrate how the presence of external matter fields can significantly influence the stability of marginally circular orbits of test particles and QPO epicyclic frequencies (see, for examples, \cite{GMM1,GMM2,GMM3,GMM4,GMM5,GMM6,GMM7,GMM8,GMM9,GMM10,GMM11}).

Scalar field perturbations provide information about the stability and dynamical response of BH spacetimes to external disturbances. The evolution of a massless scalar field is governed by the Klein-Gordon equation, which reduces to a Schr\"{o}dinger-like wave equation with an effective potential after separation of variables. The QNM spectrum, consisting of complex frequencies with negative imaginary parts indicating damped oscillations, characterizes the ringdown phase of perturbed BHs \cite{isz41,isz42}. The WKB approximation \cite{isz43,isz44} provides semi-analytical expressions for QNM frequencies, while the connection between QNMs and BH shadows in the eikonal limit \cite{isz45} relates the real part of the QNM frequency to the shadow radius. Greybody factors quantify the transmission probability of waves through the effective potential barrier and determine the spectrum of Hawking radiation modified by backscattering \cite{isz46,isz47,Aydiner:2025eii,Sucu:2025fwa,Sucu:2025olo,Gursel:2025wan}.

The motivation for this work stems from the need to understand how multiple environmental factors-regular BH cores, electric charge, dark matter, and string clouds-collectively influence BH observables. While previous studies have examined these effects individually or in limited combinations, a unified treatment of the charged Hayward BH with CoS and PFDM has not been presented. Such a configuration is astrophysically relevant, as supermassive BHs at galactic centers are embedded in dark matter halos, and cosmic strings may have formed during phase transitions in the early universe. By analyzing the geodesic structure, QPO frequencies, and scalar perturbations of this BH solution, we aim to identify distinctive observational signatures that could differentiate it from standard vacuum BHs and other modified solutions. The results may be tested against current and future observations from the EHT, gravitational wave detectors, and X-ray timing instruments.

In this paper, we construct and analyze the charged Hayward BH coupled to a CoS and surrounded by PFDM. Our specific objectives are: (i) to derive the metric function and classify the horizon structure as a function of the parameters $\{M, g, Q, \alpha, \beta\}$; (ii) to compute the effective potential, photon sphere radius, and shadow radius for null geodesics; (iii) to determine the specific energy, specific angular momentum, and ISCO radius for timelike geodesics; (iv) to calculate the orbital, radial, vertical, and periastron frequencies relevant for QPO modeling; (v) to analyze scalar perturbations, compute the effective potential, and establish the connection between QNMs and the BH shadow; and (vi) to evaluate the greybody factors and transmission probabilities using semi-analytical bounds.

The paper is organized as follows. In Section~\ref{isec2}, we present the action formulation, derive the metric function $f(r)$, and analyze the horizon structure including extremal BH configurations. We also discuss the limiting cases and present three-dimensional visualizations of the metric function. Section~\ref{isec3} is devoted to null geodesics, where we compute the effective potential, photon sphere radius, shadow radius, geodesic angular velocity, photon trajectories, and effective radial force. In Section~\ref{isec4}, we study the dynamics of neutral test particles, deriving the effective potential, radial force, specific angular momentum, specific energy, and ISCO conditions. Section~\ref{isec5} examines QPOs through the relativistic precession model, presenting the orbital, radial, vertical, and periastron frequencies. In Section~\ref{isec6}, we analyze scalar perturbations governed by the Klein-Gordon equation, compute the transmission and reflection probabilities, apply the WKB approximation for QNMs, and establish the connection between the BH shadow and QNM frequencies. Finally, Section~\ref{isec7} summarizes our main findings and outlines directions for future research. Throughout this paper, we adopt geometric units with $G = c = 1$ and work primarily in units where the BH mass $M = 1$.

\section{Geometric Background and Thermodynamic Framework} \label{isec2}

We consider a static, spherically symmetric charged Hayward BH coupled to a CoS and surrounded by PFDM. The theoretical foundation rests upon GR minimally coupled to NED, the electromagnetic field, the string cloud source, and the dark matter component. The total action governing this spacetime geometry takes the form \cite{KAB2001,RI2022,isz27,HXZ2021}:
\begin{equation}
S= \int d^{4}x \sqrt{-g}\, \left(\frac{R}{2}+\mathcal{L}(F) +\mathcal{L}_{\rm EM}+\mathcal{L}_{\text{CoS}}+\mathcal{L}_{\text{DM}}\right).
\label{act1}
\end{equation}
Here, $R$ denotes the Ricci scalar and $g$ is the determinant of the metric tensor $g_{\mu\nu}$.

The NED Lagrangian associated with the Hayward regularization mechanism reads \cite{isz10,isz09}
\begin{equation}
\mathcal{L}(F) =
\frac{6 \left(2 \ell^{2} F\right)^{3/2}}
{\kappa \ell^{2} \left[1 + \left(2 \ell^{2} F\right)^{3/4}\right]^{2}},
\label{act2}
\end{equation}
where the electromagnetic invariant is $F=F^{\mu\nu} F_{\mu\nu}$, with $F_{\mu\nu}$ being the Maxwell-Faraday tensor. For a spherically symmetric configuration carrying only magnetic charge, the sole nonzero component of $F_{\mu\nu}$ is \cite{KAB2001}
\begin{equation}
    F_{23}=F_{\theta \phi}=q_m \sin \theta,\label{act3}
\end{equation}
yielding the invariant $F=2 q^2_m/r^4$. The magnetic charge $q_m$ and the Hayward regularization parameter $g$ are related through \cite{isz17}
\begin{equation}
q_m=\frac{\sqrt{3}}{4}M^2 g^2,
\label{mag_charge_relation}
\end{equation}
where $m$ is the mass parameter. This relation ensures the de Sitter (dS) core behavior at $r \to 0$, which is the hallmark of regular BH solutions \cite{sec2is04}.

The standard Maxwell Lagrangian for the electric field component is \cite{RI2022}
\begin{equation}
\mathcal{L}_{\rm EM} = -\frac{1}{4} F_{\alpha\beta} F^{\alpha\beta},
\label{act4}
\end{equation}
with the radial electric field given by
\begin{equation}
F_{01} = F_{10} = -F^{01} \equiv E_{r} = \frac{Q}{r^{2}},
\label{act5}
\end{equation}
where $Q$ represents the electric charge.

The CoS contribution originates from the Nambu-Goto action for string-like objects \cite{isz27,sec2is05}:
\begin{equation}
S_{\text{CoS}} = \int d^{4}x \sqrt{-g}\, \mathcal{L}_{\text{CoS}}
= \int (-\gamma)^{1/2} \mathcal{M}\, d\lambda^{0} d\lambda^{1},
\label{act6}
\end{equation}
where $\gamma$ denotes the determinant of the induced metric on the string worldsheet, and $\mathcal{M}$ is the string tension density. The corresponding Lagrangian density is
\begin{equation}
\mathcal{L}_{\text{CoS}} = \mathcal{M} \sqrt{-\gamma}
= \mathcal{M} \left(-\frac{1}{2} \Sigma^{\mu\nu} \Sigma_{\mu\nu} \right)^{1/2},
\label{act7}
\end{equation}
with $\Sigma^{\mu\nu}$ being the bivector associated with the string worldsheet.

Variation of the action (\ref{act1}) with respect to the metric yields the gravitational field equations
\begin{equation}
    G_{\mu\nu}=\kappa \left(T^{\rm NED}_{\mu\nu}+T^{\rm EM}_{\mu\nu}+T^{\rm DM}_{\mu\nu}\right)+T^{\rm CoS}_{\mu\nu}.\label{EFE}
\end{equation}
The energy-momentum tensor for PFDM adopts the diagonal form \cite{HXZ2021,isz19}
\begin{equation}
T^{\text{DM}}_{\,\,\mu\nu} = \mathrm{diag}\left(-\mathcal{E}_{\rm DM},\, P_{r\,\rm DM},\, P_{\theta\,\rm DM},\, P_{\phi\,\rm DM}\right),
\label{tensor}
\end{equation}
where the energy density and pressures satisfy the equation of state
\begin{equation}
\mathcal{E}_{\rm DM} = -P_{r\,\rm DM}= -\frac{\beta}{\kappa r^3}, \quad
P_{\theta\,\rm DM}= P_{\phi\,\rm DM} = -\frac{\beta}{2 \kappa r^3}.
\label{energy_density}
\end{equation}
The parameter $\beta$ characterizes the PFDM intensity and introduces a logarithmic modification to the spacetime geometry \cite{sec2is07}.

The solution to the field equations (\ref{EFE}) yields a static, spherically symmetric line element
\begin{equation}
ds^2 = -f(r)\,dt^2 + \frac{dr^2}{f(r)} + r^2\left(d\theta^2 + \sin^2\theta\,d\phi^2\right),\label{metric}
\end{equation}
where the lapse function $f(r)$ for the charged Hayward BH with CoS and PFDM reads
\begin{equation}
{f(r) = 1 - \alpha - \frac{2 M r^2}{r^3 + g^3} + \frac{Q^2}{r^2} + \frac{\beta}{r}\ln\!\frac{r}{\left|\beta\right|}}.\label{function}
\end{equation}

The physical parameters appearing in the metric function are as follows. The mass parameter $m$ sets the overall gravitational scale, and we adopt units where $M = m = 1$ throughout. The Hayward regularization parameter $g$ controls the deviation from the Schwarzschild singularity and is related to the magnetic charge distribution arising from the NED sector. The electric charge $Q$ enters through the standard Maxwell contribution to the stress-energy tensor. The CoS parameter $\alpha$, constrained to the range $0 \leq \alpha < 1$, characterizes the string cloud density and induces a solid angle deficit in the asymptotic geometry. Finally, the PFDM parameter $\beta$ governs the dark matter halo intensity and introduces logarithmic corrections to the metric function.

The metric function (\ref{function}) encompasses several well-known BH solutions as limiting cases:

\begin{enumerate}
\item \textbf{Hayward BH with CoS} ($\beta=0$, $Q=0$): The metric reduces to
\begin{equation}
f(r) = 1 - \alpha - \frac{2Mr^2}{r^3 + g^3},\label{function1}
\end{equation}
which describes the Hayward regular BH immersed in a CoS, as studied in Ref.~\cite{FFN2024}.

\item \textbf{Letelier-PFDM BH} ($g=0$): Without the Hayward regularization, we obtain
\begin{equation}
f(r) = 1 - \alpha - \frac{2M}{r}+\frac{Q^2}{r^2}+\frac{\beta}{r}\ln\!\frac{r}{\left|\beta\right|},\label{function2}
\end{equation}
corresponding to the RN BH with CoS surrounded by PFDM \cite{AS2024}.

\item \textbf{Schwarzschild BH} ($\alpha=0$, $\beta=0$, $Q=0$, $g=0$):
\begin{equation}
f(r) = 1 - \frac{2M}{r},\label{function_sch}
\end{equation}
recovering the standard Schwarzschild solution with horizon at $r_h = 2M$.

\item \textbf{Pure Hayward BH} ($\alpha=0$, $\beta=0$, $Q=0$):
\begin{equation}
f(r) = 1 - \frac{2Mr^2}{r^3 + g^3},\label{function_hayward}
\end{equation}
describing the original Hayward regular BH \cite{isz10}.
\end{enumerate}

A comparison between the Hayward and Bardeen regularization mechanisms illuminates their distinct physical origins. The effective mass functions are:
\begin{equation}
\text{Bardeen:} \quad M_{\rm B}(r) = \frac{M r^3}{\left(q^2 + r^2\right)^{3/2}}, \qquad
\text{Hayward:} \quad M_{\rm H}(r) = \frac{M r^3}{r^3 + g^3}.
\label{mass_comparison}
\end{equation}
Both mass functions exhibit the following asymptotic behaviors:
\begin{itemize}
\item As $r \to \infty$: $M_{\rm B}(r) \to M$ and $M_{\rm H}(r) \to M$ (Schwarzschild limit);
\item As $r \to 0$: Both yield $M(r) \propto r^3$, ensuring a dS core with finite curvature.
\end{itemize}
However, the intermediate-range behavior differs: the Bardeen mass function approaches the asymptotic value more slowly due to its $(3/2)$-power dependence, while the Hayward function transitions more rapidly. This distinction leads to measurable differences in the photon sphere radius, shadow size, and QNM spectrum \cite{sec2is08}.

The event horizon(s) are located at the roots of $f(r_h) = 0$:
\begin{equation}
1 - \alpha - \frac{2Mr_h^2}{r_h^3 + g^3} + \frac{Q^2}{r_h^2} + \frac{\beta}{r_h}\ln\left(\frac{r_h}{\left|\beta\right|}\right) = 0.
\label{horizon_eq}
\end{equation}
Due to the transcendental nature of Eq.~(\ref{horizon_eq}), arising from the logarithmic PFDM term, analytical solutions are generally not available. We therefore employ numerical root-finding techniques to classify the horizon structure.

Depending on the parameter values $\{M, g, Q, \alpha, \beta\}$, several distinct configurations emerge. For non-extremal BHs, two distinct horizons exist: an inner (Cauchy) horizon $r_-$ and an outer (event) horizon $r_+$, with $f(r) > 0$ in the region $r > r_+$. In the extremal case, the two horizons merge into a single degenerate horizon $r_{\rm ext}$, which simultaneously satisfies $f(r_{\rm ext}) = 0$ and $f'(r_{\rm ext}) = 0$. Single-horizon BHs possess only one horizon and correspond to the Schwarzschild or Schwarzschild-CoS limits where the Hayward parameter and charges vanish. Finally, when no horizon exists and $f(r) > 0$ everywhere outside the origin, the configuration represents a naked singularity, thereby violating the cosmic censorship conjecture.

Table~\ref{tab:horizons_hayward} presents the numerically computed horizon radii for representative parameter combinations, classifying each configuration. The extremal configurations (highlighted in yellow) were obtained by simultaneously solving $f(r)=0$ and $f'(r)=0$, yielding parameter combinations where the PFDM parameter $\beta$ is finely tuned to produce a degenerate horizon.

\setlength{\tabcolsep}{6pt}
\renewcommand{\arraystretch}{1.5}
\begin{longtable}{|c|c|c|c|c|c|}
\hline
\rowcolor{orange!50}
\textbf{$\alpha$} & \textbf{$\beta/M$} & \textbf{$g/M$} & \textbf{$Q/M$} & \textbf{Horizon(s) $[r_h/M]$} & \textbf{Configuration} \\
\hline
\endfirsthead
\hline
\rowcolor{orange!50}
\textbf{$\alpha$} & \textbf{$\beta/M$} & \textbf{$g/M$} & \textbf{$Q/M$} & \textbf{Horizon(s) $[r_h/M]$} & \textbf{Configuration} \\
\hline
\endhead
0.00 & 0.0000 & 0.00 & 0.00 & $[2.00000]$ & Schwarzschild \\
\hline
0.10 & 0.0000 & 0.00 & 0.00 & $[2.22222]$ & Schwarzschild-CoS \\
\hline
0.00 & 0.0000 & 0.30 & 0.00 & $[0.11984,\ 1.99320]$ & Non-extremal BH \\
\hline
0.10 & 0.0000 & 0.30 & 0.00 & $[0.11314,\ 2.21673]$ & Non-extremal BH \\
\hline
0.00 & 0.5000 & 0.20 & 0.40 & $[0.15192,\ 1.37239]$ & Non-extremal BH \\
\hline
0.00 & 0.8000 & 0.25 & 0.50 & $[0.16198,\ 1.37356]$ & Non-extremal BH \\
\hline
0.10 & 0.6000 & 0.20 & 0.45 & $[0.15488,\ 1.46715]$ & Non-extremal BH \\
\hline
0.10 & 1.0000 & 0.25 & 0.55 & $[0.14172,\ 1.52384]$ & Non-extremal BH \\
\hline
\rowcolor{yellow!30}
0.00 & 2.0441 & 0.10 & 0.30 & $[1.99966]$ & Extremal BH \\
\hline
\rowcolor{yellow!30}
0.00 & 2.1180 & 0.20 & 0.50 & $[1.99718]$ & Extremal BH \\
\hline
\rowcolor{yellow!30}
0.00 & 2.2214 & 0.30 & 0.70 & $[1.99067]$ & Extremal BH \\
\hline
\rowcolor{yellow!30}
0.05 & 2.1298 & 0.15 & 0.45 & $[2.06771]$ & Extremal BH \\
\hline
\rowcolor{yellow!30}
0.10 & 2.1831 & 0.20 & 0.50 & $[2.13952]$ & Extremal BH \\
\hline
\rowcolor{yellow!30}
0.10 & 2.2295 & 0.25 & 0.60 & $[2.13672]$ & Extremal BH \\
\hline
\rowcolor{yellow!30}
0.15 & 2.2428 & 0.20 & 0.55 & $[2.21902]$ & Extremal BH \\
\hline
0.20 & 2.5000 & 0.30 & 0.80 & $[0.07258,\ 2.34896]$ & Non-extremal BH \\
\hline
0.25 & 2.8000 & 0.35 & 0.90 & $[0.08233,\ 2.56549]$ & Non-extremal BH \\
\hline
0.00 & 0.3000 & 0.80 & 1.20 & $[\,]$ & Naked singularity \\
\hline
0.05 & 0.4000 & 0.90 & 1.50 & $[\,]$ & Naked singularity \\
\hline
\caption{Horizon radii $r_h/M$ for the charged Hayward BH with CoS and PFDM. The Schwarzschild and Schwarzschild-CoS cases exhibit single horizons at $r_h = 2M/(1-\alpha)$. Non-extremal configurations display two distinct horizons $[r_-,\, r_+]$ corresponding to inner (Cauchy) and outer (event) horizons. Extremal configurations (highlighted in yellow) satisfy both $f(r_{\rm ext})=0$ and $f'(r_{\rm ext})=0$, showing a single degenerate horizon. For large Hayward parameter $g$ and charge $Q$, no horizon exists, resulting in a naked singularity.}
\label{tab:horizons_hayward}
\end{longtable}

Several important features emerge from Table~\ref{tab:horizons_hayward}. The CoS parameter $\alpha$ increases the outer horizon radius: for the Schwarzschild case, $r_h$ grows from $2M$ at $\alpha=0$ to $2.222M$ at $\alpha=0.1$, following the relation $r_h \approx 2M/(1-\alpha)$. The PFDM parameter $\beta$ introduces a logarithmic correction that, when sufficiently large ($\beta/M \gtrsim 2$), can produce extremal configurations with $r_{\rm ext} \approx 2M$. Achieving extremal BHs requires a delicate balance among the parameters: increasing $\alpha$ demands correspondingly larger $\beta$ to maintain extremality, as evidenced by the progression from $\beta = 2.044$ at $\alpha=0$ to $\beta = 2.243$ at $\alpha=0.15$. Finally, naked singularities arise when the Hayward parameter $g$ and charge $Q$ become large relative to the mass, thereby preventing horizon formation.

\begin{figure}[ht!]
\centering
\includegraphics[width=0.88\textwidth]{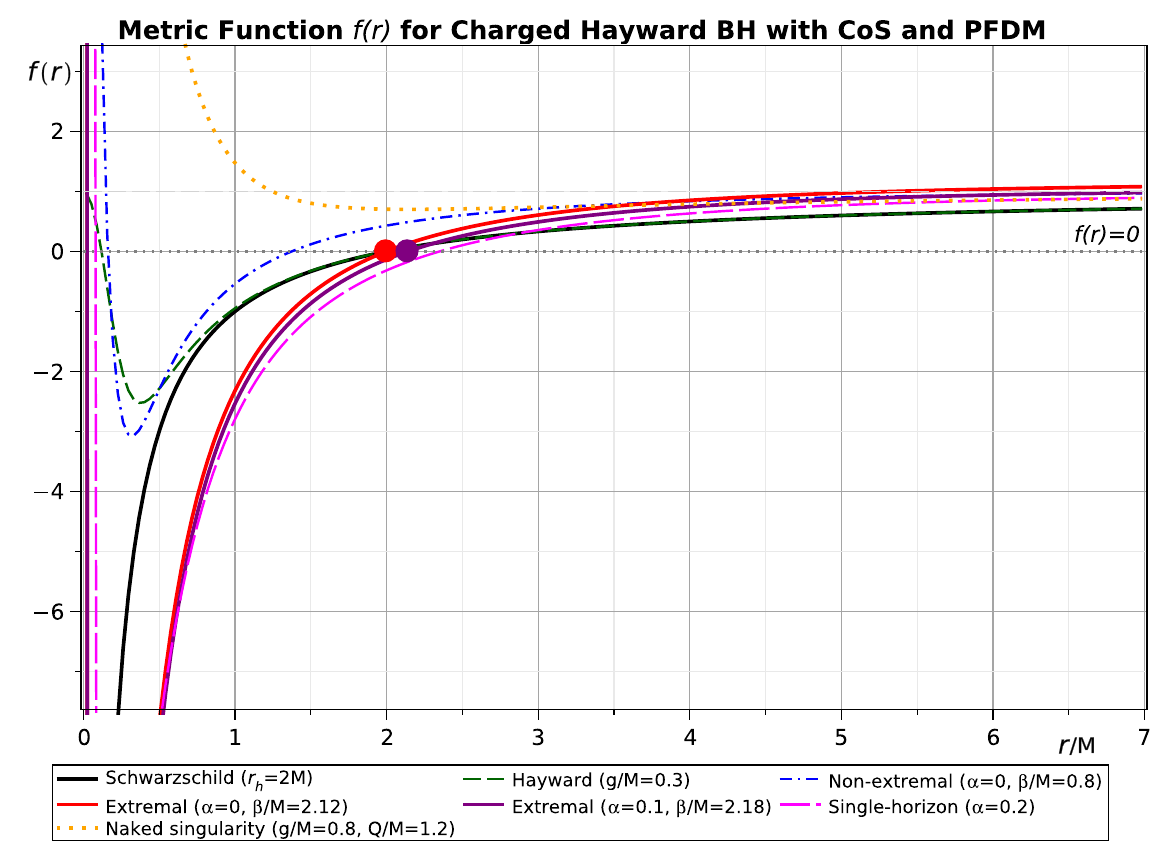}
\caption{Metric function $f(r)$ versus $r/M$ for the charged Hayward BH with CoS and PFDM. The curves illustrate distinct horizon configurations: \textbf{Schwarzschild} (black solid, $r_h=2M$) serves as the reference; \textbf{Hayward} (dark green dashed, $g/M=0.3$) shows the regular BH with two horizons; \textbf{Non-extremal} (blue dash-dotted, $\alpha=0$, $\beta/M=0.8$) crosses $f(r)=0$ at two distinct radii $r_-$ and $r_+$; \textbf{Extremal} cases (red and purple solid curves) touch $f(r)=0$ tangentially at $r_{\rm ext} \approx 2M$, with filled circles marking the degenerate horizons; \textbf{Single-horizon} (magenta long-dashed, $\alpha=0.2$) exhibits one crossing; \textbf{Naked singularity} (orange dotted, $g/M=0.8$, $Q/M=1.2$) remains positive for all $r>0$. The gray horizontal line marks $f(r)=0$. Note that curves with $\alpha>0$ asymptote to $f(\infty) = 1-\alpha < 1$, reflecting the non-flat nature of CoS spacetimes.}
\label{fig:metric_function}
\end{figure}

The behavior of the metric function $f(r)$ for selected configurations is displayed in Fig.~\ref{fig:metric_function}. 

The plot demonstrates how the parameters modify the horizon structure: increasing $\alpha$ shifts the asymptotic value of $f(r)$ from unity to $(1-\alpha)$, while increasing $\beta$ introduces logarithmic corrections that can either create or destroy horizons depending on the charge parameters. The extremal cases (red and purple curves) are distinguished by their tangential contact with the $f(r)=0$ line, in contrast to the transverse crossings of non-extremal configurations.

The metric function exhibits the following asymptotic behavior:
\begin{equation}
\lim_{r \to \infty} f(r) = 1 - \alpha \neq 1 \quad (\text{for } \alpha > 0).
\label{asymptotic}
\end{equation}
This indicates that the spacetime is asymptotically non-flat when a CoS is present. The deficit solid angle induced by the string cloud modifies the spatial geometry at radial infinity, with implications for the shadow calculation and distant observer measurements \cite{isz34}.

Near the origin, we have
\begin{equation}
\lim_{r \to 0} f(r) = 1 - \alpha + \frac{Q^2}{r^2} + \frac{\beta}{r}\ln\frac{r}{|\beta|} - \frac{2M}{g^3}r^2 + \mathcal{O}(r^3).
\label{near_origin}
\end{equation}
The $r^2$ dependence from the Hayward term ensures a dS-like core, while the electric charge and PFDM contributions introduce divergent terms that must be balanced for physical solutions.

To examine the curvature regularity, we compute the Kretschmann scalar $K = R_{\mu\nu\rho\sigma}R^{\mu\nu\rho\sigma}$. For the metric (\ref{metric}), this takes the form \cite{sec2is10}
\begin{equation}
K = \frac{4}{r^4}\left[(rf'(r) - 2f(r) + 2)^2 + (rf'(r))^2 + r^2(f''(r))^2\right].
\label{kretschmann}
\end{equation}
Substituting the metric function (\ref{function}) and taking the limit $r \to 0$, we find
\begin{equation}
\lim_{r \to 0} K = \frac{96 M^2}{g^6} < \infty \quad (\text{for } g > 0),
\label{kretschmann_limit}
\end{equation}
confirming that the Hayward regularization parameter $g$ removes the central singularity present in the Schwarzschild and RN solutions. The spacetime is geodesically complete for appropriate parameter choices \cite{sec2is04}.

\begin{figure}[ht!]
\centering
\begin{tabular}{cc}
\includegraphics[width=0.4\textwidth]{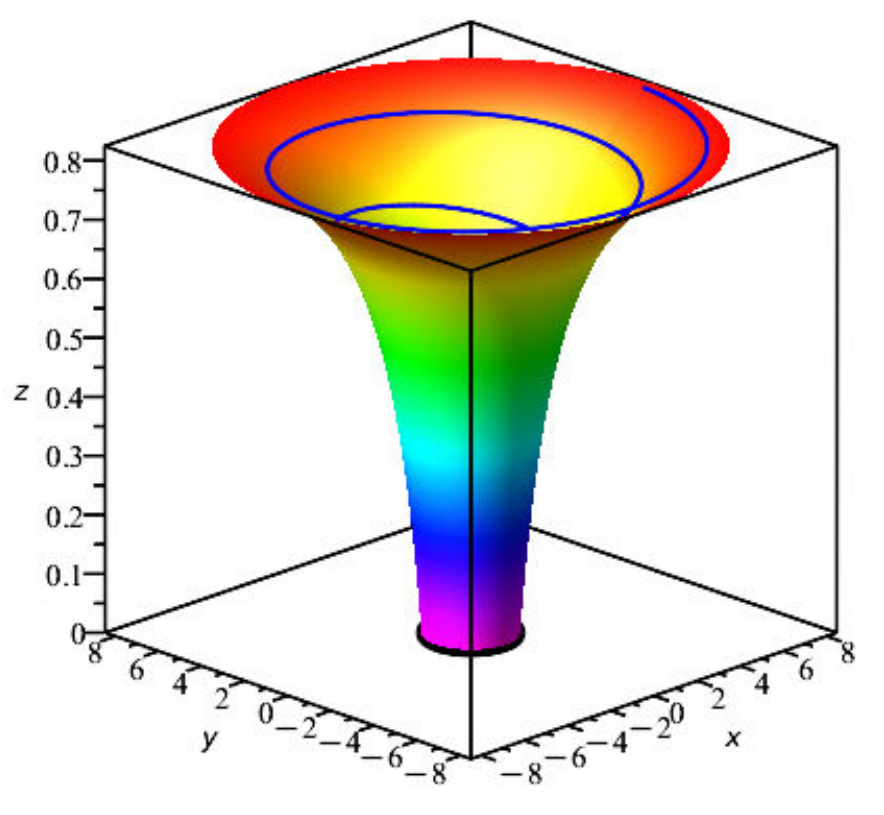} &
\includegraphics[width=0.4\textwidth]{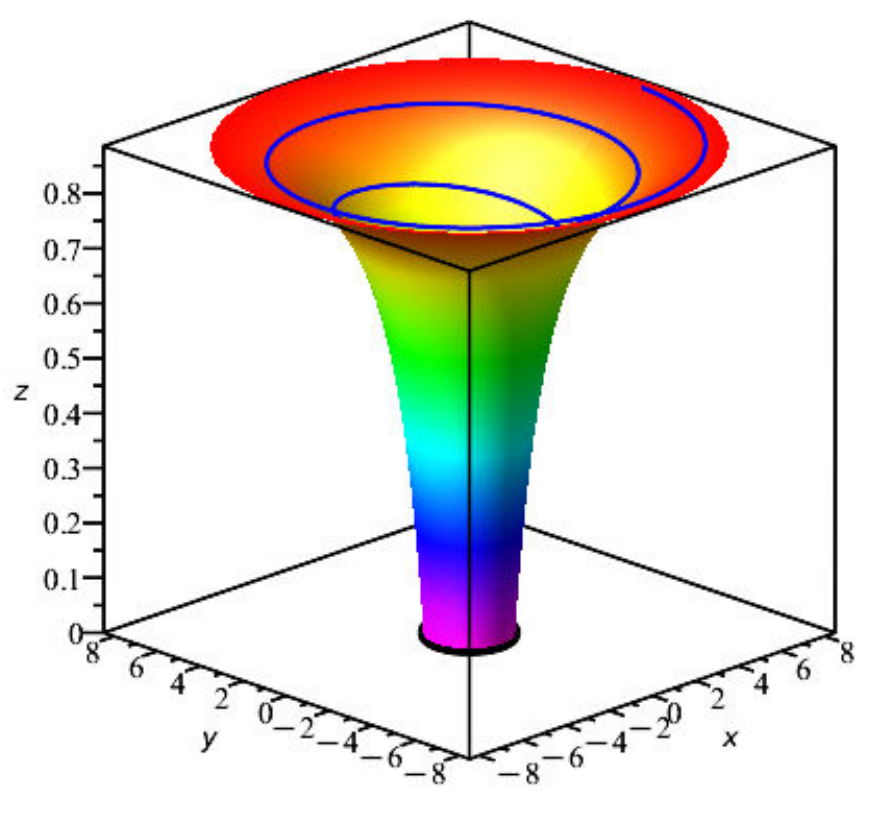} \\
(a) & (b) \\[2ex]
\includegraphics[width=0.4\textwidth]{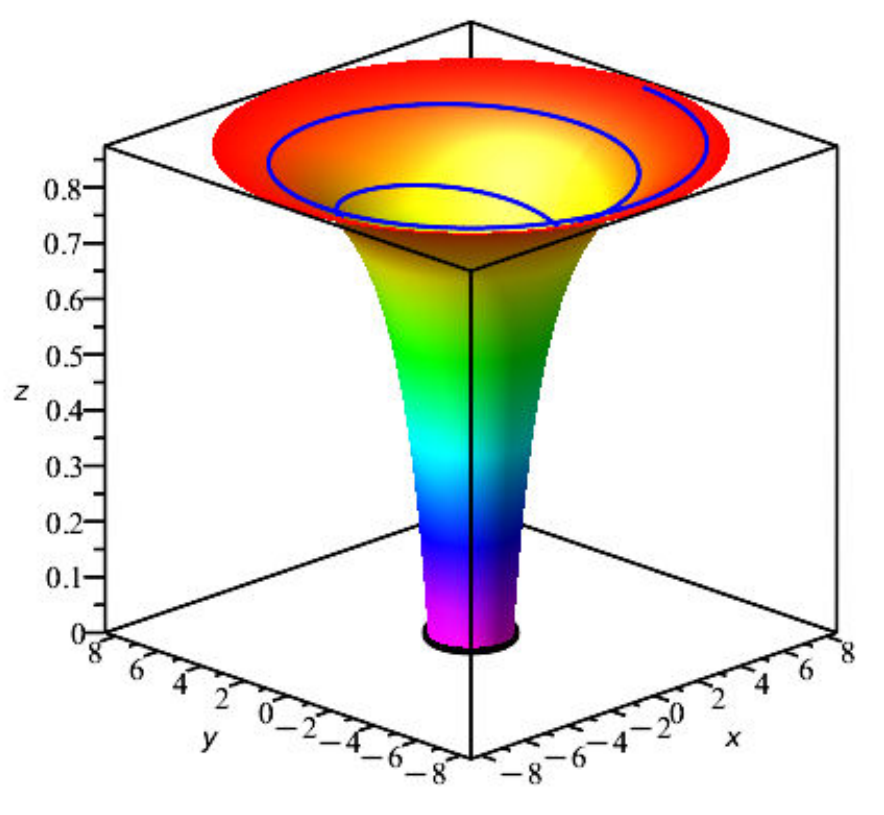} &
\includegraphics[width=0.4\textwidth]{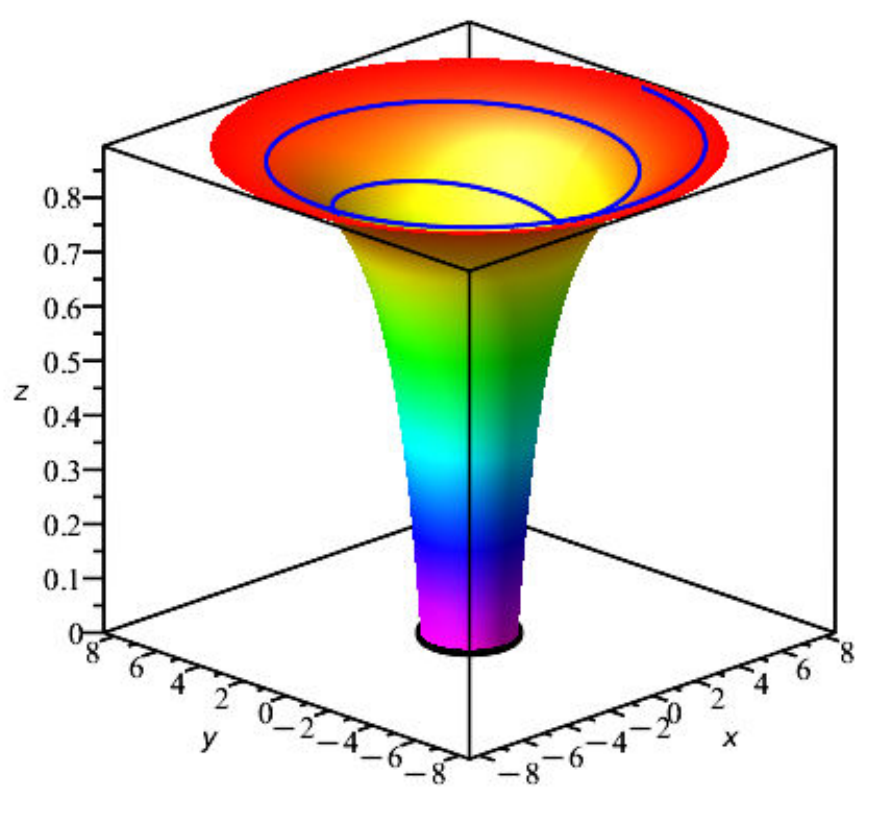} \\
(c) & (d) \\
\end{tabular}
\caption{Three-dimensional visualization of the metric function $f(r)$ for the charged Hayward BH with CoS and PFDM. The panels correspond to: \textbf{(a)} $\alpha=0.05$, $\beta/M=0.3$, $g/M=0.2$, $Q/M=0.3$; \textbf{(b)} $\alpha=0.1$, $\beta/M=0.8$, $g/M=0.25$, $Q/M=0.6$; \textbf{(c)} $\alpha=0.1$, $\beta/M=0.72$, $g/M=0.25$, $Q/M=0.68$; \textbf{(d)} $\alpha=0.15$, $\beta/M=1.2$, $g/M=0.3$, $Q/M=0.8$. The event horizon is located where the surface intersects the $z=0$ plane. The blue spiral curves depict representative infalling trajectories.}
\label{fig:3d_metric}
\end{figure}

To gain deeper insight into the spacetime structure, we present three-dimensional visualizations of the metric function $f(r)$ for the charged Hayward BH with CoS and PFDM. Figure~\ref{fig:3d_metric} displays the metric function as a surface of revolution, where the radial coordinate $r$ is mapped to the horizontal plane $(x,y) = (r\cos\phi, r\sin\phi)$ and the vertical axis represents $z = f(r)$. This representation provides an intuitive picture of how the gravitational potential varies with radial distance across different parameter regimes.

The three-dimensional representations reveal several characteristic features of the spacetime geometry. The surface intersects the $z=0$ plane at the event horizon radius $r = r_+$, where $f(r_+) = 0$, forming a characteristic throat structure. As the radial coordinate increases, the surface rises and asymptotically approaches $z = 1 - \alpha$, reflecting the non-flat nature of spacetimes with a CoS. The curvature of the surface near the throat becomes more pronounced with increasing electric charge $Q$ and PFDM parameter $\beta$, indicating stronger deviations from Schwarzschild geometry. Comparing panels (a) and (d), one observes that larger values of $\alpha$ reduce the asymptotic height, consistent with the solid angle deficit induced by the string cloud. The blue spiral trajectories represent sample infalling geodesics, illustrating how test particles approach the horizon with increasingly tight orbital winding for larger angular momentum.

The surface gravity at the outer horizon $r_+$ is defined as
\begin{equation}
\kappa_+ = \frac{{\color{black} C}}{2}\left|f'(r_+)\right|,\qquad {\color{black} C=1/\sqrt{\lim_{r \to \infty} f(r)}},
\label{surface_gravity}
\end{equation}
yielding the Hawking temperature
\begin{equation}
T_H = \frac{\kappa_+}{2\pi} = \frac{{\color{black} C}}{4\pi}\left|f'(r_+)\right|.
\label{hawking_temp}
\end{equation}
Evaluating the derivative of the metric function (\ref{function}) at $r = r_+$, we obtain
\begin{equation}
T_H = \frac{{\color{black} (1-\alpha)^{-1/2}}}{4\pi}\left|\frac{2M r_+(r_+^3 - 2g^3)}{(r_+^3 + g^3)^2} - \frac{2Q^2}{r_+^3} + \frac{\beta}{r_+^2}\left(1 - \ln\frac{r_+}{|\beta|}\right)\right|.
\label{hawking_temp_explicit}
\end{equation}
For extremal BHs, $T_H = 0$, corresponding to the condition $f'(r_{\rm ext}) = 0$. The extremal configurations identified in Table~\ref{tab:horizons_hayward} satisfy this criterion, with the PFDM logarithmic term providing the necessary balance to achieve zero temperature. The interplay between the Hayward parameter $g$, electric charge $Q$, and PFDM parameter $\beta$ determines whether a given configuration can achieve extremality. 

{\color{black}
The horizon area for a static distant observer is given by
\begin{equation}
    \mathcal{A}=\lim_{r \to r_{+}}\frac{1}{C^2}\int \int \sqrt{g_{\theta\theta}\,g_{\phi\phi}} d\theta d\phi=\frac{4 \pi r^2_{+}}{C^2}=4\pi (1-\alpha)\,r^2_{+}<4\pi r^2_{+}.
\end{equation}
which is less than the Bekestein-Hawking area law due to the presence of string cloud in the space-time since the lapse function at radial infinity is asymptotically bounded rather than flat space (\ref{asymptotic}).
}

\section{Optical Properties of BH: Photon Sphere and Shadow} \label{isec3}

In this section, we investigate the null geodesics around the charged Hayward BH spacetime with CoS and PFDM, with particular focus on several key features: the effective potential, the radial force experienced by photons, the photon sphere, the BH shadow, and the trajectories of photons \cite{isz32,FrolovZelnikov2011,sec2is10}. Our aim is to illustrate how the various parameters characterizing the BH geometry influence these optical features and observational signatures.

\subsection{Equations of Motion}

The study of null geodesics is carried out within the framework of the Lagrangian formalism. In this approach, the motion of a test particle (in this case, a photon) is derived from the Lagrangian density, which is expressed in terms of the spacetime metric tensor \(g_{\mu\nu}\) as \cite{gm1,gm2,gm3,gm4}
\begin{equation}
\mathbb{L} = \frac{1}{2} g_{\mu\nu} \dot{x}^\mu \dot{x}^\nu,\label{bb1}
\end{equation}
where the overdot denotes differentiation with respect to an affine parameter along the geodesic. This formalism allows us to obtain the conserved quantities associated with the spacetime symmetries and to derive the equations governing photon motion, thereby providing information about how the BH parameters affect the properties of null trajectories.

The Lagrangian for geodesic motion takes the form
\begin{equation}
2\mathbb{L} = -f(r)\dot{t}^2 + \frac{\dot{r}^2}{f(r)} + r^2\dot{\theta}^2 + r^2\sin^2\theta\,\dot{\phi}^2.\label{bb2}
\end{equation}

For null geodesics confined to the equatorial plane ($\theta = \pi/2$), there exist two conserved quantities associated with the coordinates ($t, \phi$). These conserved quantities are given by
\begin{equation}
\mathrm{E} = f(r)\dot{t}, \qquad \mathrm{L} = r^2\dot{\phi},\label{bb3}
\end{equation}
representing the energy and angular momentum of the photon, respectively.

With these conserved quantities, the equation of motion for the radial coordinate $r$ in the case of null geodesics is given by
\begin{equation}
    \dot r^2+V_{\rm eff}=\mathrm{E}^2,\label{bb3a}
\end{equation}
where the effective potential that governs the photon dynamics reads
\begin{equation}
    V_{\rm eff}=\frac{\mathrm{L}^2}{r^2} f(r)=\frac{\mathrm{L}^2}{r^2}\left(1 - \alpha - \frac{2mr^2}{r^3 + g^3} + \frac{Q^2}{r^2} + \frac{\beta}{r}\ln\!\frac{r}{\left|\beta\right|}\right).\label{bb3b}
\end{equation}

\begin{figure}[ht!]
    \centering
    \includegraphics[width=0.45\linewidth]{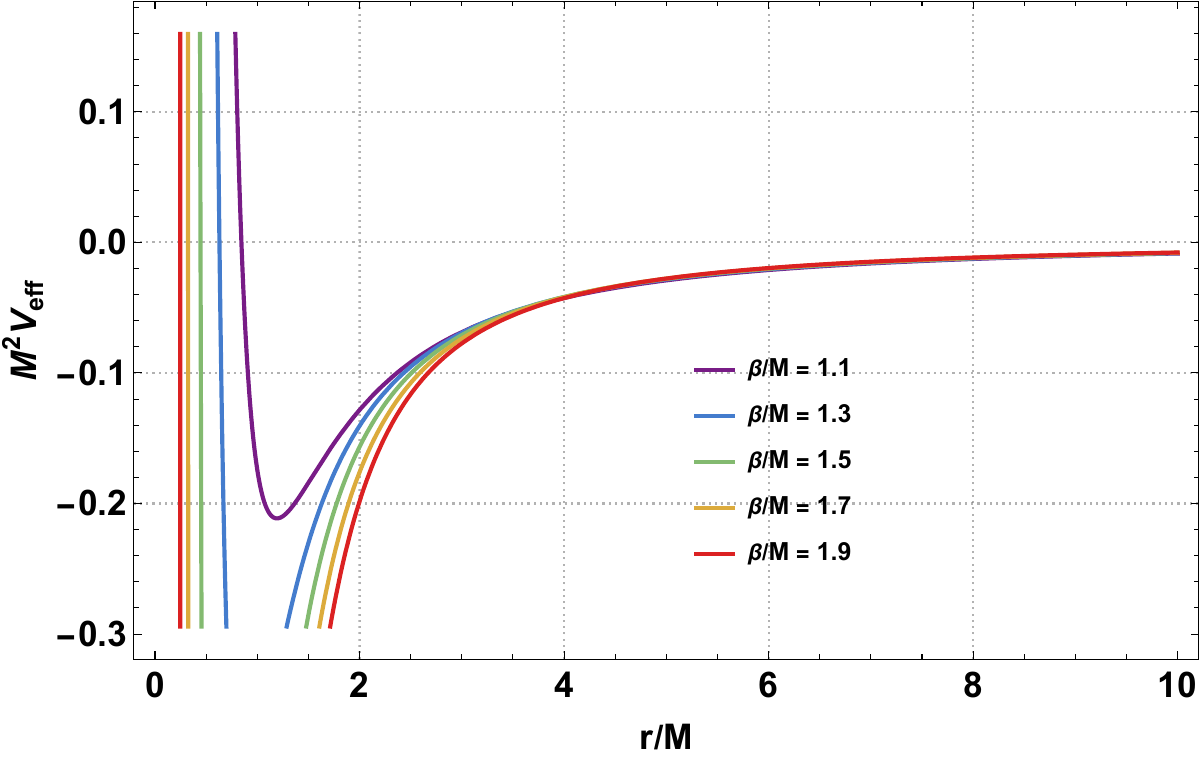}\qquad
    \includegraphics[width=0.45\linewidth]{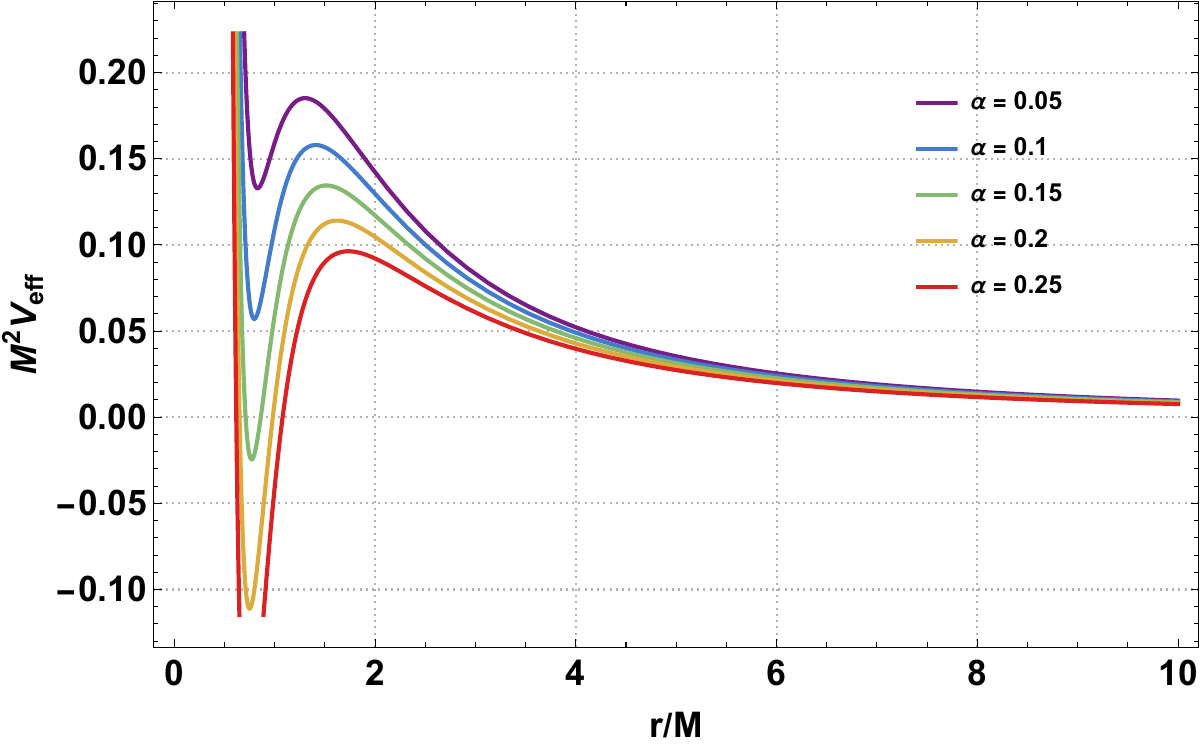}\\
    (a) $\alpha=0.1$ \hspace{6cm} (b) $\beta/M=0.8$
    \caption{Behavior of the effective potential governing photon dynamics as a function of dimensionless radial distance $r/M$ for various $\beta/M$ and $\alpha$. Here $Q/M=1,\,g/M=0.25,\,\mathrm{L}/M=1$.}
    \label{fig:null}
\end{figure}

From Eq.~(\ref{bb3b}), it is evident that the effective potential governing photon dynamics depends explicitly on various geometric parameters that modify the spacetime curvature. These include the CoS parameter $\alpha$, the PFDM parameter $\beta$, the Hayward parameter $g$, as well as the electric charge $Q$ and mass $M$ of the BH. Each of these parameters alters the spacetime geometry and, consequently, reshapes the effective potential experienced by photons. As a result, variations in these parameters lead to notable changes in the location and stability of the photon sphere, the corresponding shadow radius, and the overall photon trajectories in the gravitational field. In particular, the combined effects of the Hayward parameter, CoS, and surrounding PFDM can either enhance or suppress the gravitational attraction, thereby affecting light bending, capture cross sections, and shadow observables.

In Figure~\ref{fig:null}, we plot the effective potential given in Eq.~(\ref{bb3b}) as a function of the dimensionless radial distance $r/M$, varying the PFDM parameter $\beta$ and the CoS parameter $\alpha$, while keeping the charges and angular momentum fixed. We observe that increasing the values of $\beta/M$ and $\alpha$ leads to a suppression of the peak of the potential, indicating a reduction in the gravitational barrier experienced by photons. This behavior suggests that photons encounter a weaker potential barrier in spacetimes with stronger PFDM influence or higher string cloud density.

\subsection{Photon Sphere}

The photon sphere is a critical region around a BH where gravity is strong enough to force photons to move along unstable circular orbits. Its radius is determined by the spacetime geometry and the BH parameters, and it plays a key role in shaping the BH shadow and influencing the trajectories of light. The stability of these orbits is delicate: any small radial perturbation can cause the photon to either fall into the BH or escape to infinity. Therefore, studying the photon sphere provides essential information about gravitational lensing and observational signatures of BHs.

For circular null orbits, we impose the conditions $\dot r=0$ and $\ddot r=0$. With these conditions, one can determine the photon sphere radius $r_{s}$ from the following relation \cite{isz32,FrolovZelnikov2011,sec2is10}:
\begin{equation}
\frac{d}{dr}\left(\frac{r^2}{f(r)}\right)\bigg|_{r=r_{s}} = 0,\label{bb4}
\end{equation}
which yields the condition
\begin{equation}
2\,f(r_{s}) - r_{s}\,f'(r_{s}) = 0.\label{bb5}
\end{equation}

Substituting the metric function and after simplification, we obtain the following polynomial relation in $r$:
\begin{equation}
    1-\alpha-\frac{3 M r^5}{(r^3+g^3)^2}+\frac{2 Q^2}{r^2}+\frac{\beta}{2 r} \left(3 \ln\!\frac{r}{|\beta|}-1\right)=0.\label{bb6}
\end{equation}
An analytical solution of the above polynomial equation would, in principle, provide the photon sphere radius \(r = r_s\). However, obtaining an exact analytical solution is highly challenging due to the presence of the logarithmic term arising from the PFDM contribution. Nevertheless, one can determine the photon sphere radius numerically by choosing appropriate values of the CoS parameter \(\alpha\), the PFDM parameter \(\beta\), the Hayward parameter \(g\), and the electric charge \(Q\). This numerical approach allows us to explore how these parameters influence the location of the photon sphere and the associated properties of null geodesics.

\begin{table}[ht!]
\centering
\caption{Numerical values of photon sphere radius per unit mass $r_s/M$ for different $\alpha$ and $\beta$. Here $Q/M=1,\,g/M=0.2$.}
\begin{tabular}{|c|cccccc|}
\hline
$\alpha (\downarrow) \backslash \beta/M (\rightarrow)$ & 0.8 & 1.0 & 1.2 & 1.4 & 1.6 & 1.8 \\
\hline
0.05 & 0.75885 & 0.55740 & 1.83090 & 2.03018 & 2.21982 & 2.40375 \\
\hline
0.10 & 0.73392 & 1.68865 & 1.90251 & 2.10067 & 2.29072 & 2.47586 \\
\hline
0.15 & 0.71358 & 1.77043 & 1.97946 & 2.17636 & 2.36673 & 2.55305 \\
\hline
0.20 & 0.69632 & 1.85845 & 2.06247 & 2.25791 & 2.44848 & 2.63592 \\
\hline
0.25 & 1.74530 & 1.95371 & 2.15242 & 2.34613 & 2.53673 & 2.72520 \\
\hline
\end{tabular}
\label{tab:1}
\end{table}

\begin{table}[ht!]
\centering
\caption{Numerical values of photon sphere radius per unit mass $r_s/M$ for different $\alpha$ and $\beta$. Here $Q/M=1,\,g/M=0.4$.}
\begin{tabular}{|c|cccccc|}
\hline
$\alpha (\downarrow) \backslash \beta/M (\rightarrow)$ & 0.8 & 1.0 & 1.2 & 1.4 & 1.6 & 1.8 \\
\hline
0.05 & 1.20786 & 0.82598 & 1.78716 & 2.00248 & 2.20056 & 2.38958 \\
\hline
0.10 & 1.20786 & 1.61498 & 1.86257 & 2.07485 & 2.27257 & 2.46242 \\
\hline
0.15 & 1.11153 & 0.78973 & 1.94300 & 2.15232 & 2.34964 & 2.54031 \\
\hline
0.20 & 1.01986 & 1.80267 & 2.02922 & 2.23556 & 2.43243 & 2.62387 \\
\hline
0.25 & 0.97072 & 1.90486 & 2.12213 & 2.32540 & 2.52168 & 2.71382 \\
\hline
\end{tabular}
\label{tab:2}
\end{table}

\begin{figure}[ht!]
    \centering
    \includegraphics[width=0.45\linewidth]{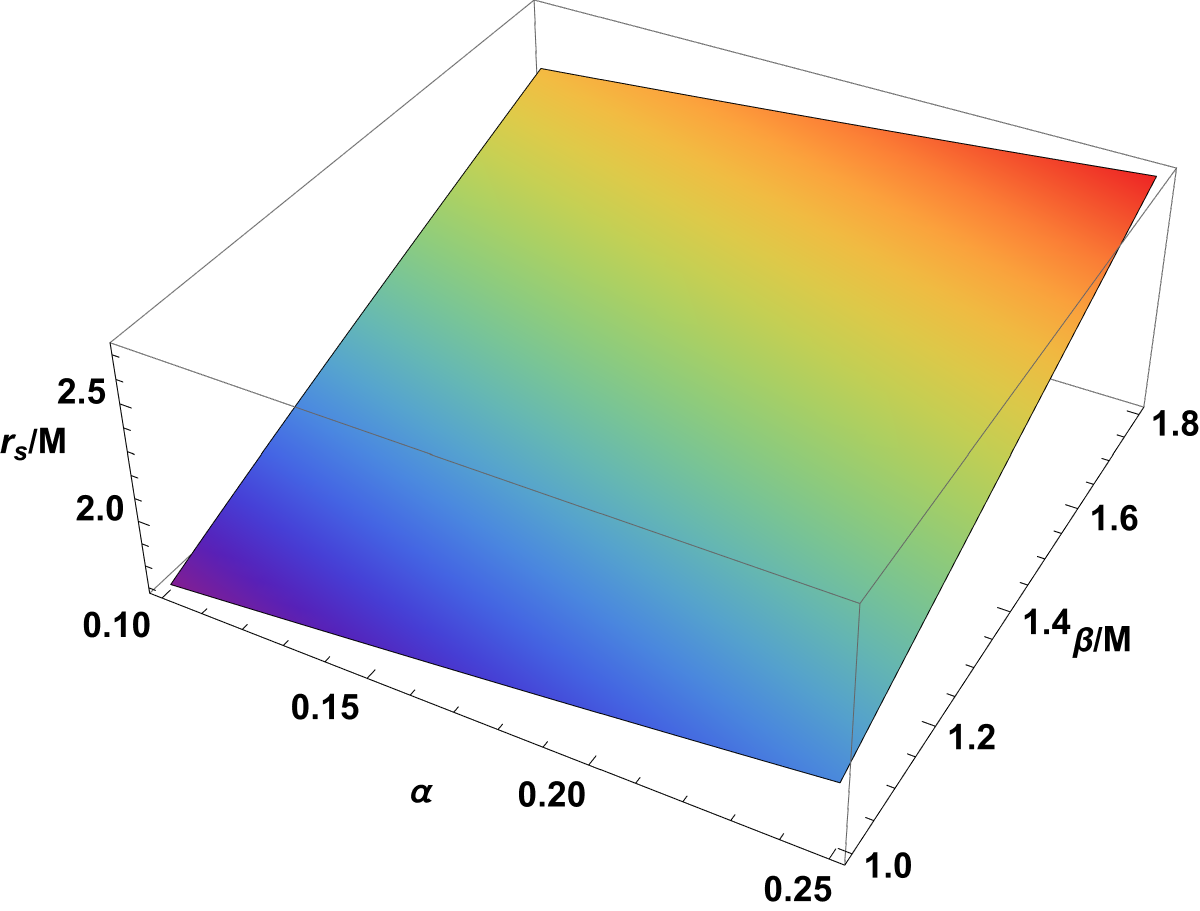}\qquad
    \includegraphics[width=0.45\linewidth]{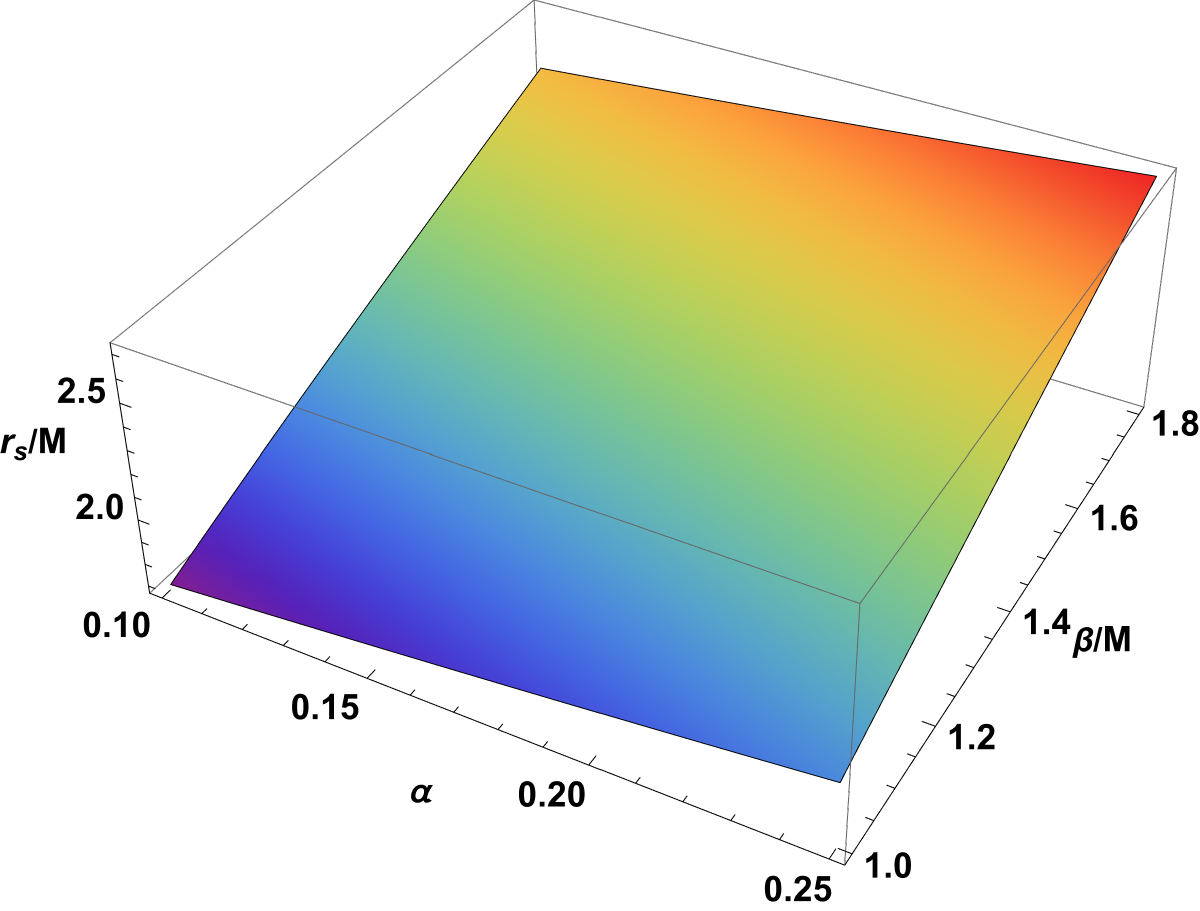}\\
    (a) $g/M=0.1$ \hspace{6cm} (b) $g/M=0.2$
    \caption{Photon sphere radius as a function of $\alpha$ and $\beta/M$ for two values of the Hayward parameter $g$. Here $Q/M=1$.}
    \label{fig:photon-sphere}
\end{figure}

In Tables~\ref{tab:1}--\ref{tab:2}, we present numerical values of the photon sphere radius per unit mass $r_s/M$ for two values of the Hayward parameter $g/M=0.2$ and $0.4$ by selecting suitable values of the aforementioned parameters. The tables reveal that the photon sphere radius generally increases with both $\alpha$ and $\beta$, though the dependence is non-monotonic in certain parameter regimes due to the competing effects of the various terms in the metric function.

In Figure~\ref{fig:photon-sphere}, we present a 3D plot of the dimensionless photon sphere radius $r_{s}/M$ as a function of the CoS parameter $\alpha \in [0.05, 0.25]$ and the PFDM parameter per unit mass $\beta/M \in [1, 1.8]$ for two values of the Hayward parameter $g/M=0.1$ and $0.2$, while keeping all other parameters fixed. The surface plots clearly demonstrate how the combined effects of CoS and PFDM modify the photon sphere location compared to the standard Hayward BH case.

\subsection{Shadow Radius}

The shadow of a BH is the dark region on the observer's sky, resulting from the capture of photons by the event horizon. Its shape and size are determined by the BH's mass, spin, charge, and the surrounding matter or fields. The EHT collaboration released the iconic image of the supermassive BH M87*, showing a bright emission ring surrounding a central dark region corresponding to the BH shadow \cite{EHTL1,EHTL4,EHTL6}. More recently, the EHT collaboration presented the first image of Sgr~A*, the supermassive BH at the center of our Milky Way, confirming the presence of a shadow consistent with predictions from GR \cite{EHTL12,EHTL14,EHTL15,EHTL16,EHTL17}. These observations not only provide direct visual evidence for the existence of BHs, but also allow detailed tests of gravitational physics in the strong-field regime and constrain the properties of the BH, including mass, spin, and the surrounding accretion flow.

To analyze the BH shadow, we first examine the behavior of the metric function \(f(r)\) at large distances. In the limit \(r \to \infty\), the metric function approaches
\begin{equation}
    \lim_{r \to \infty} f(r) = 1 - \alpha \neq 1, \label{bb7}
\end{equation}
indicating that the spacetime is asymptotically non-flat when CoS is present. In such a scenario, the BH shadow can be determined for an observer located at a finite radial position \(r_O\), and its properties are described following the formalism developed in \cite{isz34}. The shadow radius is given by \cite{isz34} 
\begin{equation}
    R_{\rm sh}=r_{s} \sqrt{\frac{f(r_O)}{f(r_s)}}.\label{bb8}
\end{equation}
Substituting the metric function, we find
\begin{equation}
    R_{\rm sh}=r_{s} \sqrt{\frac{1 - \alpha - \frac{2Mr^2_O}{g^3 + r^2_O} + \frac{Q^2}{r^2_O} + \frac{\beta}{r_O}\ln\!\frac{r_O}{|\beta|}}{1 - \alpha - \frac{2Mr^2_{s}}{g^3 + r^3_{s}} + \frac{Q^2}{r^2_{s}} + \frac{\beta}{r_{s}}\ln\!\frac{r_{s}}{|\beta|}}}.\label{bb9}
\end{equation}
For a static distant observer ($r_O \to \infty$), the BH shadow radius simplifies to
\begin{equation}
    R_{\rm sh}=(1 - \alpha)^{1/2}\,\frac{r_{s} }{\sqrt{1 - \alpha - \frac{2Mr^2_{s}}{g^3 + r^3_{s}} + \frac{Q^2}{r^2_{s}} + \frac{\beta}{r_{s}}\ln\!\frac{r_{s}}{|\beta|}}}.\label{bb10}
\end{equation}

For M87*, the angular diameter of the shadow is measured to be $\theta_{\rm sh}^{\rm M87*} = (42 \pm 3)\,\mu{\rm as}$, while its distance and mass are $D_{\rm M87*} = 16.8\,{\rm Mpc}$ and $M_{\rm M87*} = (6.5 \pm 0.7)\times10^{9}\,M_\odot$, respectively \cite{EHTL1,EHTL4,EHTL6}. Using these values, the dimensionless shadow diameter can be defined as $d_{\rm sh} = D\,\theta_{\rm sh}/M$, yielding $d_{\rm sh}^{\rm M87*} \simeq 11.0 \pm 1.5$. For the Galactic center source Sgr~A*, EHT observations report an angular shadow diameter of $\theta_{\rm sh}^{\rm Sgr\,A*} = 51.8 \pm 2.3\,\mu{\rm as}$, together with a mass $M_{\rm Sgr\,A*} = (4.297 \pm 0.013)\times10^{6}\,M_\odot$ and a distance $D_{\rm Sgr\,A*} = 8277 \pm 9 \pm 33\,{\rm pc}$ \cite{EHTL12,EHTL14,EHTL15,EHTL16,EHTL17}. Applying the same definition for the dimensionless shadow diameter, one obtains $d_{\rm sh}^{\rm Sgr\,A*} \simeq 9.5 \pm 1.4$.

\begin{figure}[ht!]
    \centering
    \includegraphics[width=0.45\linewidth]{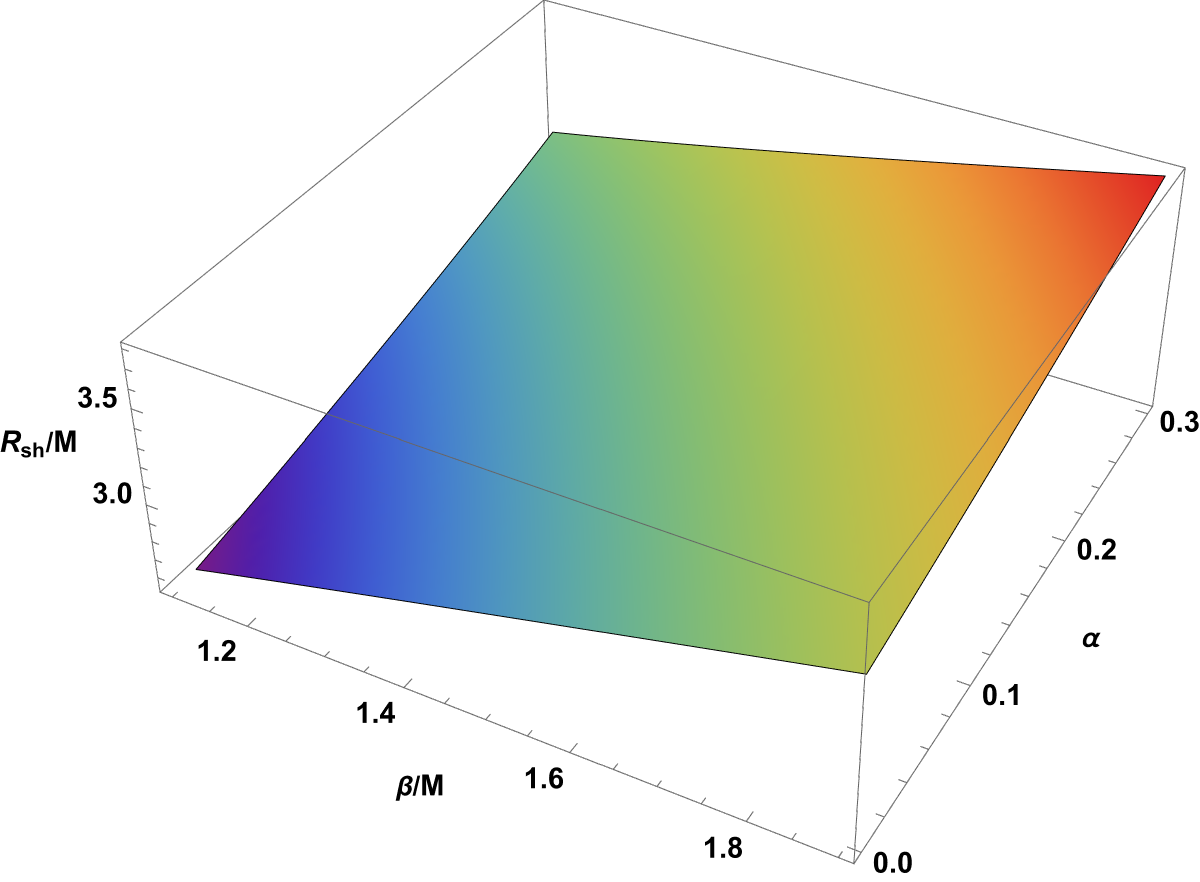}\qquad
    \includegraphics[width=0.45\linewidth]{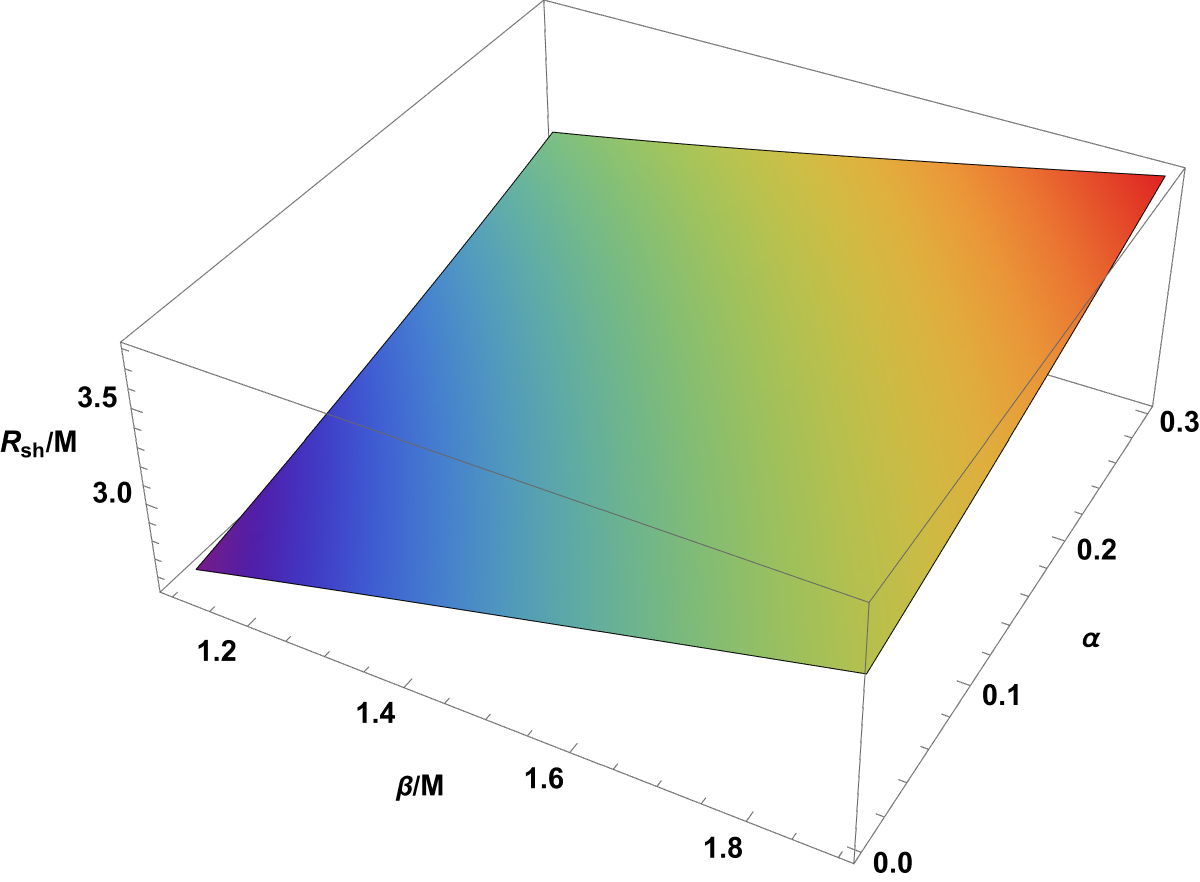}\\
    (a) $g/M=0.1$ \hspace{6cm} (b) $g/M=0.2$
    \caption{Shadow radius as a function of $\alpha$ and $\beta/M$ for two values of the Hayward parameter $g$. Here $Q/M=1$.}
    \label{fig:shadow}
\end{figure}

Figure~\ref{fig:shadow} displays the shadow radius $R_{\rm sh}$ as a function of $\alpha$ and $\beta/M$ for two values of the Hayward parameter. The shadow size increases with both $\alpha$ and $\beta$, reflecting the modifications to the spacetime geometry induced by the CoS and PFDM. These results can be used to constrain the BH parameters through comparison with EHT observations.

\begin{table}[ht]
\centering
\caption{Shadow radius $R_{\rm sh}$ for different values of $\alpha$ and $\beta$. Here $Q/M=1,\,g/M=0.1$.}
\begin{tabular}{|c|cccccc|}
\hline
$\alpha (\downarrow) \backslash \beta/M (\rightarrow )$ & 1.0 & 1.2 & 1.4 & 1.6 & 1.8 & 2.0 \\
\hline
0.10 & 2.60963 & 2.79876 & 2.98718 & 3.17594 & 3.36519 & 3.55489 \\
0.15 & 2.71202 & 2.89135 & 3.07404 & 3.25922 & 3.44621 & 3.63449 \\
0.20 & 2.82175 & 2.99020 & 3.16631 & 3.34728 & 3.53148 & 3.71793 \\
0.25 & 2.93982 & 3.09604 & 3.26455 & 3.44052 & 3.62132 & 3.80543 \\
0.30 & 3.06738 & 3.20970 & 3.36936 & 3.53937 & 3.71602 & 3.89718 \\
\hline
\end{tabular}
\label{tab:sha1}
\end{table}

\begin{table}[ht]
\centering
\caption{Shadow radius $R_{\rm sh}$ for different values of $\alpha$ and $\beta$. Here $Q/M=1,\,g/M=0.2$.}
\begin{tabular}{|c|cccccc|}
\hline
$\alpha (\downarrow) \backslash \beta/M (\rightarrow )$ & 1.0 & 1.2 & 1.4 & 1.6 & 1.8 & 2.0 \\
\hline
0.10 & 2.60379 & 2.79521 & 2.98478 & 3.17422 & 3.36391 & 3.55390 \\
0.15 & 2.70677 & 2.88807 & 3.07180 & 3.25761 & 3.44499 & 3.63355 \\
0.20 & 2.81705 & 2.98719 & 3.16423 & 3.34577 & 3.53034 & 3.71704 \\
0.25 & 2.93560 & 3.09328 & 3.26262 & 3.43910 & 3.62024 & 3.80459 \\
0.30 & 3.06362 & 3.20717 & 3.36757 & 3.53804 & 3.71500 & 3.89639 \\
\hline
\end{tabular}
\label{tab:sha2}
\end{table}

\subsection{Geodesic Angular Velocity}

We determine the geodesic angular velocity (orbital velocity) on circular null orbits and analyze how various parameters alter this quantity. Mathematically, it is defined by \footnote[4]{The orbital or azimuthal angular velocity:\\ 
\begin{equation*}
\Omega^{\rm null}_{\phi}(r)=\frac{\dot \phi}{\dot t}=\sqrt{\frac{\dot \phi^2}{\dot t^2}}=\sqrt{\frac{\mathrm{L}^2}{r^4}\,\frac{f^2(r)}{\mathrm{E}^2}}\\
=\sqrt{\frac{r^2}{f(r)}\,\frac{f^2(r)}{r^4}}=\frac{\sqrt{f(r)}}{r}
\end{equation*}
}
\begin{equation}
    \Omega^{\rm null}_{\phi}(r)=\frac{\dot \phi}{\dot t}=\frac{\sqrt{f(r)}}{r}.\label{vel1}
\end{equation}
Substituting the metric function $f(r)$ given in Eq.~(\ref{function}), we find the geodesic angular velocity at the photon sphere radius $r=r_{s}$ to be
\begin{equation}
    \Omega^{\rm null}_{\phi}(r_{s})=\frac{\sqrt{1 - \alpha - \frac{2Mr^2_{s}}{g^3 + r^3_{s}} + \frac{Q^2}{r^2_{s}} + \frac{\beta}{r_{s}}\ln\!\frac{r_{s}}{|\beta|}}}{r_{s}}=\frac{(1-\alpha)^{1/2}}{R_{\rm sh}},\label{vel2}
\end{equation}
where we have used the relation (\ref{bb10}).

From the above expression Eq.~(\ref{vel2}), one can notice that the geodesic angular velocity depends explicitly on several geometric parameters. These include the CoS parameter $\alpha$, the PFDM parameter $\beta$, the Hayward parameter $g$, the electric charge $Q$, and the mass $M$ of the BH for a given photon sphere radius $r_{s}$. Each of these parameters modifies the spacetime curvature and, consequently, alters the velocity of photons on unstable circular orbits. The inverse relationship between $\Omega^{\rm null}_{\phi}$ and the shadow radius $R_{\rm sh}$ demonstrates that larger shadows correspond to slower orbital velocities at the photon sphere.

\subsection{Photon Trajectories}

Photon trajectories around a BH are determined by the spacetime geometry and are described by null geodesics. These trajectories can bend, orbit, or escape depending on the impact parameter, giving rise to phenomena such as gravitational lensing, photon spheres, and BH shadows.

The equation of orbit using Eqs.~(\ref{bb3}) and (\ref{bb3a}) is given by
\begin{equation}
    \frac{\dot r^2}{\dot \phi^2}=\left(\frac{dr}{d\phi}\right)^2=r^4\,\left[\frac{\mathrm{E}^2}{\mathrm{L}^2}-\frac{1}{r^2}\,\left(1 - \alpha - \frac{2mr^2}{r^3 + g^3} + \frac{Q^2}{r^2} + \frac{\beta}{r}\ln\!\frac{r}{\left|\beta\right|}\right)\right].\label{bb13}
\end{equation}

Using a new variable $r(\phi)=\frac{1}{u(\phi)}$ and after simplification, we obtain
\begin{equation}
    \left(\frac{du}{d\phi}\right)^2+(1-\alpha) u^2=\frac{1}{\gamma^2}+\frac{2 M u^3}{1+g^3 u^3}-Q^2 u^4+\beta u^3 \ln\!\, (|\beta| u).\label{bb14} 
\end{equation}
Differentiating both sides with respect to $\phi$ and after rearranging, we obtain the second-order trajectory equation
\begin{equation}
    \frac{d^2u}{d\phi^2}+(1-\alpha) u=\frac{3 M u^2}{(1+g^3 u^3)^2}-2 Q^2 u^3+\frac{1}{2}\beta u^2 \left(3\,\ln\!\,(|\beta| u)+1\right).\label{bb15}
\end{equation}
Equation~(\ref{bb15}) represents the second-order differential equation describing the photon trajectories in the gravitational field of a charged Hayward BH with CoS surrounded by PFDM. As discussed earlier, these trajectories are influenced by the CoS parameter $\alpha$, the PFDM parameter $\beta$, the Hayward parameter $g$, as well as the electric charge $Q$ and the mass $M$ of the BH. The presence of the logarithmic term from the PFDM contribution makes this equation particularly interesting, as it introduces non-trivial modifications to the standard geodesic equation.

\subsection{Effective Radial Force}

The effective radial force represents the force experienced by photons moving in the gravitational field along the radial direction in the equatorial plane ($\theta=\pi/2$). This force is given by the negative gradient of the effective potential that governs the photon dynamics. Mathematically, it is defined as
\begin{equation}
    F_{\rm rad}=-\frac{1}{2} \frac{\partial V_{\rm eff}}{\partial r}.\label{bb11}
\end{equation}

\begin{figure}[ht!]
    \centering
    \includegraphics[width=0.45\linewidth]{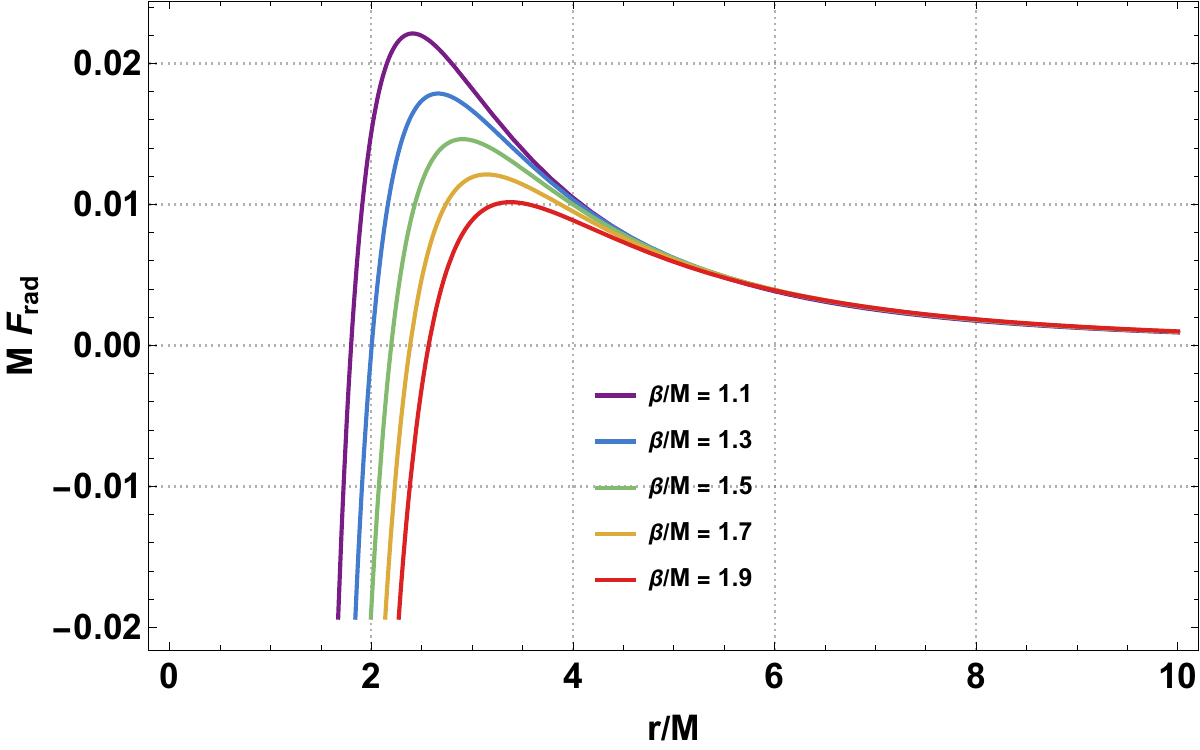}\qquad
    \includegraphics[width=0.45\linewidth]{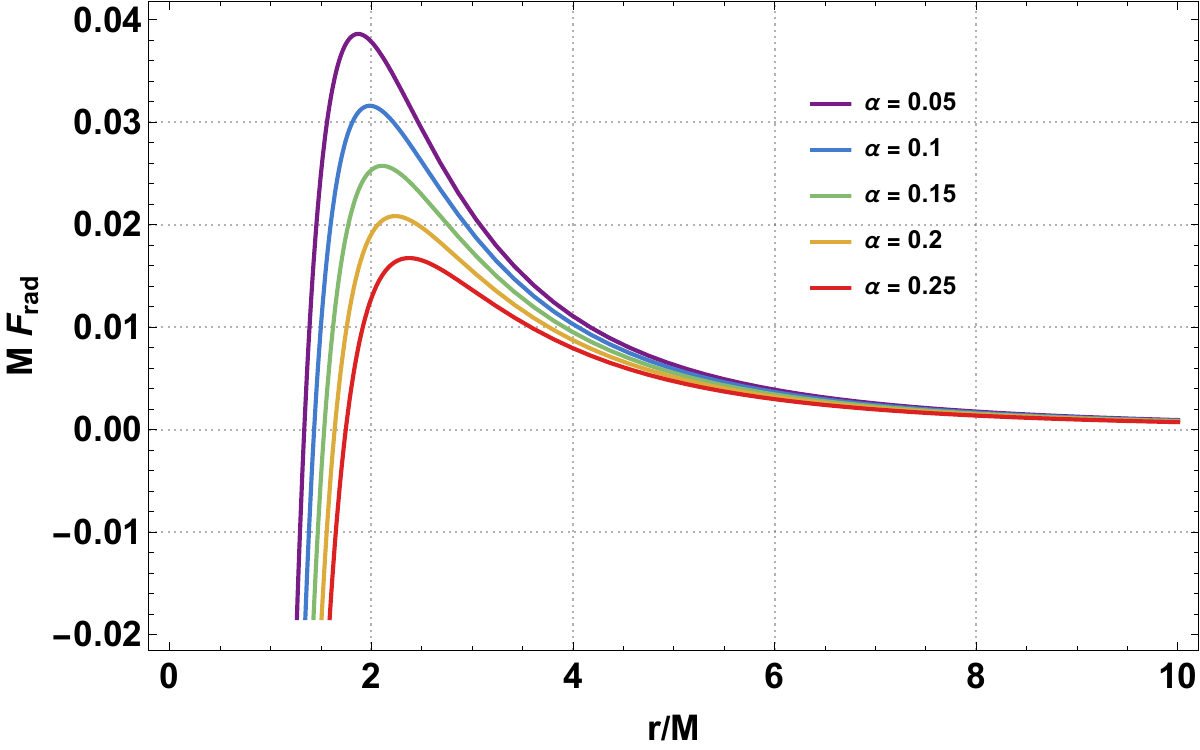}\\
    (a) $\alpha=0.1$ \hspace{6cm} (b) $\beta/M=0.8$
    \caption{Behavior of the effective radial force experienced by photons as a function of dimensionless radial distance $r/M$ for various $\beta/M$ and $\alpha$. Here $Q/M=1,\,g/M=0.2,\,\mathrm{L}/M=1$.}
    \label{fig:force}
\end{figure}

Substituting the potential given in Eq.~(\ref{bb3b}), we find the following expression for the radial force:
\begin{equation}
    F_{\rm rad}=\frac{\mathrm{L}^2}{r^3}\left[1-\alpha-\frac{3 M r^5}{(r^3+g^3)^2}+\frac{2 Q^2}{r^2}+\frac{\beta}{2 r} \left(3 \ln\!\frac{r}{|\beta|}-1\right)\right].\label{bb12}
\end{equation}
From Eq.~(\ref{bb12}), it is evident that the effective radial force experienced by photons is influenced by the CoS parameter $\alpha$, the PFDM parameter $\beta$, the Hayward parameter $g$, as well as the electric charge $Q$ and mass $M$ of the BH. The force profile determines whether photons are attracted toward or repelled from the BH at different radial distances.

In Figure~\ref{fig:force}, we illustrate the effective radial force given in Eq.~(\ref{bb12}) as a function of the dimensionless radial distance $r/M$, varying the PFDM parameter $\beta$ and the CoS parameter $\alpha$, while keeping the charges and angular momentum fixed. We observe that increasing the values of $\beta/M$ and $\alpha$ leads to a suppression of the peak of the force, indicating that photons are more loosely bound by the gravitational field for higher values of these parameters. This behavior is consistent with the suppression of the effective potential peak observed earlier, confirming that both CoS and PFDM weaken the gravitational confinement of photons near the BH.

\section{Dynamics of Neutral Particles around BH} \label{isec4}

In this section, we investigate the dynamics of neutral test particles around the charged Hayward BH coupled to a CoS and surrounded by PFDM. We derive the effective potential and examine how the BH parameters influence the specific energy and specific angular momentum of test particles in circular orbits. Furthermore, we analyze how these factors affect the location and properties of the innermost stable circular orbits (ISCOs).

For a neutral particle's motion, we consider the Hamiltonian in the form
\begin{equation}
    H=\frac{1}{2} g^{\mu\nu} p_{\mu} p_{\nu}+\frac{1}{2} m^2,\label{cc1}
\end{equation}
where $m$ is the mass of the particle, $p^{\mu} =m u^{\mu}$ denotes its 4-momentum, while $u^{\mu}=dx^{\mu}/d\tau$ and $\tau$ represent the 4-velocity and proper time, respectively.

The Hamiltonian equations of motion take the form
\begin{equation}
\frac{dx^\mu}{d\zeta} \equiv m u^\mu = \frac{\partial H}{\partial p_\mu}, 
\quad 
\frac{dp_\mu}{d\zeta} = - \frac{\partial H}{\partial x^\mu}, \label{cc2}
\end{equation}
where the affine parameter is chosen as \(\zeta = \tau / m\).  

The BH geometry possesses two constants of motion: the energy \(\mathcal{E}\) and angular momentum \(\mathcal{L}\) per unit mass, which can be expressed as
\begin{align}
p_t / m &=- f(r) \frac{dt}{d\tau}= -\mathcal{E}, \label{cc3} \\
p_\phi / m &= r^2 \sin^2\theta \frac{d\phi}{d\tau} = \mathcal{L}. \label{cc4}
\end{align}

Hence, the time $u^t$, azimuthal $u^\phi$, polar $u^\theta$, and radial $u^r$ components of the 4-velocity $u^\mu$ are given by
\begin{align}
    &\frac{dt}{d\tau}=\frac{\mathcal{E}}{f(r)},\label{cc5}\\
    &\frac{d\phi}{d\tau}=\frac{\mathcal{L}}{r^2 \sin^2\theta},\label{cc6}\\
    &\frac{d\theta}{d\tau}=\frac{p_{\theta}}{m r^2},\label{cc7}\\
    &\left(\frac{dr}{d\tau}\right)^2+\left(\epsilon+\frac{\mathcal{L}^2}{r^2 \sin^2\theta}+\frac{p^2_{\theta}}{m r^2}\right) f(r)=\mathcal{E}^2,\label{cc8}
\end{align}
with $\epsilon=1$ for timelike particles and $\epsilon=0$ for lightlike particles.

The Hamiltonian (\ref{cc1}) can be rewritten as
\begin{equation}
    H=\frac{f(r)}{2} p^2_r +\frac{p^2_{\theta}}{r^2}+\frac{1}{2} \frac{m^2}{f(r)} \left(U_{\rm eff} (r, \theta)-\mathcal{E}^2\right),\label{cc9}
\end{equation}
where the effective potential $U_{\rm eff} (r,\theta)$ is defined as
\begin{equation}
    U_{\rm eff} (r,\theta)=\left(1+\frac{\mathcal{L}^2}{r^2 \sin^2\theta}\right) f(r).\label{cc10}
\end{equation}

From Eq.~(\ref{cc10}), we observe that the effective potential governing particle dynamics depends explicitly on several geometric parameters. These include the CoS parameter $\alpha$, the PFDM parameter $\beta$, the Hayward parameter $g$, as well as the electric charge $Q$ and mass $M$ of the BH. Moreover, the angular momentum per unit mass $\mathcal{L}$ and polar angle $\theta$ also modify this effective potential. Each of these parameters alters the spacetime geometry and, consequently, reshapes the effective potential experienced by neutral test particles. As a result, variations in these parameters lead to notable changes in the location and stability of circular orbits.

\begin{figure}[ht!]
    \centering
    \includegraphics[width=0.45\linewidth]{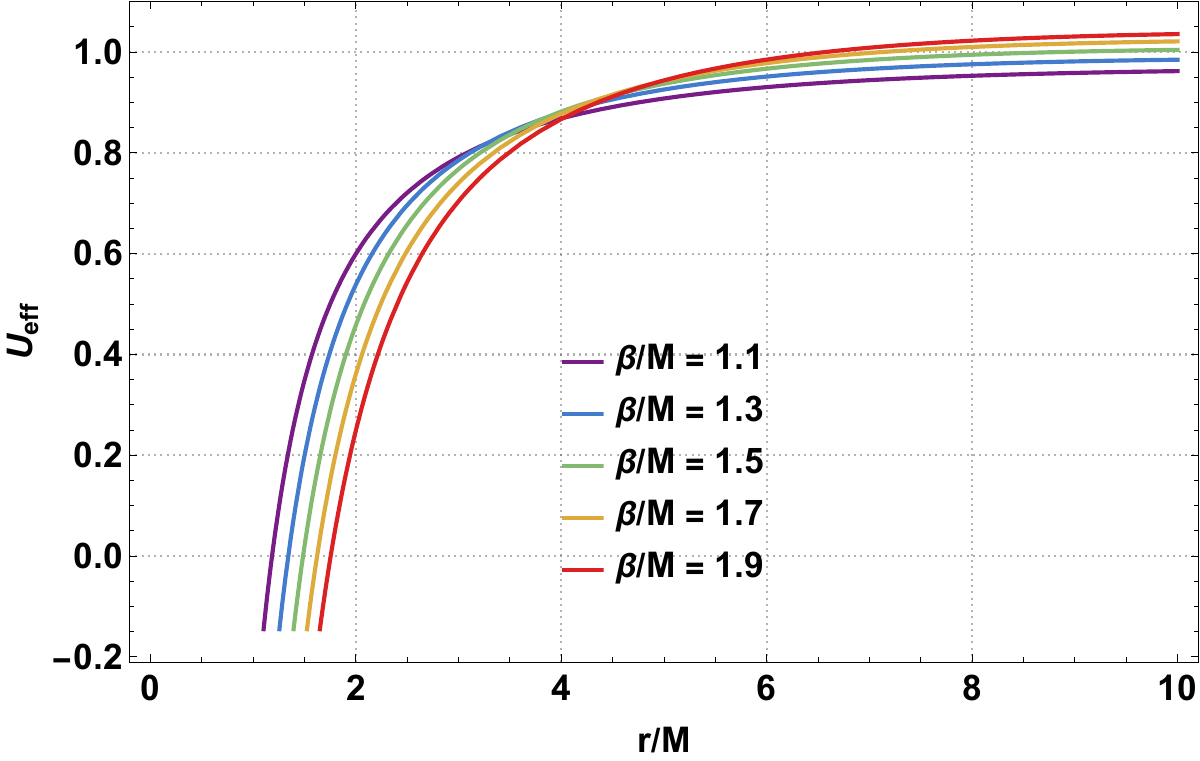}\qquad
    \includegraphics[width=0.45\linewidth]{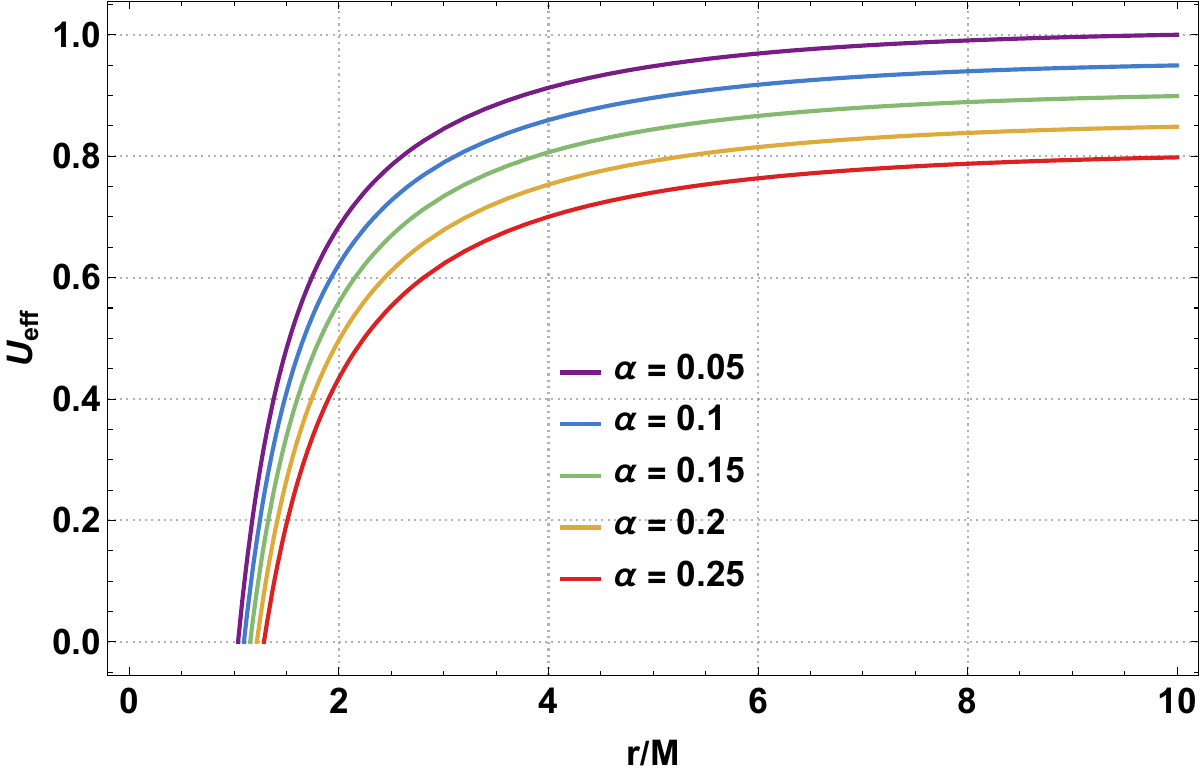}\\
     (a) $\alpha=0.1$ \hspace{6cm} (b) $\beta/M=1$
    \caption{Effective potential governing particle dynamics as a function of dimensionless radial distance $r/M$. Here $Q/M=1,\,g/M=0.2, \mathcal{L}/M=1, \theta=\pi/2$.}
    \label{fig:timelike}
\end{figure}

In Fig.~\ref{fig:timelike}, we present the effective potential given in Eq.~(\ref{cc10}) on the equatorial plane ($\theta=\pi/2$) as a function of the dimensionless radial coordinate $r/M$. The analysis is carried out by varying the PFDM parameter $\beta$ and the CoS parameter $\alpha$, while keeping the electric charge, Hayward parameter, and angular momentum fixed. Panel~(a) shows that increasing $\beta/M$ leads to an enhancement of the effective potential at larger radial distances, indicating that PFDM strengthens the gravitational barrier in the outer region. In contrast, as illustrated in panel~(b), increasing $\alpha$ suppresses the effective potential, indicating a weakening of the gravitational barrier experienced by neutral test particles. This suppression arises from the solid angle deficit induced by the CoS, which effectively reduces the gravitational binding.

\subsection{Effective Radial Force}

The effective radial force experienced by a neutral test particle provides information about its motion, indicating whether the particle is attracted toward the BH or repelled away from it. In general, the BH generates gravitational forces that can both attract and repel particles depending on the particle's position and the spacetime parameters. We analyze the motion of test particles and compute the corresponding effective force using Eq.~(\ref{cc10}) as follows:
\begin{equation}
    \mathcal{F}=-\frac{1}{2} \frac{\partial U_{\rm eff}}{\partial r}.\label{force1}
\end{equation}
Substituting the potential given in Eq.~(\ref{cc10}) and after simplification, we obtain
\begin{eqnarray}
    \mathcal{F}_{\rm eff}&=&-\frac{f'(r)}{2}+\frac{\mathcal{L}^2}{2 r^3 \sin^2\theta} (2 f (r)-r f'(r))\nonumber\\
    &=&-\frac{M\left(r^{4}-2r g^{3}\right)}{\left(r^{3}+g^{3}\right)^{2}}
+\frac{Q^{2}}{r^{3}}
-\frac{\beta}{2r^{2}}\left(1-\ln\!\frac{r}{\left|\beta\right|}\right)+\frac{\mathcal{L}^2}{r^3 \sin^2\theta} \left[1-\alpha-\frac{3 M r^5}{(r^3+g^3)^2}+\frac{2 Q^2}{r^2}+\frac{\beta}{2 r} \left(3 \ln\!\frac{r}{|\beta|}-1\right)\right].\quad\label{force2}
\end{eqnarray}

\begin{figure}[ht!]
    \centering
    \includegraphics[width=0.45\linewidth]{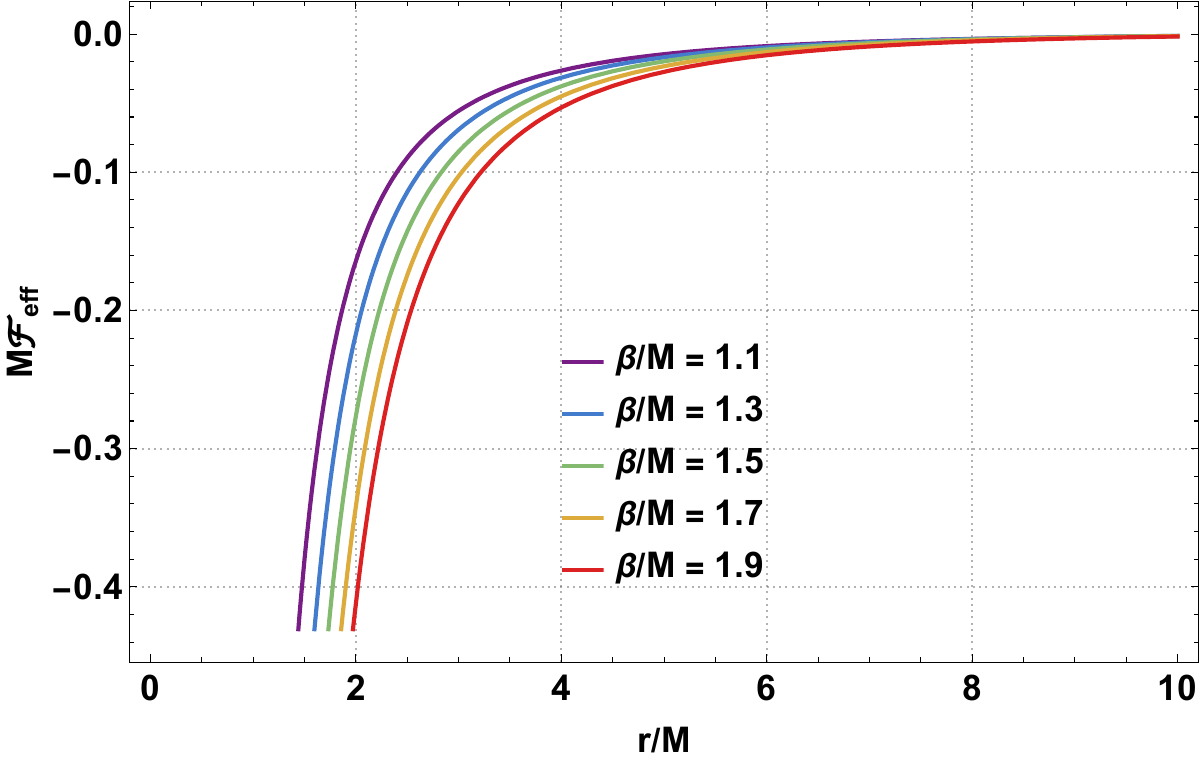}\qquad
    \includegraphics[width=0.45\linewidth]{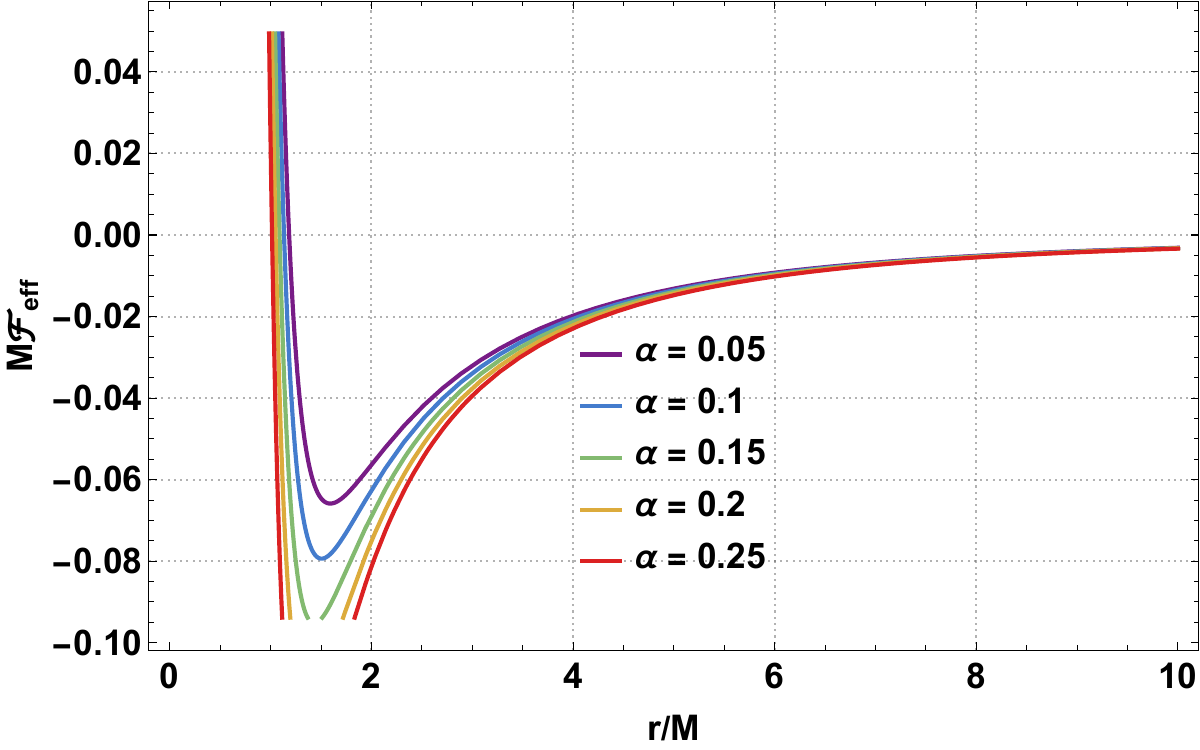}\\
    (a) $\alpha=0.1$ \hspace{6cm} (b) $\beta/M=0.5$
    \caption{Radial force experienced by neutral test particles as a function of dimensionless radial distance $r/M$. Here $Q/M=1,\,g/M=0.2,\mathcal{L}/M=1$.}
    \label{fig:force-timelike}
\end{figure}

Similar to the effective potential in Eq.~(\ref{cc10}), the effective radial force given in Eq.~(\ref{force2}) depends explicitly on the CoS parameter $\alpha$, the PFDM parameter $\beta$, the Hayward parameter $g$, as well as the electric charge $Q$ and mass $M$ of the BH. The angular momentum per unit mass $\mathcal{L}$ also modifies the force profile. Consequently, variations in these parameters lead to notable changes in the radial force experienced by neutral test particles, affecting their orbital dynamics and stability.

In Fig.~\ref{fig:force-timelike}, we depict the effective radial force defined in Eq.~(\ref{force2}) on the equatorial plane ($\theta=\pi/2$) as a function of the dimensionless radial coordinate $r/M$. The force profile is examined by varying the PFDM parameter $\beta$ and the CoS parameter $\alpha$, while keeping the electric charge, Hayward parameter, and angular momentum fixed. As shown in both panels, increasing the values of $\beta/M$ and $\alpha$ leads to a suppression of the effective force. This behavior indicates a reduction in the strength of the gravitational attraction acting on neutral test particles. The weakened gravitational pull lowers the effective potential barrier and facilitates particle motion at smaller radial distances for higher values of these parameters. Physically, this suggests that both PFDM and CoS act to diminish the gravitational confinement of particles near the BH.

\subsection{Circular Orbits: Specific Angular Momentum and Energy}

For circular orbits in the equatorial plane, the following conditions must be satisfied:
\begin{equation}
    U_{\rm eff} (r,\theta)=\mathcal{E}^2,\label{cc11}
\end{equation}
and
\begin{equation}
    \frac{\partial U_{\rm eff}}{\partial r}=0.\label{cc12}
\end{equation}

Using the effective potential in Eq.~(\ref{cc10}) together with Eqs.~(\ref{cc11})--(\ref{cc12}) and after simplification, we find the specific angular momentum and specific energy, respectively, as
\begin{equation}
    \mathcal{L}^2_{\rm sp}=\frac{r^3 f'(r)}{2 f (r)-r f'(r)}=r^2\,\left(\frac{\frac{M\left(r^5-2r^2 g^3\right)}{\left(r^3+g^3\right)^2}
-\frac{Q^2}{r^2}
+\frac{\beta}{2r}\left(1-\ln\!\frac{r}{|\beta|}\right)}{1-\alpha-\frac{3 M r^5}{(r^3+g^3)^2}+\frac{2 Q^2}{r^2}+\frac{\beta}{2 r} \left(3 \ln\!\frac{r}{|\beta|}-1\right)}\right),\label{cc13}
\end{equation}
and
\begin{equation}
    \mathcal{E}^2_{\rm sp}=\frac{2 f^2(r)}{2 f (r)-r f'(r)}=\frac{\left(1 - \alpha - \frac{2 M r^2}{r^3 + g^3} + \frac{Q^2}{r^2} + \frac{\beta}{r}\ln\!\frac{r}{\left|\beta\right|}\right)^2}{1-\alpha-\frac{3 M r^5}{(r^3+g^3)^2}+\frac{2 Q^2}{r^2}+\frac{\beta}{2 r} \left(3 \ln\!\frac{r}{|\beta|}-1\right)}.\label{cc14}
\end{equation}

\begin{figure}[ht!]
    \centering
    \includegraphics[width=0.45\linewidth]{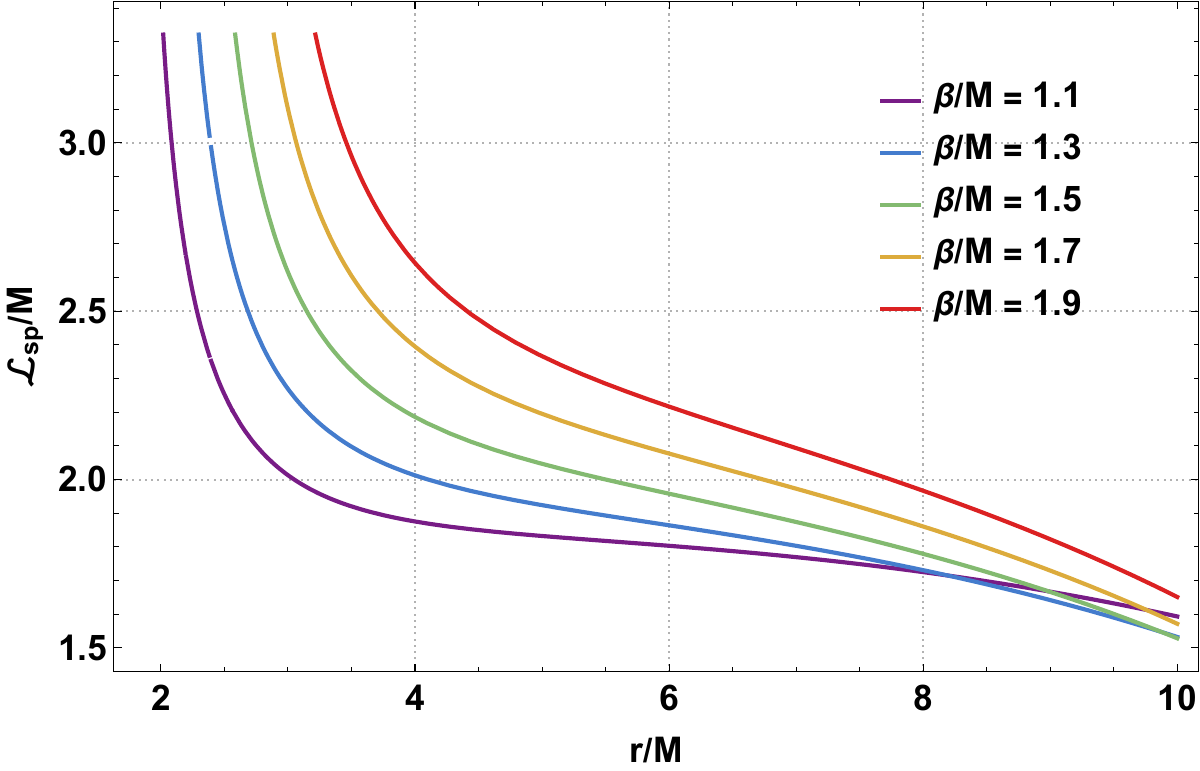}\qquad
    \includegraphics[width=0.45\linewidth]{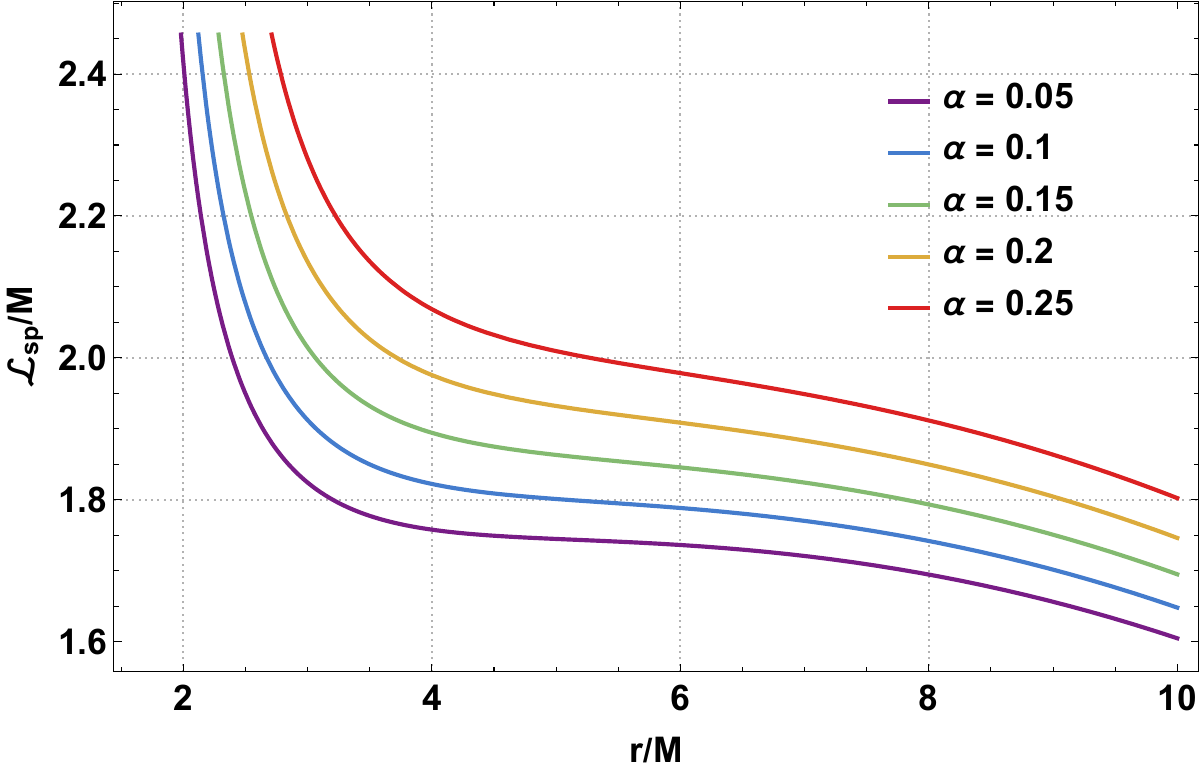}\\
    (a) $\alpha=0.1$ \hspace{6cm} (b) $\beta/M=1$
    \caption{Specific angular momentum per unit mass $\mathcal{L}_{\rm sp}/M$ as a function of the dimensionless radial distance $r/M$ for various $\beta/M$ and $\alpha$. Here $Q/M=1,\,g/M=0.2$.}
    \label{fig:specific-momentum}
\end{figure}

\begin{figure}[ht!]
    \centering
    \includegraphics[width=0.45\linewidth]{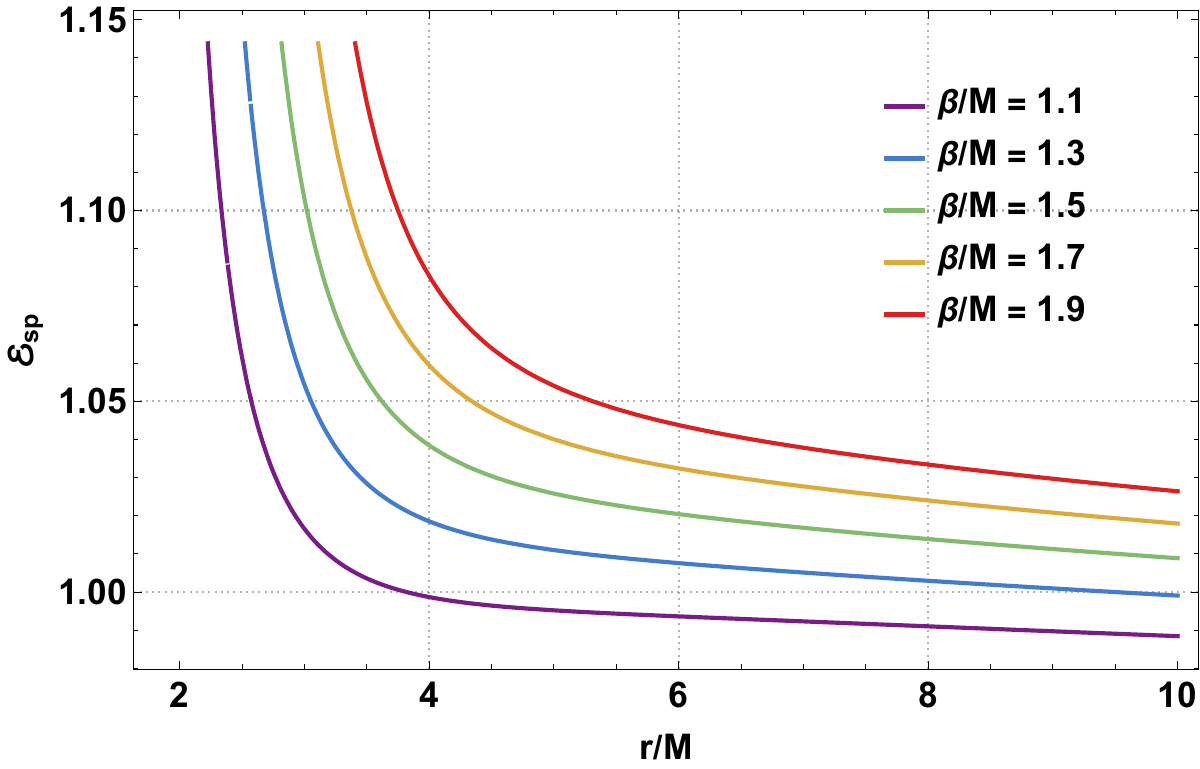}\qquad
    \includegraphics[width=0.45\linewidth]{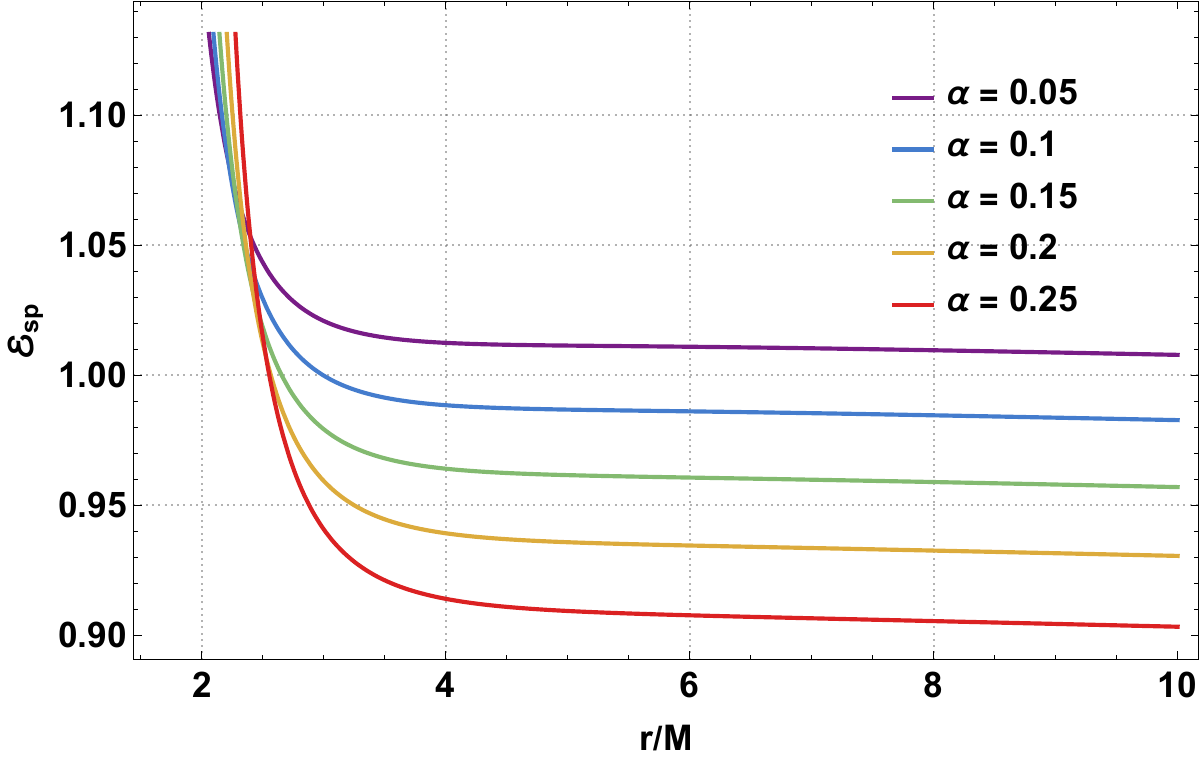}\\
    (a) $\alpha=0.1$ \hspace{6cm} (b) $\beta/M=1$
    \caption{Specific energy $\mathcal{E}_{\rm sp}$ as a function of the dimensionless radial distance $r/M$ for various $\beta/M$ and $\alpha$. Here $Q/M=1,\,g/M=0.2$.}
    \label{fig:specific-energy}
\end{figure}

From Eqs.~(\ref{cc13})--(\ref{cc14}), we observe that the specific angular momentum and specific energy associated with test particles orbiting in circular paths at fixed radius depend explicitly on several geometric parameters that modify the spacetime curvature. These include the CoS parameter $\alpha$, the PFDM parameter $\beta$, the Hayward parameter $g$, as well as the electric charge $Q$ and mass $M$ of the BH. As a result, variations in these parameters lead to notable changes in the orbital motion and energy of test particles in circular orbits.

In Fig.~\ref{fig:specific-momentum}, we show the specific angular momentum per unit mass, defined in Eq.~(\ref{cc13}), on the equatorial plane ($\theta=\pi/2$) as a function of the dimensionless radial coordinate $r/M$. The profile is analyzed by varying the PFDM parameter $\beta$ and the CoS parameter $\alpha$, while keeping the electric charge and Hayward parameter fixed. As illustrated, increasing the values of $\beta/M$ and $\alpha$ results in an increase in the specific angular momentum. This indicates that higher values of these parameters require test particles to possess greater angular momentum to maintain circular motion at a given radius. The physical interpretation is that both PFDM and CoS weaken the effective gravitational attraction, necessitating larger centrifugal support for orbital balance.

Similarly, in Fig.~\ref{fig:specific-energy}, we present the specific energy, defined in Eq.~(\ref{cc14}), on the equatorial plane ($\theta=\pi/2$) as a function of $r/M$, for varying $\beta$ and $\alpha$. In panel~(a), increasing $\beta/M$ leads to a rise in the specific energy, reflecting a stronger contribution of the PFDM to the particle's total energy. In contrast, increasing $\alpha$ results in a decrease of the specific energy, as shown in panel~(b). This indicates that the presence of CoS reduces the energy required for a test particle to remain in circular orbit at a given radius. The opposing effects of PFDM and CoS on the specific energy demonstrate that these two components influence the orbital dynamics in distinct ways.

\subsection{Innermost Stable Circular Orbits (ISCOs)}

A stable circular orbit is located at a minimum of the effective potential, whereas an unstable orbit corresponds to a maximum. In Newtonian gravitational theory, the effective potential always has a minimum regardless of the angular momentum, and therefore there is no well-defined ISCO with a finite radius. 

However, when the shape of the effective potential depends on parameters such as the particle's angular momentum, the situation changes. In GR, for any given angular momentum, the effective potential for particles orbiting near a Schwarzschild BH exhibits two extrema. Only for a specific value of the angular momentum do these two extrema coincide at \(r = 3 r_g\), which defines the ISCO, where \(r_g\) is the Schwarzschild radius. 

For the charged Hayward BH with CoS and PFDM, the ISCO can be determined from the following conditions:
\begin{equation}
U_\text{eff}(r) = \mathcal{E}^2, \quad \frac{\partial U_\text{eff}(r)}{\partial r} = 0, \quad \frac{\partial^2 U_\text{eff}(r)}{\partial r^2} = 0.\label{cc25}
\end{equation}

Using the effective potential given in Eq.~(\ref{cc10}) in the equatorial plane, we find the following relation in terms of the metric function $f(r)$:
\begin{equation}
    \frac{3}{r}\,f(r)\,f'(r)-2 (f'(r))^2+f(r)\,f''(r)=0.\label{cc26}
\end{equation}
Substituting the metric function $f(r)$ and after manipulation, we obtain
\begin{eqnarray}
    &\left[
3(1-\alpha)
-\frac{6 M r^2}{r^3+g^3}
-\frac{4 M \left(r^5-2r^2 g^3\right)}{(r^3+g^3)^2}
+\frac{7Q^2}{r^2}
+\frac{5\beta}{r}\ln\!\frac{r}{|\beta|}
-\frac{2\beta}{r}
\right]
\left[\frac{2M\left(r^4-2rg^3\right)}{(r^3+g^3)^2}
-\frac{2Q^2}{r^3}
+\frac{\beta}{r^2}\left(1-\ln\!\frac{r}{|\beta|}\right)
\right]+\nonumber\\ &
r \left[1-\alpha-\frac{2Mr^2}{r^3+g^3}+\frac{Q^2}{r^2}
+\frac{\beta}{r}\ln\!\frac{r}{|\beta|}\right]\left[-\frac{4M\left(r^6-7r^3 g^3+g^6\right)}{(r^3+g^3)^3}
+\frac{6Q^2}{r^4}
+\frac{\beta}{r^3}\left(2\ln\!\frac{r}{|\beta|}-3\right)\right]=0.\label{cc27}
\end{eqnarray}

An analytical solution of Eq.~(\ref{cc27}) would, in principle, provide the ISCO radius \(r = r_{\rm ISCO}\). However, obtaining an exact analytical solution is highly challenging due to the presence of the logarithmic term arising from the PFDM contribution. Nevertheless, the ISCO radius can be determined numerically by choosing appropriate values of the CoS parameter \(\alpha\), the PFDM parameter \(\beta\), the Hayward parameter \(g\), and the electric charge \(Q\). The ISCO plays a fundamental role in astrophysics, as it determines the inner edge of accretion disks around BHs and influences the electromagnetic emission from accreting systems. Modifications to the ISCO radius induced by CoS and PFDM could potentially be constrained through observations of X-ray binaries and active galactic nuclei.

\section{Quasi-periodic Oscillations (QPOs)} \label{isec5}

In this section, we study the QPO frequencies and examine how the presence of string clouds and PFDM affects the azimuthal, radial, and vertical epicyclic frequencies of test particle motion. QPOs observed in X-ray binaries serve as powerful probes of the strong-field regime of gravity, as they are closely connected to the spacetime geometry near black holes and the dynamics of accretion disks. Accurate measurements of these oscillations enable constraints on black hole parameters and provide a means to test deviations from general relativity \cite{OD2025}. To this end, we employ the relativistic precession model to analyze QPOs around a charged Hayward black hole immersed in CoS and PFDM. We derive the orbital, radial, vertical, and periastron precession frequencies and investigate how the black hole parameters and surrounding matter distributions modify these observationally relevant quantities. Some recent investigations of particle dynamics around black holes and QPOs epicyclic frequencies were reported in \cite{GMM1,GMM2,GMM3,GMM4,GMM5,GMM6,GMM7,GMM8,GMM9,GMM10,GMM11} and related references there in. 

\subsection{Orbital Frequency}

The Keplerian frequency, which characterizes the orbital motion of test particles around the BH, is given by
\begin{equation}
   \nu _{\phi,r,\theta} = \frac{1}{2\,\pi }\frac{c^3}{G\,M}\, \Omega _{r,\theta, \phi }\ , [{\textrm{Hz}}]\ . \label{pp1}
\end{equation}
The orbital or azimuthal frequency \(\Omega_{\phi}\), representing the frequency of Keplerian orbits in the azimuthal direction, takes the form \footnote[5]{The squared orbital frequency:\\ 
\begin{equation*}
\Omega^2_{\phi}=\frac{\dot \phi^2}{\dot t^2}=\frac{\mathrm{L}^2_{\rm sp}}{r^4}\,\frac{f^2(r)}{\mathrm{E}^2_{\rm sp}}=\frac{f^2(r)}{r^4}\,\frac{r^3\,f'(r)}{2\,f(r)-r\,f'(r)}\,\frac{2\,f(r)-r\,f'(r)}{2\,f^2(r)}=\frac{f'(r)}{2\,r}
\end{equation*}
}
\begin{equation}
   \nu_{\phi}=\Omega_{\phi}=\frac{d\phi}{dt}=\frac{\dot{\phi}}{\dot{t}}=\frac{\omega_{\phi}}{\dot{t}}=\sqrt{\frac{f'(r)}{2\,r}}=\sqrt{\frac{M \left(r^3 - 2 g^3\right)}{\left(r^3 + g^3\right)^2} 
- \frac{Q^2}{r^4} 
+ \frac{\beta}{2 r^3} \left(1 - \ln \frac{r}{\left|\beta\right|}\right)}.\label{pp2}
\end{equation}

\begin{figure}[ht!]
    \centering
    \includegraphics[width=0.5\linewidth]{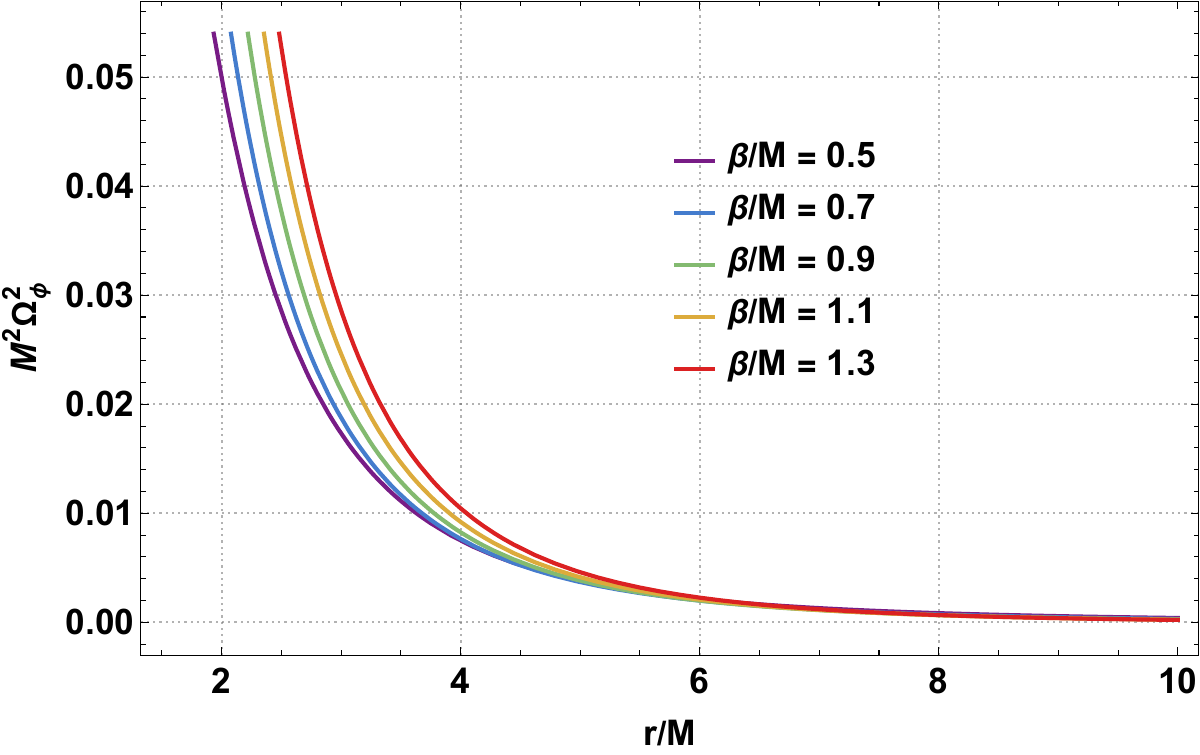}
    \caption{Squared azimuthal frequency as a function of dimensionless radial distance $r/M$ for various $\beta/M$. Here $Q/M=1,\,g/M=0.2$.}
    \label{fig:azimuthal-frequency}
\end{figure}

From Eq.~(\ref{pp2}), we notice that the orbital frequency depends explicitly on the PFDM parameter $\beta$, the Hayward parameter $g$, the electric charge $Q$, and mass $M$ of the BH. Notably, the CoS parameter $\alpha$ does not appear in the azimuthal frequency expression. This absence arises because $\alpha$ enters the metric function as a constant term $(1-\alpha)$, which vanishes upon differentiation. Consequently, the CoS affects the spacetime geometry but does not directly modify the orbital frequency of test particles.

In Fig.~\ref{fig:azimuthal-frequency}, we show the squared azimuthal frequency $M^2 \Omega_{\phi}^2$, given in Eq.~(\ref{pp2}), as a function of the dimensionless radial coordinate $r/M$. The behavior is analyzed by varying the PFDM parameter $\beta$ only, while keeping the electric charge and Hayward parameter fixed. We observe that increasing $\beta/M$ leads to a gradual decrease in $M^2 \Omega_\phi^2$, reflecting a reduction in the orbital angular velocity of test particles. At large distances, the azimuthal frequency approaches zero, indicating that the gravitational influence of the BH diminishes far from the central object. This behavior is consistent with the Keplerian limit where orbital frequencies scale as $r^{-3/2}$ at large radii.

\subsection{Radial and Vertical Frequencies}

The radial \(\nu_r=\Omega_r/(2\pi M)\) and vertical \(\nu_{\theta}=\Omega_{\theta}/(2\pi M)\) epicyclic frequencies characterize small oscillations in the radial and vertical ($\theta$) directions around stable circular orbits. These frequencies are obtained from the second derivatives of the effective potential $U_{\rm eff}(r, \theta)$ given in Eq.~(\ref{cc10}) with respect to the $r$ and $\theta$ coordinates, determining how perturbations affect the orbit's stability. The epicyclic frequencies are critical for analyzing QPO phenomena in BH systems, as they are directly related to the observed high-frequency QPO pairs in X-ray binaries.

Test particles orbiting in the equatorial plane around a BH in stable orbits oscillate along the radial, angular, and vertical directions due to small displacements from their equilibrium positions, denoted as $\delta r$ and $\delta \theta$. The following harmonic oscillator equations describe the radial and vertical oscillations~\cite{ZS2020}:
\begin{align} 
\frac{d^2\delta r}{dt^2}+\Omega _r^2 \delta r=0, \qquad \frac{d^2\delta \theta }{dt^2}+\Omega _\theta ^2 \delta \theta =0,\label{original}
\end{align}
where the frequencies $\Omega _r$ and $\Omega _\theta$ as measured by a static distant observer are given by 

{\small
\begin{align} 
\Omega_r^2&=-\frac{1}{2\, g_{rr}\,(u^t)^2}\, \frac{\partial^2 U_{\text {eff}}}{\partial r^2}=-\frac{1}{2}\left(f(r)\,f''(r)-2\,(f'(r))^2+\frac{3}{r}\,f(r)\,f'(r)\right)\,f(r)\nonumber\\
&=-\frac{1}{2}\Bigg[\left\{
3(1-\alpha)
-\frac{6 M r^2}{r^3+g^3}
-\frac{4 M \left(r^5-2r^2 g^3\right)}{(r^3+g^3)^2}
+\frac{7Q^2}{r^2}
+\frac{5\beta}{r}\ln\!\frac{r}{|\beta|}
-\frac{2\beta}{r}
\right\}\left\{\frac{2M\left(r^4-2rg^3\right)}{(r^3+g^3)^2}
-\frac{2Q^2}{r^3}
+\frac{\beta}{r^2}\left(1-\ln\!\frac{r}{|\beta|}\right)
\right\}+\nonumber\\ &
r \left\{1-\alpha-\frac{2Mr^2}{r^3+g^3}+\frac{Q^2}{r^2}
+\frac{\beta}{r}\ln\!\frac{r}{|\beta|}\right\}\left\{-\frac{4M\left(r^6-7r^3 g^3+g^6\right)}{(r^3+g^3)^3}
+\frac{6Q^2}{r^4}
+\frac{\beta}{r^3}\left(2\ln\!\frac{r}{|\beta|}-3\right)\right\}
\Bigg]\times\nonumber\\
&\left(1 - \alpha - \frac{2 M r^2}{r^3 + g^3} + \frac{Q^2}{r^2} + \frac{\beta}{r}\ln\!\frac{r}{\left|\beta\right|}\right).\label{pp3} 
\end{align}
}
And
\begin{align} 
\Omega^2_{\theta}&=- \frac{1}{2\, g_{\theta \theta }\, (u^t)^2} \frac{\partial ^2 U_{\text {eff}}}{\partial \theta ^2}=-\frac{f(r)\,f'(r)}{2 r}\nonumber\\
&=-\frac{1}{2}\left(1 - \alpha - \frac{2 M r^2}{r^3 + g^3} + \frac{Q^2}{r^2} + \frac{\beta}{r}\ln\!\frac{r}{\left|\beta\right|}\right)\,\left(\frac{2 M \left(r^3 - 2 g^3\right)}{(r^3 + g^3)^2} - \frac{2 Q^2}{r^4} + \frac{\beta}{r^3} \Biggl( 1 - \ln\frac{r}{|\beta|}\right). \label{pp4}
\end{align}

\begin{figure}[ht!]
    \centering
    \includegraphics[width=0.45\linewidth]{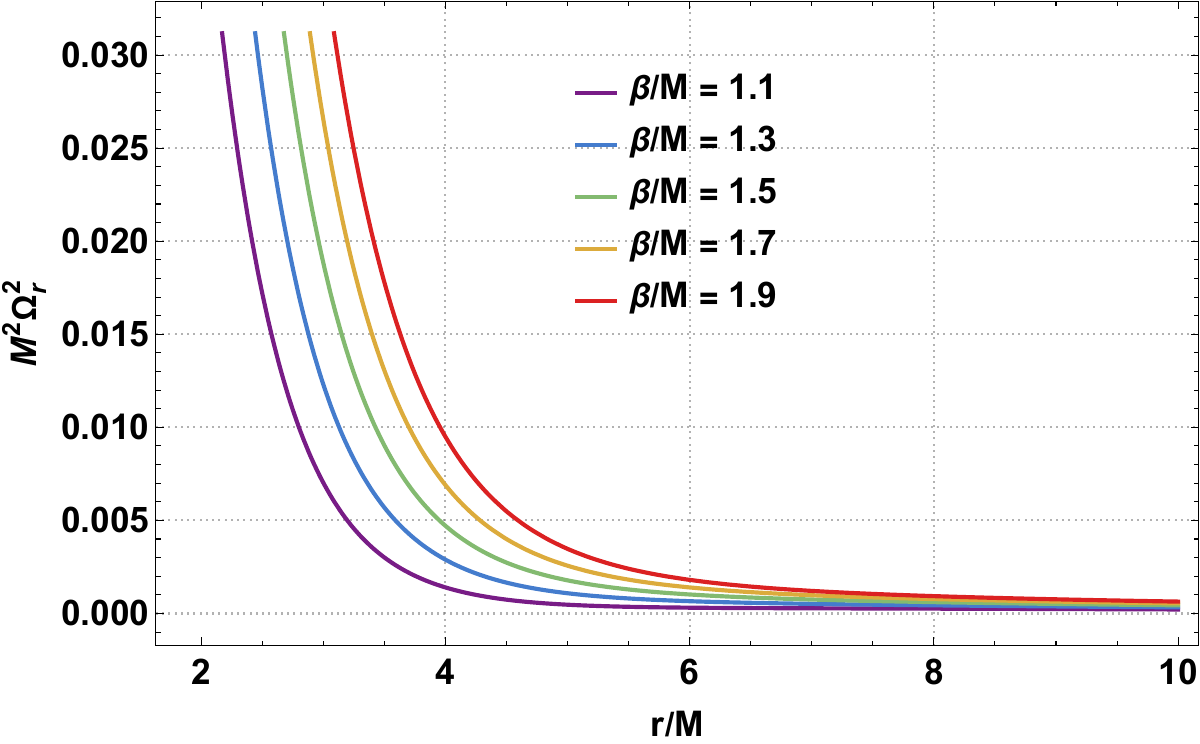}\qquad
    \includegraphics[width=0.45\linewidth]{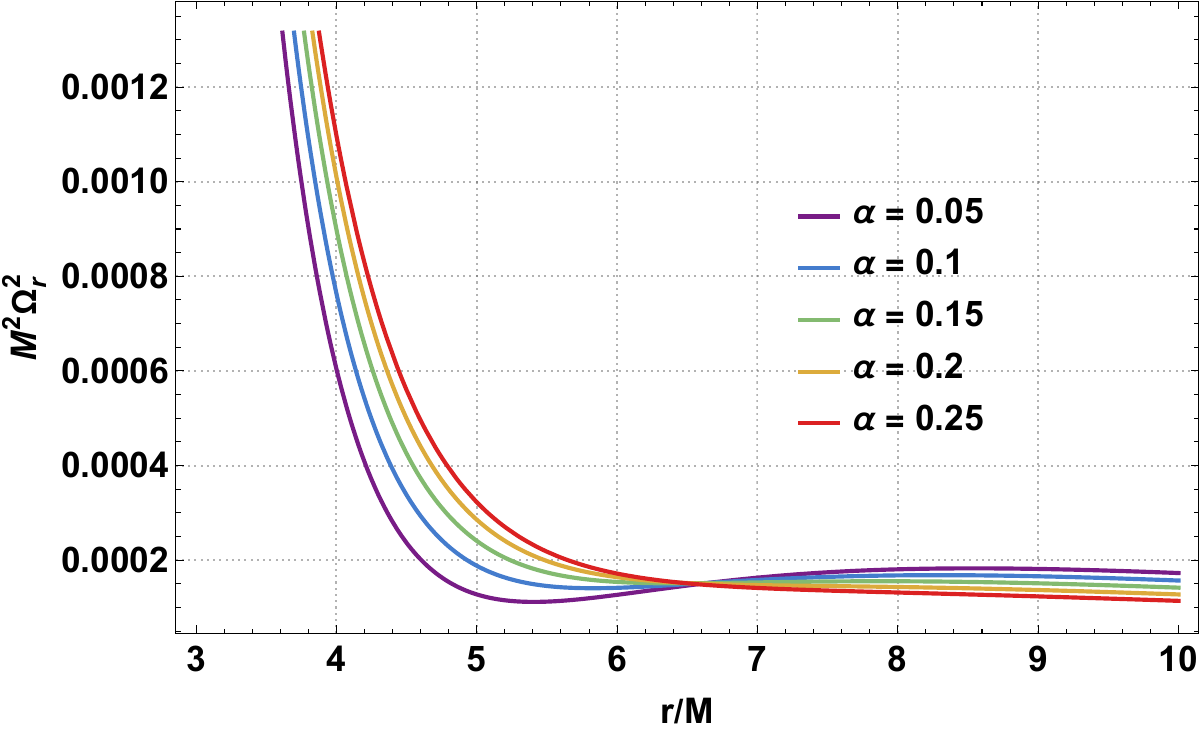}\\
    (a) $\alpha=0.1$ \hspace{8cm} (b) $\beta/M=1$
    \caption{Squared radial frequency $M^2 \Omega^2_r$ as a function of dimensionless radial distance $r/M$ for various $\beta/M$ and $\alpha$. Here $Q/M=1,\,g/M=0.2$.}
    \label{fig:radial-frequency}
\end{figure}

\begin{figure}[ht!]
    \centering
    \includegraphics[width=0.45\linewidth]{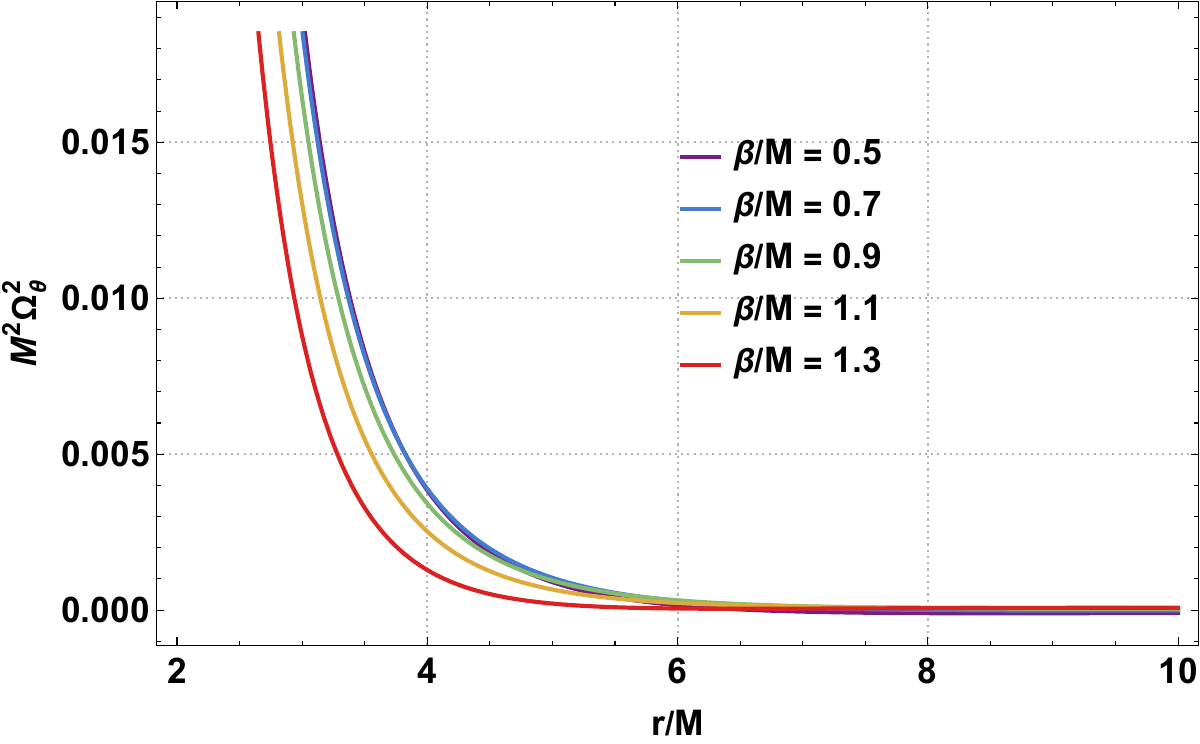}\qquad
    \includegraphics[width=0.45\linewidth]{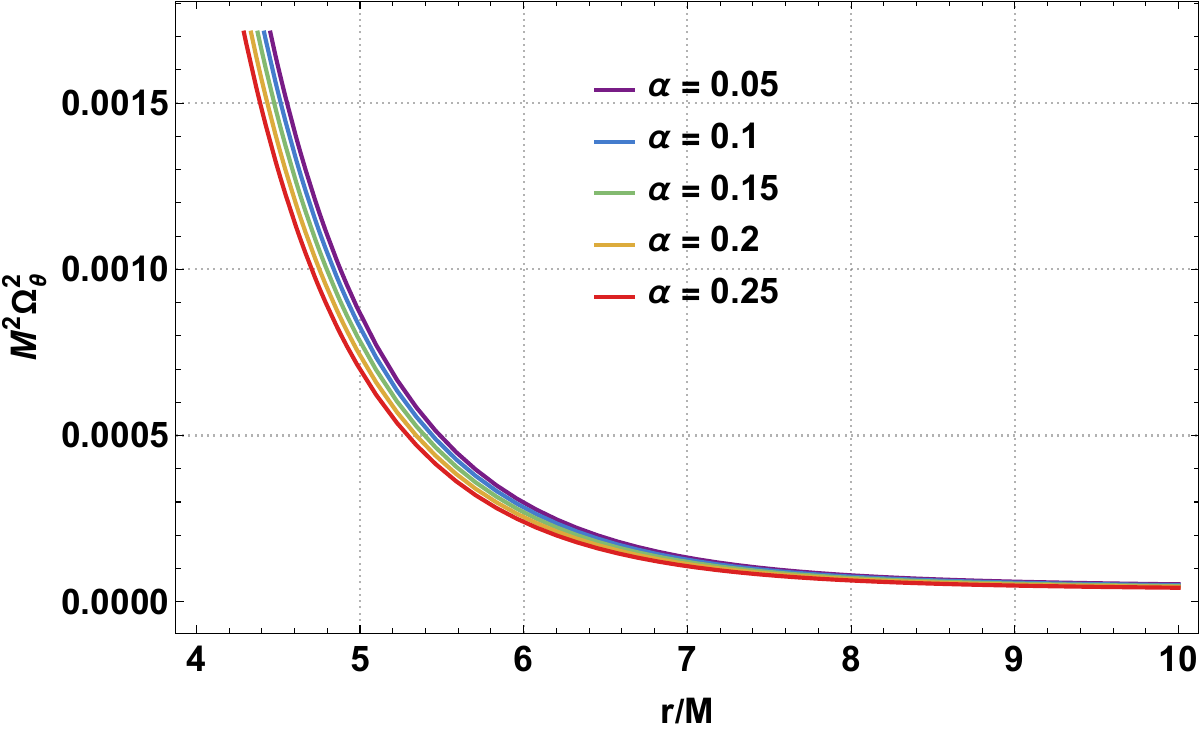}\\
    (a) $\alpha=0.1$ \hspace{8cm} (b) $\beta/M=1$
    \caption{Squared vertical frequency $M^2 \Omega^2_{\theta}$ as a function of dimensionless radial distance $r/M$ for various $\beta/M$ and $\alpha$. Here $Q/M=2,\,g/M=0.2$.}
    \label{fig:vertical-frequency}
\end{figure}

For a local observer, the frequencies of harmonic oscillatory motion are given by
\begin{align}
   \omega_r^2&=-\frac{1}{2\, g_{rr}}\, \frac{\partial^2 U_{\text {eff}}}{\partial r^2}=-\frac{f(r)}{2}\left(f(r)\,f''(r)-2\,(f'(r))^2+\frac{3}{r}\,f(r)\,f'(r)\right)\,\left(\frac{2 f(r)-r\,f'(r)}{2}\right)^{-1},\label{pp5} \\
   \omega^2_{\theta}&=-\frac{1}{2\, g_{\theta \theta }} \frac{\partial ^2 U_{\text {eff}}}{\partial \theta ^2}=-\frac{f(r)\,f'(r)}{2 r}\,\left(\frac{2 f(r)-r\,f'(r)}{2}\right)^{-1}. \label{pp6}
\end{align}
The relation between the frequencies measured by a local observer and a static distant observer is $\Omega^2_i=\omega^2_i/(u^t)^2$, where $u^t=\dot{t}=\sqrt{2/(2\,f(r)-r\,f'(r))}$ is the time component of the 4-velocity.

By substituting the metric function $f(r)$ into Eqs.~(\ref{pp3})--(\ref{pp4}) or Eqs.~(\ref{pp5})--(\ref{pp6}), one can verify that the radial and vertical frequencies are influenced by the geometric parameters of the spacetime. These include the CoS parameter $\alpha$, the PFDM parameter $\beta$, the Hayward parameter $g$, the electric charge $Q$, and mass $M$ of the BH. Unlike the azimuthal frequency, the epicyclic frequencies depend on $\alpha$ through the metric function and its derivatives, as the second derivatives of $f(r)$ retain contributions from all terms.

In Fig.~\ref{fig:radial-frequency}, we present the squared radial epicyclic frequency $M^2\Omega_r^2$, defined in Eq.~(\ref{pp3}), on the equatorial plane ($\theta=\pi/2$) as a function of the dimensionless radial coordinate $r/M$. The behavior is analyzed by varying the PFDM parameter $\beta$ and the CoS parameter $\alpha$, while keeping the electric charge and Hayward parameter fixed. We find that increasing the values of $\beta/M$ and $\alpha$ leads to an enhancement of the radial frequency, indicating stronger radial oscillations of test particles around circular orbits as these parameters increase. The radial frequency vanishes at the ISCO, where the orbit becomes marginally stable.

Similarly, Fig.~\ref{fig:vertical-frequency} shows the squared vertical epicyclic frequency $M^2\Omega_\theta^2$, given in Eq.~(\ref{pp4}), on the equatorial plane as a function of $r/M$ for different values of $\beta$ and $\alpha$. In contrast to the radial case, increasing $\beta/M$ and $\alpha$ results in a suppression of the vertical frequency. This behavior suggests that the restoring force associated with vertical perturbations weakens in the presence of stronger PFDM and CoS effects, thereby reducing the stability of out-of-plane oscillations. The opposing trends in radial and vertical frequencies have implications for QPO models that relate observed frequency ratios to the underlying spacetime geometry.

\subsection{Periastron Frequency}

The periastron frequency characterizes the precession of elliptical orbits around the BH. For a neutral test particle orbiting the charged Hayward BH with CoS and PFDM, we consider small perturbations near the equatorial plane at $\theta = \pi/2$. When the particle is perturbed slightly from its stable circular orbit, it undergoes oscillations about the equilibrium position characterized by the radial frequency $\Omega_r$. The periastron frequency $\Omega_p$ is defined as the difference between the orbital (azimuthal) frequency $\Omega_\phi$ and the radial frequency $\Omega_r$:
\begin{equation}
  \Omega_p = \Omega_{\phi} - \Omega_r=\sqrt{\frac{f'(r)}{2r}}-\sqrt{-\frac{1}{2}\left(f(r)\,f''(r)-2\,(f'(r))^2+\frac{3}{r}\,f(r)\,f'(r)\right)\,f(r)}.\label{pp10} 
\end{equation}

\begin{figure}[ht!]
    \centering
    \includegraphics[width=0.45\linewidth]{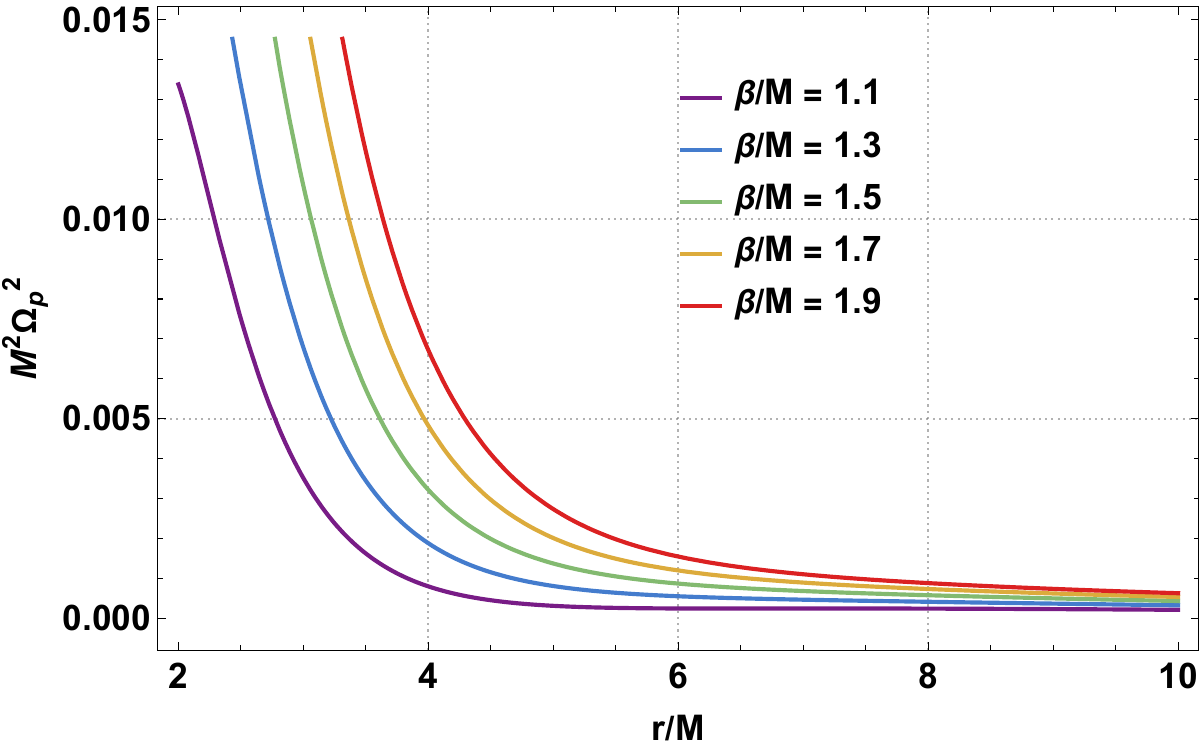}\qquad
    \includegraphics[width=0.45\linewidth]{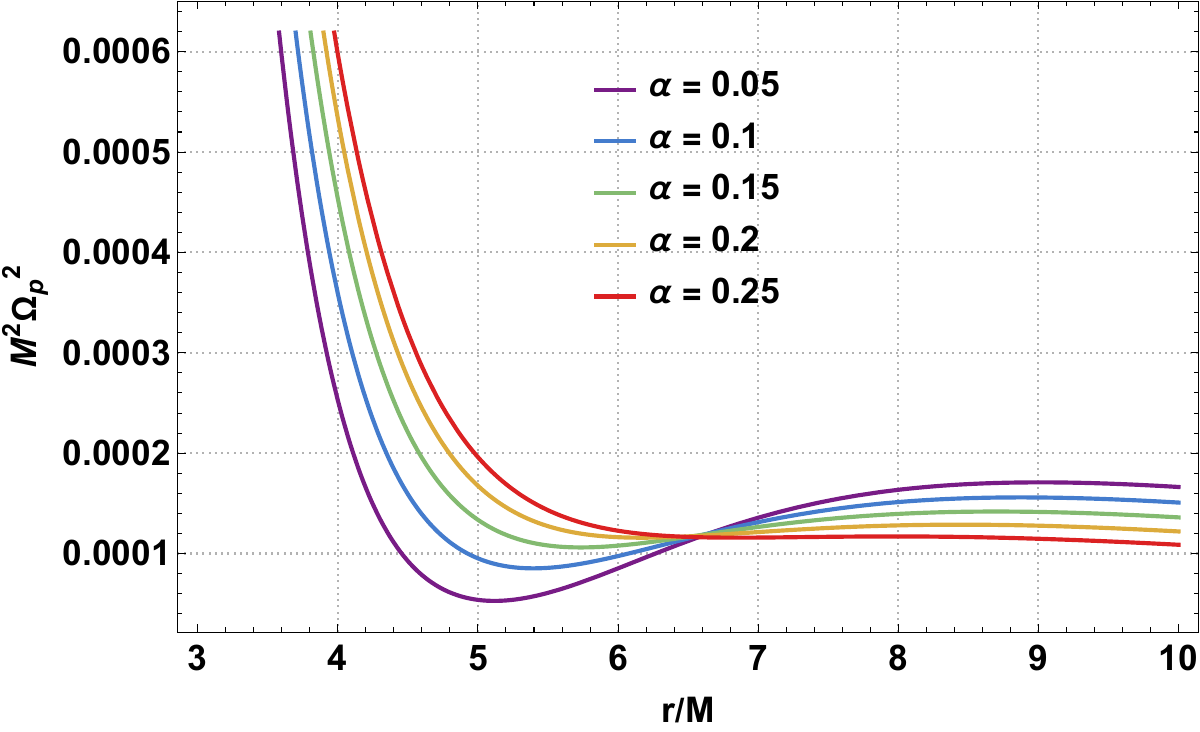}\\
    (a) $\alpha=0.1$ \hspace{8cm} (b) $\beta/M=1$
    \caption{Squared periastron frequency $M^2 \Omega^2_p$ as a function of dimensionless radial distance $r/M$ for various $\beta/M$ and $\alpha$. Here $Q/M=2,\,g/M=0.2$.}
    \label{fig:periastron-frequency}
\end{figure}

By substituting the metric function $f(r)$ into Eq.~(\ref{pp10}), we observe that the periastron frequency on the equatorial plane is influenced by the CoS parameter $\alpha$, the PFDM parameter $\beta$, the Hayward parameter $g$, the electric charge $Q$, and mass $M$ of the BH. Each of these parameters modifies the spacetime curvature and, consequently, alters both the epicyclic frequencies and the periastron frequency. The periastron precession is particularly relevant for testing GR in strong-field regimes, as deviations from the Schwarzschild prediction can reveal the presence of additional matter fields or modified gravity effects.

In Fig.~\ref{fig:periastron-frequency}, we present the squared periastron frequency $M^2\Omega_p^2$, defined in Eq.~(\ref{pp10}), on the equatorial plane ($\theta=\pi/2$) as a function of the dimensionless radial coordinate $r/M$. The behavior is analyzed by varying the PFDM parameter $\beta$ and the CoS parameter $\alpha$, while keeping the electric charge and Hayward parameter fixed. We find that increasing the values of $\beta/M$ and $\alpha$ leads to an enhancement of the periastron frequency, indicating stronger precession of test particle orbits as these parameters increase. This enhancement could potentially be detected through timing observations of X-ray binaries, providing constraints on the PFDM and CoS parameters. The combination of orbital, radial, vertical, and periastron frequencies forms a complete set of observables that can be used to test the charged Hayward BH model against astrophysical data from QPO measurements.

\section{Scalar Perturbations and Greybody Factors} \label{isec6}

Scalar perturbations provide an effective probe of the stability and dynamical properties of BH spacetimes under external disturbances. When a BH is surrounded by PFDM and coupled to a CoS, the background geometry deviates from the standard vacuum case, modifying the evolution of scalar fields governed by the Klein-Gordon equation. After separation of variables, the perturbation equation reduces to a Schr\"{o}dinger-like wave equation with an effective potential that explicitly depends on the PFDM and CoS parameters, thereby determining the QNM spectrum \cite{isz42,Becar2023,QT2025}. Previous studies have shown that PFDM alters the height and width of the effective potential barrier, leading to noticeable changes in both the real and imaginary parts of the QNM frequencies and hence in the oscillation and damping rates of perturbations \cite{Becar2023,QT2025}. Similarly, the inclusion of a CoS modifies the effective potential and influences the QNM spectrum, while preserving stability under scalar perturbations as indicated by the negative imaginary part of the frequencies \cite{YL2023}. The combined effects of PFDM and CoS therefore enrich the perturbative dynamics of BHs and may leave observable imprints in the ringdown phase of gravitational waves, making QNM analyses—via methods such as the WKB approximation and time-domain evolution—valuable tools for probing exotic matter distributions in astrophysical environments \cite{isz42,QT2025}.

\subsection{Klein-Gordon Equation}

The dynamics of a massless scalar field $\Phi$ in the BH background is governed by the Klein-Gordon equation:
\begin{equation}
\frac{1}{\sqrt{-g}}\partial_\mu\left(\sqrt{-g}g^{\mu\nu}\partial_\nu\Phi\right) = 0,\label{ss1}
\end{equation}
where $g_{\mu\nu}$ is the metric tensor. For the charged Hayward BH with CoS and PFDM, we have
\begin{equation}
    g_{\mu\nu}=\mbox{diag}\left(-f(r),\,\frac{1}{f(r)},\,r^2,\,r^2 \sin^2 \theta\right),\qquad g^{\mu\nu}=\mbox{diag}\left(-1/f(r),\,f(r),\,1/r^2,\,1/r^2 \sin^2 \theta\right)\label{ss2}
\end{equation}
with the determinant $g=\mbox{det}(g_{\mu\nu})=-r^4 \sin^2 \theta$.

Using the scalar field ansatz 
\begin{equation}
\Phi = \frac{\psi(r)}{r}Y^{m}_{\ell}(\theta,\phi)e^{-i\omega t},\label{ss3}    
\end{equation}
where $\omega$ is the frequency, $Y^m_\ell(\theta, \phi)$ denotes the spherical harmonics with multipole number $\ell$ and azimuthal number $m$, and $\psi(r)$ is the radial function.

Substituting the metric tensor in Eq.~(\ref{ss2}) and the scalar field ansatz in Eq.~(\ref{ss3}) into the massless Klein-Gordon equation (\ref{ss1}), we obtain the Schr\"{o}dinger-like wave equation:
\begin{equation}
\frac{d^2\psi}{dr_*^2} + \left[\omega^2 - V_{\text{s}}(r)\right]\psi = 0,\label{ss4}
\end{equation}
where 
\begin{equation}
    r_* =\int \frac{dr}{f(r)}\label{ss5}
\end{equation}
is the tortoise coordinate. The effective potential for scalar perturbations is given by
\begin{equation}
V_{\text{s}}(r) = \frac{f(r)}{r^2}\left[\ell(\ell+1)+r f'(r)\right].\label{ss6}
\end{equation}

Substituting the metric function $f(r)$ and after simplification, we obtain
\begin{equation}
    V_{\text{s}}(r) =\frac{1 - \alpha - \frac{2Mr^2}{r^3 + g^3} + \frac{Q^2}{r^2} + \frac{\beta}{r}\ln\!\frac{r}{|\beta|}}{r^2}\left[\ell(\ell+1)+ 2 M r^2 \frac{(r^3 - 2 g^3)}{(r^3 + g^3)^2} - \frac{2 Q^2}{r^2} + \frac{\beta}{r} \left(1 - \ln\frac{r}{|\beta|}\right)\right].\label{ss7}
\end{equation}

\begin{figure}[ht!]
    \centering
    \includegraphics[width=0.45\linewidth]{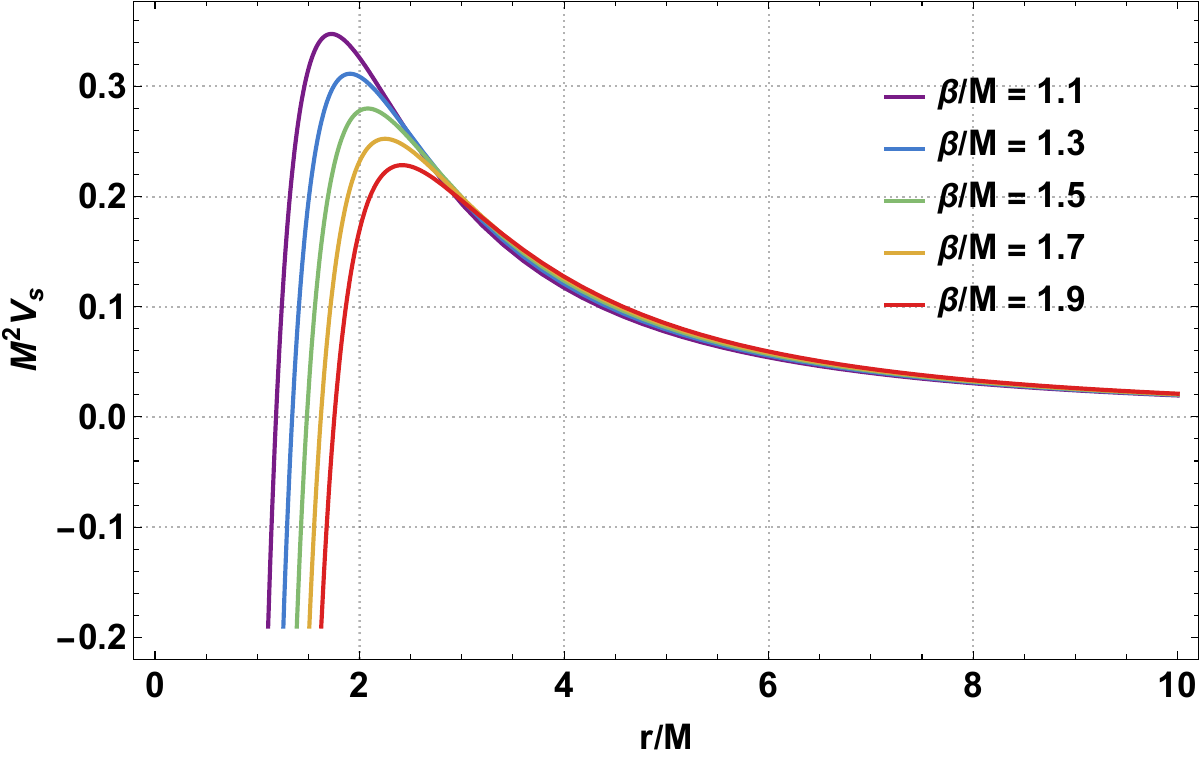}\quad
    \includegraphics[width=0.45\linewidth]{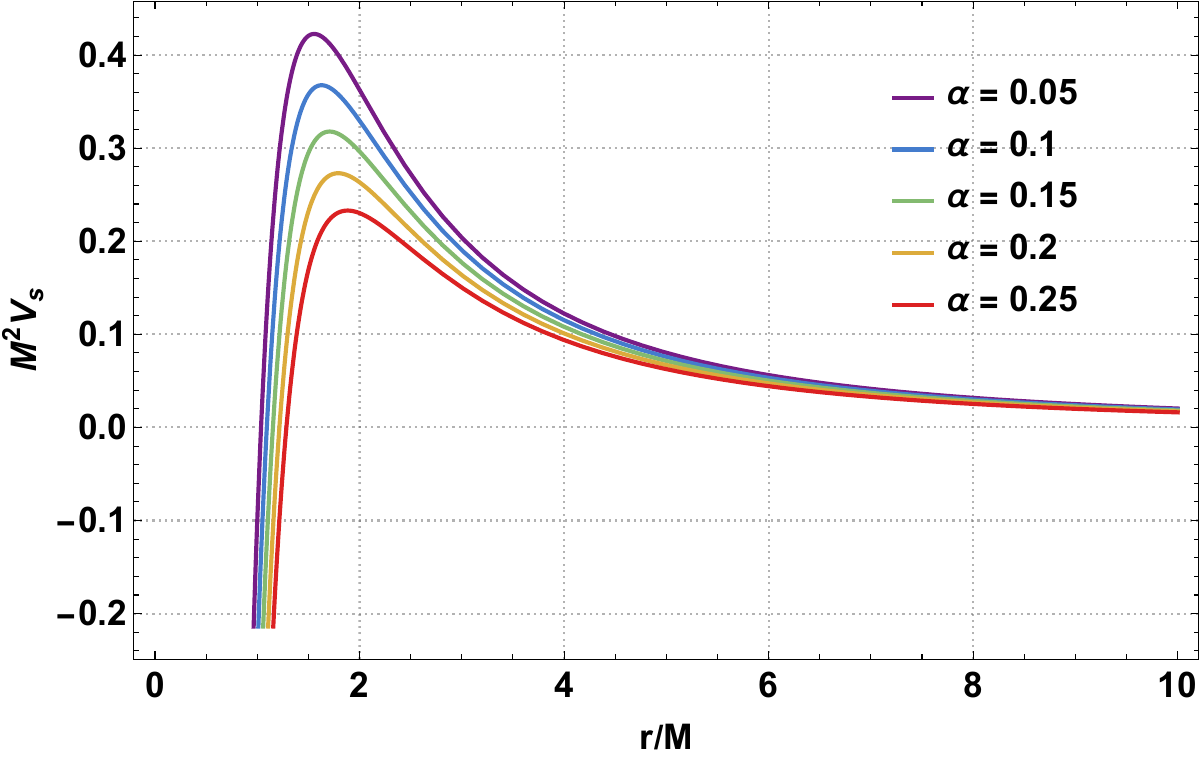}\\
    (a) $\alpha=0.1$ \hspace{6cm} (b) $\beta/M=1.0$
    \caption{Scalar perturbation potential for various $\alpha$ and $\beta/M$. Here $Q/M=1,\,g/M=0.2,\,\ell=1$.}
    \label{fig:scalar}
\end{figure}

From Eq.~(\ref{ss7}), we observe that the scalar perturbation potential governing the propagation of a massless scalar field in the BH background depends explicitly on several geometric parameters. These include the CoS parameter $\alpha$, the PFDM parameter $\beta$, the Hayward parameter $g$, and the BH's charge $Q$ and mass $M$. Each of these parameters influences the spacetime curvature and, consequently, alters the perturbative potential experienced by the scalar field. As a result, variations in these parameters can affect the height, shape, and location of the potential peak, thereby modifying the spectrum of QNMs and the stability of scalar perturbations in the BH spacetime.

In Figure~\ref{fig:scalar}, we illustrate the scalar perturbation potential given in Eq.~(\ref{ss7}) as a function of the dimensionless radial distance $r/M$, varying the PFDM parameter $\beta$ and the CoS parameter $\alpha$, while keeping the other parameters fixed. We observe that increasing $\beta/M$ and $\alpha$ lowers the height of the potential peak, indicating a reduction in the effective gravitational barrier encountered by the massless scalar field. This suppression of the potential peak suggests that scalar waves can propagate more easily in the BH spacetime for higher values of $\beta$ and $\alpha$, which may lead to longer-lived QNMs and slower decay of scalar perturbations in such backgrounds.

\subsection{Transmission and Reflection Probabilities}

In this subsection, we investigate the transmission and reflection probabilities of the charged Hayward BH surrounded by a CoS and PFDM by employing a semi-analytic approach. This method provides rigorous bounds on the transmission and reflection coefficients for one-dimensional potential scattering problems \cite{isz46,gf3,gf4}.  

Following Refs.~\cite{isz46,gf3,gf4,gf5,gf6}, the lower bound on the greybody factor (transmission probability) is given by
\begin{equation}\label{kk1}
T(\omega)\ge \text{sech}^2\left(\int_{-\infty}^{+\infty} \wp \, dr_* \right),
\end{equation}
while the corresponding upper bound on the reflection probability reads
\begin{equation}\label{kk2}
R(\omega)\le \tanh^2\left(\int_{-\infty}^{+\infty} \wp \, dr_* \right),
\end{equation}
where the function $\wp$ is defined as
\begin{equation}\label{kk3}
\wp=\frac{\sqrt{(H^{\prime})^2+\left(\omega^2-V_{\text{eff}}-H^2\right)^2}}{2H}.
\end{equation}
Here, $H$ is a positive function satisfying the conditions $H(r_*)>0$ and $H(\pm\infty)=\omega$, and $V_{\text{eff}}$ denotes the effective potential governing null geodesics, as given in Eq.~(\ref{bb5}).

Without loss of generality, we choose $H^2=\omega^2-V_{\text{eff}}$ in Eq.~(\ref{kk3}). With this choice, the bounds simplify to
\begin{equation}\label{kk4}
T(\omega)\ge \text{sech}^2\left(\frac{1}{2}\int_{-\infty}^{+\infty}\left|\frac{H'}{H}\right|\,dr_*\right),
\end{equation}
and
\begin{equation}\label{kk5}
R(\omega)\le \tanh^2\left(\frac{1}{2}\int_{-\infty}^{+\infty}\left|\frac{H'}{H}\right|\,dr_*\right).
\end{equation}

The integrand yields a logarithmic contribution of the form $\text{sech}^2\!\left[\ln\!\left(H_{\text{peak}}/H\right)\right]$. To avoid potential divergences, the integration domain is divided into three regions: $-\infty<r_*<r_{V_{\text{max}}}$, $r_{V_{\text{max}}}<r_*<r_{V_{\text{min}}}$, and $r_{V_{\text{min}}}<r_*<+\infty$. Consequently, the transmission and reflection probabilities can be expressed in terms of the peak value of the effective potential $V_{\text{peak}}$ as \cite{gf5}
\begin{equation}\label{kk6}
T(\omega)\ge\frac{4\,\omega^2\left(\omega^2-V_{\text{peak}}\right)}{\left(2\,\omega^2-V_{\text{peak}}\right)^2},
\end{equation}
and
\begin{equation}\label{kk7}
R(\omega)\le\frac{V_{\text{peak}}^2}{\left(2\,\omega^2-V_{\text{peak}}\right)^2}.
\end{equation}

\begin{figure}[ht!]
    \centering
    \includegraphics[width=0.45\linewidth]{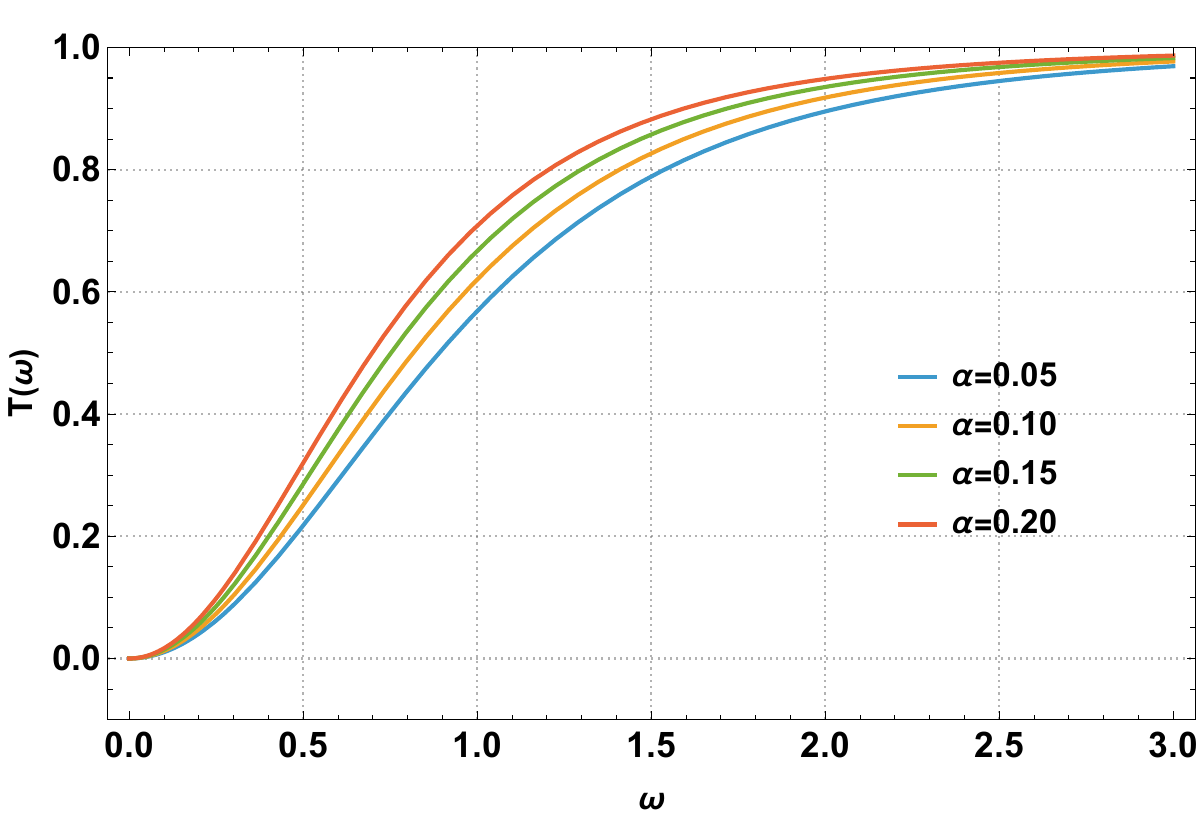}\quad
    \includegraphics[width=0.45\linewidth]{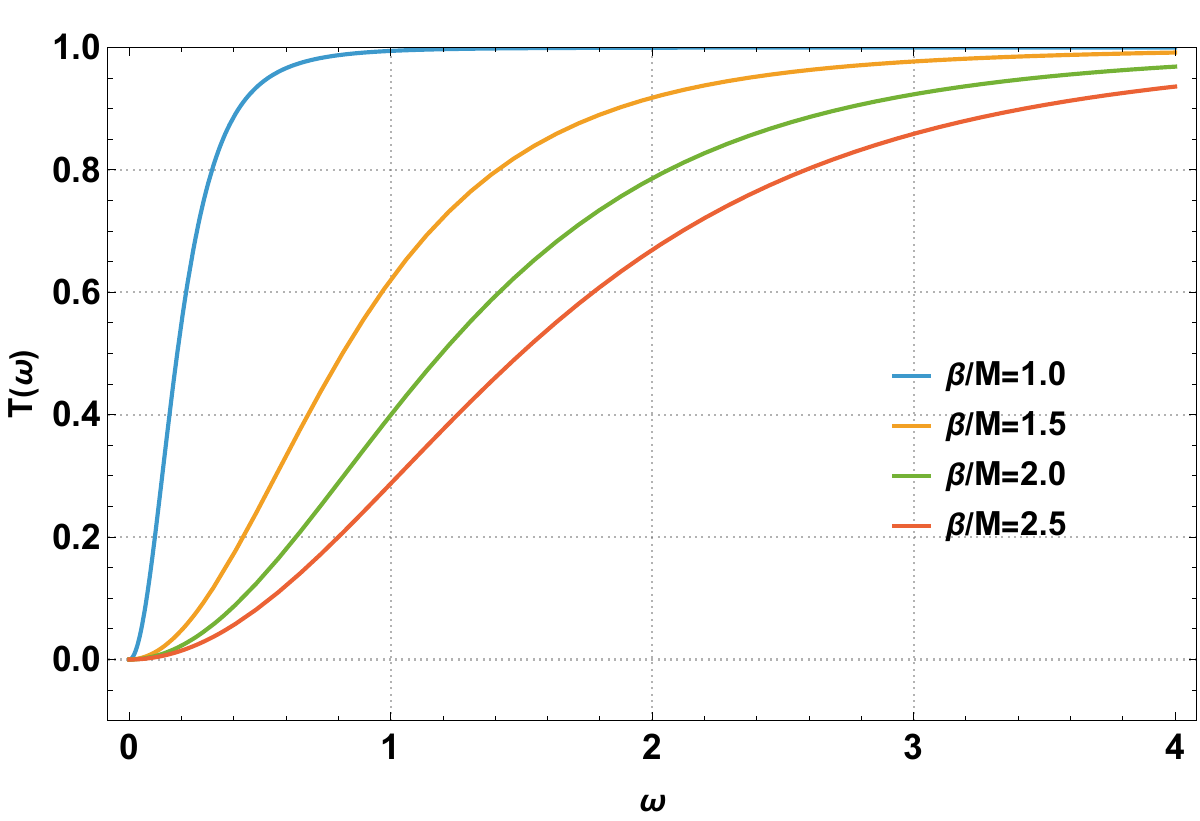}\\
    (a) $\beta /M=1.5$ \hspace{6cm} (b) $\alpha=0.1$
    \caption{Transmission probability of charged Hayward BH for various values of the CoS parameter $\alpha$ in (a) and PFDM parameter $\beta$ in (b). Here $\ell=2$, $Q/M=1$, and $g/M=0.25$.}
    \label{figA01}
\end{figure}

\begin{figure}[ht!]
    \centering
    \includegraphics[width=0.45\linewidth]{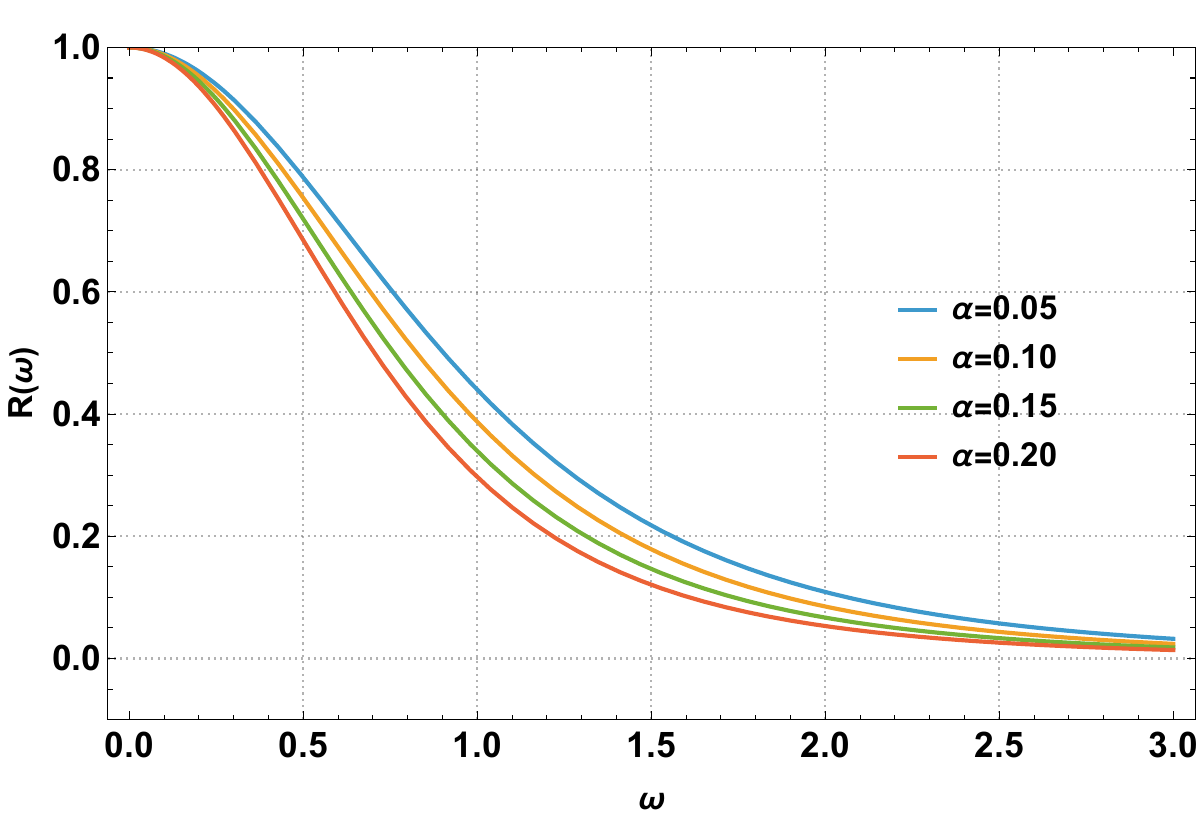}\quad
    \includegraphics[width=0.45\linewidth]{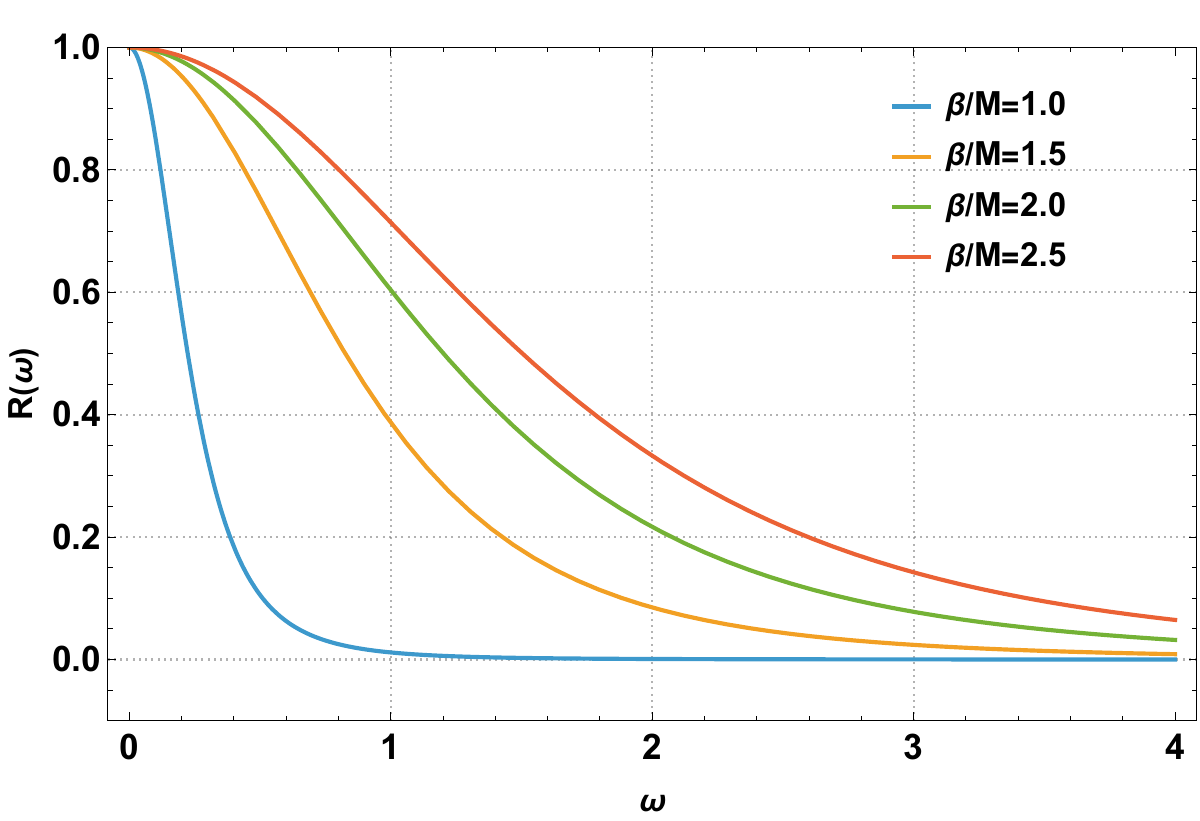}\\
    (a) $\beta/M=1.5$ \hspace{6cm} (b) $\alpha=0.1$
    \caption{Reflection probability of charged Hayward BH for various values of the CoS parameter $\alpha$ in (a) and PFDM parameter $\beta$ in (b). Here $\ell=2$, $Q/M=1$, and $g/M=0.25$.}
    \label{figA02}
\end{figure}

Figure~\ref{figA01} illustrates the influence of the BH parameters on the transmission probability. We observe that increasing the PFDM parameter $\beta$ leads to a suppression of the transmission probability, whereas increasing the CoS parameter $\alpha$ enhances transmission. The suppression caused by $\beta$ can be attributed to the modification of the effective potential barrier by the PFDM, which increases the scattering of scalar waves. Conversely, the CoS reduces the effective gravitational barrier through the solid angle deficit, allowing more radiation to escape. As expected, parameters that reduce the transmission probability correspondingly enhance the reflection probability, as shown in Figure~\ref{figA02}. These results demonstrate that the greybody factors of the charged Hayward BH are sensitive to both the PFDM and CoS parameters, offering potential observational signatures for constraining these parameters through gravitational wave observations.

\subsection{WKB Approximation}

Based on the effective potential obtained above, we study the QNM frequencies of the charged Hayward BH with CoS and PFDM. The QNMs of this BH spacetime can be calculated using the WKB method, which allows us to investigate the associated perturbative properties. QNMs arise only when appropriate boundary conditions are imposed, namely
\begin{equation}
\Psi_{\omega}(r_*) \sim e^{-i\omega r_*}, \qquad r_* \to -\infty,
\label{bc1}
\end{equation}
\begin{equation}
\Psi_{\omega}(r_*) \sim e^{+i\omega r_*}, \qquad r_* \to +\infty,
\label{bc2}
\end{equation}
where purely ingoing modes occur at $r_* \to -\infty$ (the event horizon), while purely outgoing modes occur at $r_* \to +\infty$ (spatial infinity). These boundary conditions ensure that no radiation emerges from the horizon and no radiation comes in from infinity, corresponding to the physical situation of a perturbed BH settling down to equilibrium.

The QNM spectrum cannot be obtained analytically due to the complexity of the effective potential. Therefore, a numerical approach is required. The WKB method, first proposed by Schutz and Will \cite{isz43}, was originally developed to study BH scattering problems. Subsequent improvements were introduced by Iyer and Will \cite{isz44} and later by Konoplya \cite{isz42}, extending the method up to 13th order. In this work, we employ the commonly used 6th-order WKB approximation, expressed as
\begin{equation}
\frac{i\left(\omega^2 - V_0\right)}{\sqrt{-2V_0''}} - \sum_{i=2}^{6} \Lambda_i
= n + \frac{1}{2}, \qquad (n = 0,1,2,\ldots),
\label{wkb6}
\end{equation}
where the prime denotes differentiation with respect to the tortoise coordinate $r_*$, $V_0$ is the maximum value of the effective potential, and $\Lambda_i$ are higher-order correction terms depending on the derivatives of the potential \cite{isz42}. The WKB method becomes less accurate when the multipole number $\ell$ is comparable to or smaller than the overtone number $n$. To improve numerical accuracy, we adopt the Pad\'{e}-averaged 6th-order WKB method \cite{MatyjasekOpala2017}. The associated error can be estimated as
\begin{equation}
\Delta = \frac{|\omega_7 - \omega_5|}{2},
\label{error}
\end{equation}
where $\omega_7$ and $\omega_5$ correspond to the QNM frequencies obtained from the 7th- and 5th-order WKB approximations, respectively.

\subsection{Connection between Shadow and QNMs}

In the eikonal limit ($\ell \gg 1$), the BH shadow exhibits a specific relationship with the QNM frequencies. This correspondence is intrinsically linked to the validity of the WKB approximation; when the WKB approximation becomes inapplicable, even the relationship in the limit $\ell \gg 1$ may break down \cite{RAK2023,RAK2017,SVB2024}. This pioneering connection was first proposed by Cardoso \textit{et al.}~\cite{isz45}, who showed that the real part of the QNM frequencies is related to the angular velocity $\Omega^{\rm null}_{\phi}(r_{s})$ of unstable circular null geodesics, while the imaginary part is governed by the Lyapunov exponent $\lambda_L$, which characterizes the instability timescale of these orbits. Consequently, QNM frequencies can be determined through the photon sphere radius via this correspondence.

In the eikonal limit, the QNM frequencies are given by
\begin{equation}
\omega_{\ell \gg 1} = \Omega^{\rm null}_{\phi}(r_{s})\, \ell - i \left(n + \frac{1}{2}\right) |\lambda_L|,
\label{ww1}
\end{equation}
where $n$ and $\ell$ denote the overtone number and the multipole number, respectively. The angular velocity $\Omega^{\rm null}_{\phi}(r_{s})$ at the unstable circular null geodesic is given in Eq.~(\ref{vel2}), and the Lyapunov exponent $\lambda_L$ can be expressed as
\begin{equation}
\lambda_L = \sqrt{\frac{f(r_{s})\left[2f(r_{s}) - r_{s}^2 f''(r_{s})\right]}{2r_{s}^2}} =\Omega^{\rm null}_{\phi}(r_s)\,\sqrt{f(r_{s}) - r_{s}^2 f''(r_{s})/2}.\label{ww2}
\end{equation}

The real and imaginary parts of the QNM frequencies can thus be written as functions of the photon sphere radius. By exploiting the relation between the photon sphere radius $r_{s}$ and the shadow radius $R_{\rm sh}$, one finds that the real part of the QNM frequency is directly related to the shadow size in the eikonal limit:
\begin{equation}
\omega_{\Re}=\Omega^{\rm null}_{\phi}(r_{s})\, \ell= \lim_{\ell \gg 1} \frac{\ell\, (1-\alpha)^{1/2}}{R_{\rm sh}},
\label{ww3}
\end{equation}
where we have used Eq.~(\ref{vel2}). Here, $\omega_{\Re}$ denotes the real part of the QNM frequency, while $R_{\rm sh}$ represents the BH shadow radius. This relation demonstrates that a larger shadow corresponds to lower oscillation frequencies, providing a direct link between observable shadow properties and the BH's vibrational characteristics.

Moreover, the Lyapunov exponent (\ref{ww2}) can also be expressed in terms of shadow radius as
\begin{equation}
    \lambda_L=\frac{(1-\alpha)^{1/2}}{R_{\rm sh}}\,\zeta,\qquad \zeta=\sqrt{f(r_{s}) - r_{s}^2 f''(r_{s})/2}.\label{ww4}
\end{equation}
Accordingly, Eq.~(\ref{ww1}) can be rewritten as
\begin{equation}
\omega_{\ell \gg 1} =\frac{(1-\alpha)^{1/2}}{R_{\rm sh}} \left[\ell-i \left(n + \frac{1}{2}\right) |\zeta|\right].\label{ww5}
\end{equation}
The importance of this relationship lies in the fact that the BH shadow can be directly obtained through astronomical observations. Therefore, it is feasible to use the shadow instead of the angular velocity to represent the real and imaginary parts of the QNMs. Using this method, the size of the BH shadow can be inferred from the real part of QNMs without employing the geodesic method, providing an alternative route to constrain BH parameters through gravitational wave observations of the ringdown phase.

\begin{table}[ht!]
\centering
\caption{QNM frequencies $\omega$ for $n=0,\,\ell=1$ with different values of $\alpha$ and $\beta$. Here $Q/M=1$.}
\begin{minipage}{0.48\textwidth}
\centering
\caption*{(i)\,$g/M=0.1$}
\small
\begin{tabular}{|c|c|c|}
\hline
$\alpha $ & $\beta/M$ & $\omega$ \\
\hline
0.10 & 1.0 & $0.363531-0.189521\,i$ \\
0.10 & 1.2 & $0.338965-0.192689\,i$ \\
0.10 & 1.4 & $0.317585-0.190908\,i$ \\
0.10 & 1.6 & $0.298710-0.186881\,i$ \\
0.10 & 1.8 & $0.281911-0.181799\,i$ \\
0.10 & 2.0 & $0.266867-0.176264\,i$ \\
0.15 & 1.0 & $0.339951-0.174925\,i$ \\
0.15 & 1.2 & $0.318866-0.178108\,i$ \\
0.15 & 1.4 & $0.299917-0.176907\,i$ \\
0.15 & 1.6 & $0.282875-0.173581\,i$ \\
0.15 & 1.8 & $0.267527-0.169200\,i$ \\
0.15 & 2.0 & $0.253668-0.164328\,i$ \\
0.20 & 1.0 & $0.316976-0.160673\,i$ \\
0.20 & 1.2 & $0.299119-0.163951\,i$ \\
0.20 & 1.4 & $0.282482-0.163309\,i$ \\
0.20 & 1.6 & $0.267210-0.160651\,i$ \\
0.20 & 1.8 & $0.253272-0.156939\,i$ \\
0.20 & 2.0 & $0.240571-0.152701\,i$ \\
0.25 & 1.0 & $0.294585-0.146804\,i$ \\
0.25 & 1.2 & $0.279720-0.150230\,i$ \\
0.25 & 1.4 & $0.265281-0.150122\,i$ \\
0.25 & 1.6 & $0.251714-0.148096\,i$ \\
0.25 & 1.8 & $0.239146-0.145019\,i$ \\
0.25 & 2.0 & $0.227576-0.141385\,i$ \\
0.30 & 1.0 & $0.272760-0.133354\,i$ \\
0.30 & 1.2 & $0.260666-0.136957\,i$ \\
0.30 & 1.4 & $0.248314-0.137354\,i$ \\
0.30 & 1.6 & $0.236387-0.135923\,i$ \\
0.30 & 1.8 & $0.225150-0.133446\,i$ \\
0.30 & 2.0 & $0.214683-0.130387\,i$ \\
\hline
\end{tabular}
\end{minipage}%
\hfill
\begin{minipage}{0.48\textwidth}
\centering
\caption*{(ii)\,$g/M=0.2$}
\small
\begin{tabular}{|c|c|c|}
\hline
$\alpha$ & $\beta/M$ & $\omega$ \\
\hline
0.10 & 1.0 & $0.364348-0.188379\,i$ \\
0.10 & 1.2 & $0.339397-0.192187\,i$ \\
0.10 & 1.4 & $0.317840-0.190643\,i$ \\
0.10 & 1.6 & $0.298871-0.186725\,i$ \\
0.10 & 1.8 & $0.282018-0.181701\,i$ \\
0.10 & 2.0 & $0.266942-0.176200\,i$ \\
0.15 & 1.0 & $0.340610-0.174060\,i$ \\
0.15 & 1.2 & $0.319228-0.177707\,i$ \\
0.15 & 1.4 & $0.300135-0.176689\,i$ \\
0.15 & 1.6 & $0.283016-0.173451\,i$ \\
0.15 & 1.8 & $0.267622-0.169118\,i$ \\
0.15 & 2.0 & $0.253734-0.164273\,i$ \\
0.20 & 1.0 & $0.317505-0.160020\,i$ \\
0.20 & 1.2 & $0.299421-0.163633\,i$ \\
0.20 & 1.4 & $0.282668-0.163131\,i$ \\
0.20 & 1.6 & $0.267331-0.160543\,i$ \\
0.20 & 1.8 & $0.253355-0.156869\,i$ \\
0.20 & 2.0 & $0.240629-0.152654\,i$ \\
0.25 & 1.0 & $0.295008-0.146315\,i$ \\
0.25 & 1.2 & $0.279970-0.149979\,i$ \\
0.25 & 1.4 & $0.265439-0.149979\,i$ \\
0.25 & 1.6 & $0.251818-0.148007\,i$ \\
0.25 & 1.8 & $0.239218-0.144961\,i$ \\
0.25 & 2.0 & $0.227627-0.141346\,i$ \\
0.30 & 1.0 & $0.273095-0.132991\,i$ \\
0.30 & 1.2 & $0.260872-0.136761\,i$ \\
0.30 & 1.4 & $0.248446-0.137239\,i$ \\
0.30 & 1.6 & $0.236476-0.135850\,i$ \\
0.30 & 1.8 & $0.225211-0.133398\,i$ \\
0.30 & 2.0 & $0.214727-0.130354\,i$ \\
\hline
\end{tabular}
\label{tab:qnm1}
\end{minipage}
\end{table}

\begin{table}[ht!]
\centering
\caption{QNM frequencies $\omega$ for $n=0,\,\ell=2$ with different values of $\alpha$ and $\beta$. Here $Q/M=1$.}
\begin{minipage}{0.48\textwidth}
\centering
\caption*{(i)\,$g/M=0.1$}
\small
\begin{tabular}{|c|c|c|}
\hline
$\alpha$ & $\beta/M$ & $\omega$ \\
\hline
0.10 & 1.0 & $0.727063-0.189521\,i$ \\
0.10 & 1.2 & $0.677930-0.192689\,i$ \\
0.10 & 1.4 & $0.635170-0.190908\,i$ \\
0.10 & 1.6 & $0.597419-0.186881\,i$ \\
0.10 & 1.8 & $0.563821-0.181799\,i$ \\
0.10 & 2.0 & $0.533735-0.176264\,i$ \\
0.15 & 1.0 & $0.679903-0.174925\,i$ \\
0.15 & 1.2 & $0.637733-0.178108\,i$ \\
0.15 & 1.4 & $0.599833-0.176907\,i$ \\
0.15 & 1.6 & $0.565751-0.173581\,i$ \\
0.15 & 1.8 & $0.535055-0.169200\,i$ \\
0.15 & 2.0 & $0.507337-0.164328\,i$ \\
0.20 & 1.0 & $0.633952-0.160673\,i$ \\
0.20 & 1.2 & $0.598239-0.163951\,i$ \\
0.20 & 1.4 & $0.564964-0.163309\,i$ \\
0.20 & 1.6 & $0.534420-0.160651\,i$ \\
0.20 & 1.8 & $0.506545-0.156939\,i$ \\
0.20 & 2.0 & $0.481143-0.152701\,i$ \\
0.25 & 1.0 & $0.589169-0.146804\,i$ \\
0.25 & 1.2 & $0.559441-0.150230\,i$ \\
0.25 & 1.4 & $0.530563-0.150122\,i$ \\
0.25 & 1.6 & $0.503427-0.148096\,i$ \\
0.25 & 1.8 & $0.478293-0.145019\,i$ \\
0.25 & 2.0 & $0.455152-0.141385\,i$ \\
0.30 & 1.0 & $0.545520-0.133354\,i$ \\
0.30 & 1.2 & $0.521332-0.136957\,i$ \\
0.30 & 1.4 & $0.496628-0.137354\,i$ \\
0.30 & 1.6 & $0.472774-0.135923\,i$ \\
0.30 & 1.8 & $0.450299-0.133446\,i$ \\
0.30 & 2.0 & $0.429367-0.130387\,i$ \\
\hline
\end{tabular}
\end{minipage}%
\hfill
\begin{minipage}{0.48\textwidth}
\centering
\caption*{(ii)\,$g/M=0.2$}
\small
\begin{tabular}{|c|c|c|}
\hline
$\alpha$ & $\beta/M$ & $\omega$ \\
\hline
0.10 & 1.0 & $0.728695-0.188379\,i$ \\
0.10 & 1.2 & $0.678793-0.192187\,i$ \\
0.10 & 1.4 & $0.635680-0.190643\,i$ \\
0.10 & 1.6 & $0.597742-0.186725\,i$ \\
0.10 & 1.8 & $0.564036-0.181701\,i$ \\
0.10 & 2.0 & $0.533883-0.176200\,i$ \\
0.15 & 1.0 & $0.681221-0.174060\,i$ \\
0.15 & 1.2 & $0.638457-0.177707\,i$ \\
0.15 & 1.4 & $0.600270-0.176689\,i$ \\
0.15 & 1.6 & $0.566031-0.173451\,i$ \\
0.15 & 1.8 & $0.535243-0.169118\,i$ \\
0.15 & 2.0 & $0.507468-0.164273\,i$ \\
0.20 & 1.0 & $0.635011-0.160020\,i$ \\
0.20 & 1.2 & $0.598842-0.163633\,i$ \\
0.20 & 1.4 & $0.565336-0.163131\,i$ \\
0.20 & 1.6 & $0.534662-0.160543\,i$ \\
0.20 & 1.8 & $0.506709-0.156869\,i$ \\
0.20 & 2.0 & $0.481258-0.152654\,i$ \\
0.25 & 1.0 & $0.590015-0.146315\,i$ \\
0.25 & 1.2 & $0.559941-0.149979\,i$ \\
0.25 & 1.4 & $0.530878-0.149979\,i$ \\
0.25 & 1.6 & $0.503635-0.148007\,i$ \\
0.25 & 1.8 & $0.478435-0.144961\,i$ \\
0.25 & 2.0 & $0.455253-0.141346\,i$ \\
0.30 & 1.0 & $0.546191-0.132991\,i$ \\
0.30 & 1.2 & $0.521744-0.136761\,i$ \\
0.30 & 1.4 & $0.496893-0.137239\,i$ \\
0.30 & 1.6 & $0.472951-0.135850\,i$ \\
0.30 & 1.8 & $0.450422-0.133398\,i$ \\
0.30 & 2.0 & $0.429454-0.130354\,i$ \\
\hline
\end{tabular}
\label{tab:qnm2}
\end{minipage}
\end{table}

\begin{table}[ht!]
\centering
\caption{QNM frequencies $\omega$ for $n=0,\,\ell=3$ with different values of $\alpha$ and $\beta$. Here $Q/M=1$.}

\begin{minipage}[t]{0.48\textwidth}
\centering
\caption*{(i)\,$g/M=0.1$}
\small
\begin{tabular}{|c|c|c|}
\hline
$\alpha$ & $\beta$ & $\omega$ \\
\hline
0.10 & 1.0 & $1.09059-0.189521\,i$ \\
0.10 & 1.2 & $1.0169-0.192689\,i$ \\
0.10 & 1.4 & $0.952755-0.190908\,i$ \\
0.10 & 1.6 & $0.896129-0.186881\,i$ \\
0.10 & 1.8 & $0.845732-0.181799\,i$ \\
0.10 & 2.0 & $0.800602-0.176264\,i$ \\
0.15 & 1.0 & $1.01985-0.174925\,i$ \\
0.15 & 1.2 & $0.956599-0.178108\,i$ \\
0.15 & 1.4 & $0.89975-0.176907\,i$ \\
0.15 & 1.6 & $0.848626-0.173581\,i$ \\
0.15 & 1.8 & $0.802582-0.1692\,i$ \\
0.15 & 2.0 & $0.761005-0.164328\,i$ \\
0.20 & 1.0 & $0.950928-0.160673\,i$ \\
0.20 & 1.2 & $0.897358-0.163951\,i$ \\
0.20 & 1.4 & $0.847446-0.163309\,i$ \\
0.20 & 1.6 & $0.80163-0.160651\,i$ \\
0.20 & 1.8 & $0.759817-0.156939\,i$ \\
0.20 & 2.0 & $0.721714-0.152701\,i$ \\
0.25 & 1.0 & $0.883754-0.146804\,i$ \\
0.25 & 1.2 & $0.839161-0.15023\,i$ \\
0.25 & 1.4 & $0.795844-0.150122\,i$ \\
0.25 & 1.6 & $0.755141-0.148096\,i$ \\
0.25 & 1.8 & $0.717439-0.145019\,i$ \\
0.25 & 2.0 & $0.682729-0.141385\,i$ \\
0.30 & 1.0 & $0.81828-0.133354\,i$ \\
0.30 & 1.2 & $0.781999-0.136957\,i$ \\
0.30 & 1.4 & $0.744942-0.137354\,i$ \\
0.30 & 1.6 & $0.70916-0.135923\,i$ \\
0.30 & 1.8 & $0.675449-0.133446\,i$ \\
0.30 & 2.0 & $0.64405-0.130387\,i$ \\
\hline
\end{tabular}
\end{minipage}%
\hfill
\begin{minipage}[t]{0.48\textwidth}
\centering
\caption*{(ii)\, $g/M=0.2$}
\small
\begin{tabular}{|c|c|c|}
\hline
$\alpha$ & $\beta$ & $\omega$ \\
\hline
0.10 & 1.0 & $1.09304-0.188379\,i$ \\
0.10 & 1.2 & $1.01819-0.192187\,i$ \\
0.10 & 1.4 & $0.95352-0.190643\,i$ \\
0.10 & 1.6 & $0.896613-0.186725\,i$ \\
0.10 & 1.8 & $0.846054-0.181701\,i$ \\
0.10 & 2.0 & $0.800825-0.1762\,i$ \\
0.15 & 1.0 & $1.02183-0.17406\,i$ \\
0.15 & 1.2 & $0.957685-0.177707\,i$ \\
0.15 & 1.4 & $0.900405-0.176689\,i$ \\
0.15 & 1.6 & $0.849047-0.173451\,i$ \\
0.15 & 1.8 & $0.802865-0.169118\,i$ \\
0.15 & 2.0 & $0.761202-0.164273\,i$ \\
0.20 & 1.0 & $0.952516-0.16002\,i$ \\
0.20 & 1.2 & $0.898264-0.163633\,i$ \\
0.20 & 1.4 & $0.848005-0.163131\,i$ \\
0.20 & 1.6 & $0.801993-0.160543\,i$ \\
0.20 & 1.8 & $0.760064-0.156869\,i$ \\
0.20 & 2.0 & $0.721887-0.152654\,i$ \\
0.25 & 1.0 & $0.885023-0.146315\,i$ \\
0.25 & 1.2 & $0.839911-0.149979\,i$ \\
0.25 & 1.4 & $0.796317-0.149979\,i$ \\
0.25 & 1.6 & $0.755453-0.148007\,i$ \\
0.25 & 1.8 & $0.717653-0.144961\,i$ \\
0.25 & 2.0 & $0.68288-0.141346\,i$ \\
0.30 & 1.0 & $0.819286-0.132991\,i$ \\
0.30 & 1.2 & $0.782615-0.136761\,i$ \\
0.30 & 1.4 & $0.745339-0.137239\,i$ \\
0.30 & 1.6 & $0.709427-0.13585\,i$ \\
0.30 & 1.8 & $0.675633-0.133398\,i$ \\
0.30 & 2.0 & $0.644181-0.130354\,i$ \\
\hline
\end{tabular}
\end{minipage}
\end{table}

\section{Conclusions} \label{isec7}

In this paper, we investigated the charged Hayward BH coupled to a CoS and surrounded by PFDM. The metric function incorporated five physical parameters: the mass $M$, the Hayward regularization parameter $g$, the electric charge $Q$, the CoS parameter $\alpha$, and the PFDM parameter $\beta$. We examined the horizon structure, null and timelike geodesics, QPOs, and scalar perturbations, revealing how each parameter modifies the BH properties and observational signatures.

In Section~\ref{isec2}, we derived the metric function given in Eq.~(\ref{function}) and analyzed the horizon structure through both analytical limiting cases and numerical computations. Table~\ref{tab:horizons_hayward} presented 20 parameter configurations spanning non-extremal, extremal, single-horizon, and naked singularity cases. We found that the CoS parameter $\alpha$ increased the outer horizon radius following $r_h \approx 2M/(1-\alpha)$, while the PFDM parameter $\beta$ introduced logarithmic corrections that enabled extremal configurations when $\beta/M \gtrsim 2$. The extremal BH conditions required fine-tuning: achieving $f(r_{\rm ext})=0$ and $f'(r_{\rm ext})=0$ simultaneously demanded larger $\beta$ values as $\alpha$ increased, ranging from $\beta = 2.044$ at $\alpha=0$ to $\beta = 2.243$ at $\alpha=0.15$. The 3D visualizations of the metric function in Figure~\ref{fig:3d_metric} illustrated how the surface curvature near the horizon throat intensified with increasing $Q$ and $\beta$, while larger $\alpha$ reduced the asymptotic height toward $z_\infty = 1-\alpha$. These findings extend recent work on regular BHs in modified gravity theories \cite{isz01,isz02}.

Section~\ref{isec3} focused on null geodesics, where we computed the effective potential for photon dynamics given in Eq.~(\ref{bb3b}) and determined the photon sphere radius from the condition $2f(r_s) - r_s f'(r_s) = 0$. The numerical results in Tables~\ref{tab:1}--\ref{tab:2} demonstrated that both $\alpha$ and $\beta$ generally increased the photon sphere radius, though the dependence was non-monotonic in certain parameter regimes due to competing effects among the metric terms. Figure~\ref{fig:null} showed that increasing $\beta/M$ and $\alpha$ suppressed the effective potential peak, indicating a reduced gravitational barrier for photons. For the BH shadow, we employed the formalism appropriate for asymptotically non-flat spacetimes since $\lim_{r\to\infty} f(r) = 1-\alpha \neq 1$, obtaining the shadow radius expression in Eq.~(\ref{bb10}). The geodesic angular velocity at the photon sphere satisfied $\Omega_\phi^{\rm null}(r_s) = (1-\alpha)^{1/2}/R_{\rm sh}$, establishing a direct link between the orbital frequency and the observable shadow size. The photon trajectory equation~(\ref{bb15}) incorporated contributions from all BH parameters, with the PFDM logarithmic term introducing distinctive modifications to light deflection patterns. These results are consistent with recent shadow calculations for BHs in dark matter environments \cite{EHTL1,EHTL6}.

In Section~\ref{isec4}, we studied the dynamics of neutral test particles using the Hamiltonian formalism. The effective potential in Eq.~(\ref{cc10}) governed the particle motion, and we derived expressions for the specific angular momentum and specific energy for circular orbits in Eqs.~(\ref{cc13})--(\ref{cc14}). Figure~\ref{fig:timelike} revealed that increasing $\beta/M$ enhanced the effective potential at large radii, whereas increasing $\alpha$ suppressed it, weakening the gravitational barrier. The specific angular momentum increased with both $\alpha$ and $\beta$ (Figure~\ref{fig:specific-momentum}), indicating that particles required greater angular momentum to maintain circular orbits in these modified spacetimes. The specific energy exhibited opposite trends: it increased with $\beta$ but decreased with $\alpha$ (Figure~\ref{fig:specific-energy}). For the ISCO, we derived the governing equation~(\ref{cc27}), which could only be solved numerically due to the logarithmic PFDM term. The ISCO radius determines the inner edge of accretion disks around astrophysical BHs, and modifications induced by CoS and PFDM could potentially be constrained through X-ray observations of accreting systems \cite{EHTL12}.

Section~\ref{isec5} examined QPOs using the relativistic precession model. We computed the azimuthal frequency in Eq.~(\ref{pp2}) and noted that the CoS parameter $\alpha$ did not appear in this expression, as it enters the metric as a constant term that vanishes upon differentiation. The radial and vertical epicyclic frequencies given in Eqs.~(\ref{pp3})--(\ref{pp4}) exhibited opposite behaviors: increasing $\beta/M$ and $\alpha$ enhanced the radial frequency (Figure~\ref{fig:radial-frequency}) while suppressing the vertical frequency (Figure~\ref{fig:vertical-frequency}). This asymmetry has implications for QPO models that relate observed frequency ratios to the underlying spacetime geometry. The periastron frequency defined in Eq.~(\ref{pp10}) increased with both $\alpha$ and $\beta$, indicating stronger orbital precession in the presence of CoS and PFDM. These epicyclic frequencies are directly related to high-frequency QPO pairs observed in X-ray binaries and could provide tests of the charged Hayward BH model against astrophysical data.

In Section~\ref{isec6}, we analyzed scalar perturbations governed by the Klein-Gordon equation. The effective potential for scalar waves given in Eq.~(\ref{ss7}) showed that increasing $\alpha$ and $\beta$ lowered the potential peak (Figure~\ref{fig:scalar}), suggesting that scalar perturbations decay more slowly in spacetimes with stronger CoS and PFDM effects. We computed the greybody factors using the semi-analytic bounds in Eqs.~(\ref{kk6})--(\ref{kk7}) and found that increasing $\beta$ suppressed the transmission probability while increasing $\alpha$ enhanced it (Figure~\ref{figA01}). The WKB method provided the QNM spectrum through Eq.~(\ref{wkb6}), and we established the connection between QNMs and the BH shadow in the eikonal limit via Eq.~(\ref{ww5}). This relation demonstrated that $\omega_\Re = \ell(1-\alpha)^{1/2}/R_{\rm sh}$, allowing shadow measurements to constrain QNM frequencies and vice versa. Such connections between observable quantities provide multiple independent probes of BH parameters and may help distinguish the charged Hayward BH with CoS and PFDM from other BH models through gravitational wave observations of the ringdown phase \cite{isz06}.

Our results demonstrated that the charged Hayward BH with CoS and PFDM exhibits distinctive features that differentiate it from standard Schwarzschild, RN, and Hayward BHs. The CoS parameter $\alpha$ introduced a solid angle deficit that affected the asymptotic structure and horizon radii, while the PFDM parameter $\beta$ contributed logarithmic corrections that modified the gravitational potential at all radii. The Hayward parameter $g$ regularized the central singularity, and the electric charge $Q$ provided additional electromagnetic contributions. The combined effects of these parameters created a diverse parameter space encompassing extremal configurations, enhanced photon spheres, modified accretion disk properties, and altered QNM spectra.

Several directions remain for future investigation. First, extending this analysis to rotating (Kerr-like) charged Hayward BHs with CoS and PFDM would enable comparison with EHT observations of M87* and Sgr~A*, which exhibit signatures of BH spin \cite{isz12,isconc2,isconc3}. Second, computing the full QNM spectrum numerically using time-domain integration or continued fraction methods would provide more accurate frequencies beyond the WKB approximation \cite{isconc4,isconc5,isz42}. Third, studying the thermodynamic properties including Hawking temperature, entropy, and phase transitions would reveal how CoS and PFDM affect BH thermodynamics \cite{isz30,isconc8,isconc9}. Fourth, analyzing gravitational lensing observables such as deflection angles, Einstein rings, and relativistic images would offer additional tests through optical observations \cite{isconc10,isconc11,isconc12}. Fifth, investigating the stability under gravitational perturbations would confirm whether the charged Hayward BH with CoS and PFDM remains stable against all perturbation modes \cite{isconc13,isconc14}. Finally, constraining the model parameters using observational data from the EHT, LIGO/Virgo gravitational wave detections, and X-ray timing measurements would establish whether such configurations exist in nature \cite{isconc15,isconc16,isconc17}. These future studies would further clarify the role of regular BHs, string clouds, and dark matter in strong-gravity astrophysical environments \cite{isconc18}.

\footnotesize

\section*{Acknowledgments}
F.A. gratefully acknowledges the Inter University Centre for Astronomy and Astrophysics (IUCAA), Pune, India, for the award of a visiting associateship. \.{I}.~S. is thankful for academic support provided by EMU, T\"{U}B\.{I}TAK, ANKOS, and SCOAP3, as well as for networking support received through COST Actions CA22113, CA21106, CA23130, and CA23115.

\section*{Conflict of Interest}

Authors declare(s) no such conflict of interest.

\bibliography{ref2}

\begin{thebibliography}{113}
\providecommand{\natexlab}[1]{#1}
\providecommand{\url}[1]{\texttt{#1}}
\expandafter\ifx\csname urlstyle\endcsname\relax
  \providecommand{\doi}[1]{doi: #1}\else
  \providecommand{\doi}{doi: \begingroup \urlstyle{rm}\Url}\fi

\bibitem[Abbott et~al.(2016{\natexlab{a}})]{isz01}
B.~P. Abbott et~al.
\newblock {Observation of Gravitational Waves from a Binary Black Hole Merger}.
\newblock \emph{Phys. Rev. Lett.}, 116\penalty0 (6):\penalty0 061102,
  2016{\natexlab{a}}.
\newblock \doi{10.1103/PhysRevLett.116.061102}.

\bibitem[Abbott et~al.(2017)]{isz02}
B.~P. Abbott et~al.
\newblock {GW170817: Observation of Gravitational Waves from a Binary Neutron
  Star Inspiral}.
\newblock \emph{Phys. Rev. Lett.}, 119\penalty0 (16):\penalty0 161101, 2017.
\newblock \doi{10.1103/PhysRevLett.119.161101}.

\bibitem[Akiyama et~al.(2019{\natexlab{a}})]{EHTL1}
Kazunori Akiyama et~al.
\newblock {First M87 Event Horizon Telescope Results. I. The Shadow of the
  Supermassive Black Hole}.
\newblock \emph{Astrophys. J. Lett.}, 875:\penalty0 L1, 2019{\natexlab{a}}.
\newblock \doi{10.3847/2041-8213/ab0ec7}.

\bibitem[Akiyama et~al.(2019{\natexlab{b}})]{EHTL4}
Kazunori Akiyama et~al.
\newblock {First M87 Event Horizon Telescope Results. IV. Imaging the Central
  Supermassive Black Hole}.
\newblock \emph{Astrophys. J. Lett.}, 875:\penalty0 L4, 2019{\natexlab{b}}.
\newblock \doi{10.3847/2041-8213/ab0e85}.

\bibitem[Akiyama et~al.(2019{\natexlab{c}})]{EHTL6}
Kazunori Akiyama et~al.
\newblock {First M87 Event Horizon Telescope Results. VI. The Shadow and Mass
  of the Central Black Hole}.
\newblock \emph{Astrophys. J. Lett.}, 875:\penalty0 L6, 2019{\natexlab{c}}.
\newblock \doi{10.3847/2041-8213/ab1141}.

\bibitem[Akiyama et~al.(2022{\natexlab{a}})]{EHTL12}
Kazunori Akiyama et~al.
\newblock {First Sagittarius A* Event Horizon Telescope Results. I. The Shadow
  of the Supermassive Black Hole in the Center of the Milky Way}.
\newblock \emph{Astrophys. J. Lett.}, 930\penalty0 (2):\penalty0 L12,
  2022{\natexlab{a}}.
\newblock \doi{10.3847/2041-8213/ac6674}.

\bibitem[Akiyama et~al.(2022{\natexlab{b}})]{EHTL14}
Kazunori Akiyama et~al.
\newblock {First Sagittarius A* Event Horizon Telescope Results. IV.
  Variability, Morphology, and Black Hole Mass}.
\newblock \emph{Astrophys. J. Lett.}, 930\penalty0 (2):\penalty0 L14,
  2022{\natexlab{b}}.
\newblock \doi{10.3847/2041-8213/ac6736}.

\bibitem[Akiyama et~al.(2022{\natexlab{c}})]{EHTL15}
Kazunori Akiyama et~al.
\newblock {First Sagittarius A* Event Horizon Telescope Results. V. Testing
  Astrophysical Models of the Galactic Center Black Hole}.
\newblock \emph{Astrophys. J. Lett.}, 930\penalty0 (2):\penalty0 L15,
  2022{\natexlab{c}}.
\newblock \doi{10.3847/2041-8213/ac6672}.

\bibitem[Akiyama et~al.(2022{\natexlab{d}})]{EHTL16}
Kazunori Akiyama et~al.
\newblock {First Sagittarius A* Event Horizon Telescope Results. VI. Testing
  the Black Hole Metric}.
\newblock \emph{Astrophys. J. Lett.}, 930\penalty0 (2):\penalty0 L16,
  2022{\natexlab{d}}.
\newblock \doi{10.3847/2041-8213/ac6756}.

\bibitem[Akiyama et~al.(2022{\natexlab{e}})]{EHTL17}
Kazunori Akiyama et~al.
\newblock {First Sagittarius A* Event Horizon Telescope Results. VII.
  Polarization of the Ring}.
\newblock \emph{Astrophys. J. Lett.}, 930\penalty0 (2):\penalty0 L17,
  2022{\natexlab{e}}.
\newblock \doi{10.3847/2041-8213/ac6675}.

\bibitem[Berti et~al.(2009)Berti, Cardoso, and Starinets]{isz06}
Emanuele Berti, Vitor Cardoso, and Andrei~O. Starinets.
\newblock {Quasinormal modes of black holes and black branes}.
\newblock \emph{Class. Quant. Grav.}, 26:\penalty0 163001, 2009.
\newblock \doi{10.1088/0264-9381/26/16/163001}.

\bibitem[Will(2014)]{isz07}
Clifford~M. Will.
\newblock {The Confrontation between General Relativity and Experiment}.
\newblock \emph{Living Rev. Rel.}, 17:\penalty0 4, 2014.
\newblock \doi{10.12942/lrr-2014-4}.

\bibitem[Bardeen(1968)]{isz08}
James~M. Bardeen.
\newblock {Non-singular general-relativistic gravitational collapse}.
\newblock In \emph{{Proceedings of GR5}}, page 174, Tbilisi, USSR, 1968.

\bibitem[Ayon-Beato and Garcia(2000)]{isz09}
Eloy Ayon-Beato and Alberto Garcia.
\newblock {The Bardeen model as a nonlinear magnetic monopole}.
\newblock \emph{Phys. Lett. B}, 493:\penalty0 149--152, 2000.
\newblock \doi{10.1016/S0370-2693(00)01125-4}.

\bibitem[Hayward(2006)]{isz10}
Sean~A. Hayward.
\newblock {Formation and Evaporation of Nonsingular Black Holes}.
\newblock \emph{Phys. Rev. Lett.}, 96:\penalty0 031103, 2006.
\newblock \doi{10.1103/PhysRevLett.96.031103}.

\bibitem[Halilsoy et~al.(2014)Halilsoy, Ovgun, and Mazharimousavi]{isz11}
M.~Halilsoy, A.~Ovgun, and S.~Habib Mazharimousavi.
\newblock {Thin shell wormholes from the regular Hayward black hole}.
\newblock \emph{Eur. Phys. J. C}, 74:\penalty0 2796, 2014.
\newblock \doi{10.1140/epjc/s10052-014-2796-4}.

\bibitem[Bambi and Modesto(2013)]{isz12}
Cosimo Bambi and Leonardo Modesto.
\newblock {Rotating regular black holes}.
\newblock \emph{Phys. Lett. B}, 721:\penalty0 329--334, 2013.
\newblock \doi{10.1016/j.physletb.2013.03.025}.

\bibitem[Chiba and Kimura(2017)]{isz13}
Takeshi Chiba and Masashi Kimura.
\newblock {A Note on Geodesics in the Hayward Metric}.
\newblock \emph{PTEP}, 2017\penalty0 (4):\penalty0 043E01, 2017.
\newblock \doi{10.1093/ptep/ptx037}.

\bibitem[Fernando and Correa(2012)]{isz14}
Sharmanthie Fernando and Juan Correa.
\newblock {Quasinormal Modes of the Bardeen Black Hole: Scalar Perturbations}.
\newblock \emph{Phys. Rev. D}, 86:\penalty0 064039, 2012.
\newblock \doi{10.1103/PhysRevD.86.064039}.

\bibitem[Wei and Liu(2013)]{isz15}
Shao-Wen Wei and Yu-Xiao Liu.
\newblock Observing the shadow of einstein--maxwell--dilaton--axion black hole.
\newblock \emph{JCAP}, 2013\penalty0 (11):\penalty0 063, 2013.
\newblock \doi{10.1088/1475-7516/2013/11/063}.

\bibitem[Amir and Ghosh(2016)]{isz16}
Muhammed Amir and Sushant~G. Ghosh.
\newblock {Shapes of rotating nonsingular black hole shadows}.
\newblock \emph{Phys. Rev. D}, 94\penalty0 (2):\penalty0 024054, 2016.
\newblock \doi{10.1103/PhysRevD.94.024054}.

\bibitem[Fan and Wang(2016)]{isz17}
Zhong-Ying Fan and Xiaobao Wang.
\newblock {Construction of Regular Black Holes in General Relativity}.
\newblock \emph{Phys. Rev. D}, 94\penalty0 (12):\penalty0 124027, 2016.
\newblock \doi{10.1103/PhysRevD.94.124027}.

\bibitem[Toshmatov et~al.(2017)Toshmatov, Stuchlik, and Ahmedov]{isz18}
Bobir Toshmatov, Zdenek Stuchlik, and Bobomurat Ahmedov.
\newblock {Rotating black hole solutions with quintessential energy}.
\newblock \emph{Eur. Phys. J. Plus}, 132:\penalty0 98, 2017.
\newblock \doi{10.1140/epjp/i2017-11373-4}.

\bibitem[Kiselev(2003)]{isz19}
V.~V. Kiselev.
\newblock {Quintessence and black holes}.
\newblock \emph{Class. Quant. Grav.}, 20:\penalty0 1187--1198, 2003.
\newblock \doi{10.1088/0264-9381/20/6/310}.

\bibitem[Xu et~al.(2018)Xu, Hou, and Wang]{isz20}
Zhaoyi Xu, Xian Hou, and Jiancheng Wang.
\newblock {Kerr-anti-de Sitter/de Sitter black hole in perfect fluid dark
  matter background}.
\newblock \emph{Class. Quant. Grav.}, 35\penalty0 (11):\penalty0 115003, 2018.
\newblock \doi{10.1088/1361-6382/aabcb6}.

\bibitem[Navarro et~al.(1996)Navarro, Frenk, and White]{isz21}
Julio~F. Navarro, Carlos~S. Frenk, and Simon D.~M. White.
\newblock {The Structure of cold dark matter halos}.
\newblock \emph{Astrophys. J.}, 462:\penalty0 563--575, 1996.
\newblock \doi{10.1086/177173}.

\bibitem[Burkert(1995)]{isz22}
A.~Burkert.
\newblock {The Structure of dark matter halos in dwarf galaxies}.
\newblock \emph{Astrophys. J. Lett.}, 447:\penalty0 L25, 1995.
\newblock \doi{10.1086/309560}.

\bibitem[Haroon et~al.(2019)Haroon, Jamil, Jusufi, Lin, and Mann]{isz23}
Shahid Haroon, Mubasher Jamil, Kimet Jusufi, Kai Lin, and Robert~B. Mann.
\newblock {Shadow and Deflection Angle of Rotating Black Holes in Perfect Fluid
  Dark Matter with a Cosmological Constant}.
\newblock \emph{Phys. Rev. D}, 99\penalty0 (4):\penalty0 044015, 2019.
\newblock \doi{10.1103/PhysRevD.99.044015}.

\bibitem[Jamil et~al.(2015)Jamil, Hussain, and Majeed]{isz24}
Mubasher Jamil, Saqib Hussain, and Bushra Majeed.
\newblock {Dynamics of particles around a Schwarzschild-like black hole in the
  presence of quintessence and magnetic field}.
\newblock \emph{Eur. Phys. J. C}, 75:\penalty0 24, 2015.
\newblock \doi{10.1140/epjc/s10052-014-3230-7}.

\bibitem[Konoplya(2019)]{isz25}
R.~A. Konoplya.
\newblock {Shadow of a black hole surrounded by dark matter}.
\newblock \emph{Phys. Lett. B}, 795:\penalty0 1--6, 2019.
\newblock \doi{10.1016/j.physletb.2019.05.043}.

\bibitem[Jusufi et~al.(2019)Jusufi, Jamil, Salucci, Zhu, and Haroon]{isz26}
Kimet Jusufi, Mubasher Jamil, Paolo Salucci, Tao Zhu, and Shahid Haroon.
\newblock {Black Hole Surrounded by a Dark Matter Halo in the M87 Galactic
  Center and its Identification with Shadow Images}.
\newblock \emph{Phys. Rev. D}, 100\penalty0 (4):\penalty0 044012, 2019.
\newblock \doi{10.1103/PhysRevD.100.044012}.

\bibitem[Letelier(1979)]{isz27}
Patricio~S. Letelier.
\newblock {Clouds of Strings in General Relativity}.
\newblock \emph{Phys. Rev. D}, 20:\penalty0 1294--1302, 1979.
\newblock \doi{10.1103/PhysRevD.20.1294}.

\bibitem[Ghosh and Maharaj(2014)]{isz28}
Sushant~G. Ghosh and Sunil~D. Maharaj.
\newblock {Cloud of strings for radiating black holes in Lovelock gravity}.
\newblock \emph{Phys. Rev. D}, 89\penalty0 (8):\penalty0 084027, 2014.
\newblock \doi{10.1103/PhysRevD.89.084027}.

\bibitem[Morais~Graca et~al.(2017)Morais~Graca, Salako, and Bezerra]{isz29}
J.~P. Morais~Graca, G.~I. Salako, and V.~B. Bezerra.
\newblock {Quasinormal modes of a black hole with a cloud of strings in
  Einstein-Gauss-Bonnet gravity}.
\newblock \emph{Int. J. Mod. Phys. D}, 26\penalty0 (10):\penalty0 1750113,
  2017.
\newblock \doi{10.1142/S0218271817501139}.

\bibitem[Ma and Zhao(2014)]{isz30}
Meng-Sen Ma and Ren Zhao.
\newblock {Corrected form of the first law of thermodynamics for regular black
  holes}.
\newblock \emph{Class. Quant. Grav.}, 31:\penalty0 245014, 2014.
\newblock \doi{10.1088/0264-9381/31/24/245014}.

\bibitem[Ghaffarnejad(2008)]{isz31}
Hossein Ghaffarnejad.
\newblock {Scalar-vector-tensor gravity from preferred reference frame
  effects}.
\newblock \emph{Gen. Rel. Grav.}, 40:\penalty0 2229--2239, 2008.
\newblock \doi{10.1007/s10714-009-0898-3}.
\newblock [Erratum: Gen.Rel.Grav. 41, 2941--2943 (2009)].

\bibitem[Chandrasekhar(1983)]{isz32}
Subrahmanyan Chandrasekhar.
\newblock \emph{{The Mathematical Theory of Black Holes}}.
\newblock Oxford Univ. Pr., Oxford, UK, 1983.

\bibitem[Frolov and Novikov(1998)]{isz33}
Valeri~P. Frolov and Igor~D. Novikov.
\newblock \emph{{Black Hole Physics: Basic Concepts and New Developments}},
  volume~96 of \emph{Fundamental Theories of Physics}.
\newblock Springer, Dordrecht, 1998.
\newblock \doi{10.1007/978-94-011-5139-9}.

\bibitem[Perlick and Tsupko(2022)]{isz34}
Volker Perlick and Oleg~Yu. Tsupko.
\newblock {Calculating black hole shadows: Review of analytical studies}.
\newblock \emph{Phys. Rept.}, 947:\penalty0 1--39, 2022.
\newblock \doi{10.1016/j.physrep.2021.10.004}.

\bibitem[Cunha and Herdeiro(2018)]{isz35}
Pedro V.~P. Cunha and Carlos A.~R. Herdeiro.
\newblock {Shadows and strong gravitational lensing: a brief review}.
\newblock \emph{Gen. Rel. Grav.}, 50:\penalty0 42, 2018.
\newblock \doi{10.1007/s10714-018-2361-9}.

\bibitem[Bardeen et~al.(1972)Bardeen, Press, and Teukolsky]{isz36}
James~M. Bardeen, William~H. Press, and Saul~A. Teukolsky.
\newblock {Rotating Black Holes: Locally Nonrotating Frames, Energy Extraction,
  and Scalar Synchrotron Radiation}.
\newblock \emph{Astrophys. J.}, 178:\penalty0 347--370, 1972.
\newblock \doi{10.1086/151796}.

\bibitem[Stella and Vietri(1998)]{isz37}
L.~Stella and M.~Vietri.
\newblock {Lense-Thirring precession and quasi-periodic oscillations in
  low-mass X-ray binaries}.
\newblock \emph{Astrophys. J. Lett.}, 492:\penalty0 L59, 1998.
\newblock \doi{10.1086/311075}.

\bibitem[Abramowicz and Kluzniak(2001)]{isz38}
Marek~Artur Abramowicz and Wlodek Kluzniak.
\newblock {A Precise determination of angular momentum in the black hole
  candidate GRO J1655-40}.
\newblock \emph{Astron. Astrophys.}, 374:\penalty0 L19, 2001.
\newblock \doi{10.1051/0004-6361:20010791}.

\bibitem[Torok et~al.(2005)Torok, Abramowicz, Kluzniak, and Stuchlik]{isz39}
Gabriel Torok, Marek~A. Abramowicz, Wlodek Kluzniak, and Zdenek Stuchlik.
\newblock {The Orbital resonance model for twin peak kHz quasi periodic
  oscillations in microquasars}.
\newblock \emph{Astron. Astrophys.}, 436:\penalty0 1--8, 2005.
\newblock \doi{10.1051/0004-6361:20047115}.

\bibitem[Bambi(2017)]{isz40}
Cosimo Bambi.
\newblock {Testing black hole candidates with electromagnetic radiation}.
\newblock \emph{Rev. Mod. Phys.}, 89\penalty0 (2):\penalty0 025001, 2017.
\newblock \doi{10.1103/RevModPhys.89.025001}.

\bibitem[Steiner et~al.(2011)Steiner, Reis, McClintock, Narayan, Remillard,
  Orosz, Gou, Fabian, and Torres]{JFS2011}
J.~F. Steiner, R.~C. Reis, J.~E. McClintock, R.~Narayan, R.~A. Remillard, J.~A.
  Orosz, L.~Gou, A.~C. Fabian, and M.~A. Torres.
\newblock The spin of the black hole microquasar xte j1550--564 via the
  continuum-fitting and fe-line methods.
\newblock \emph{Mon. Not. R. Astron. Soc.}, 416:\penalty0 941, 2011.
\newblock \doi{10.1111/j.1365-2966.2011.19089.x}.

\bibitem[Gou et~al.(2014)Gou, McClintock, Remillard, Steiner, Reid, Orosz,
  Narayan, Hanke, and Garcia]{LG2014}
L.~Gou, J.~E. McClintock, R.~A. Remillard, J.~F. Steiner, M.~J. Reid, J.~A.
  Orosz, R.~Narayan, M.~Hanke, and J.~Garcia.
\newblock Confirmation via the continuum-fitting method that the spin of the
  black hole in cygnus x-1 is extreme.
\newblock \emph{Astrophys. J.}, 790:\penalty0 29, 2014.
\newblock \doi{10.1088/0004-637X/790/1/29}.

\bibitem[McClintock et~al.(2015)McClintock, Narayan, and Steiner]{JEM2015}
J.~E. McClintock, R.~Narayan, and J.~F. Steiner.
\newblock Black hole spin via continuum fitting and the role of spin in
  powering transient jets.
\newblock \emph{Space Sci. Rev.}, 183:\penalty0 295, 2015.
\newblock \doi{10.1007/s11214-013-0003-9}.

\bibitem[Liu et~al.(2023)Liu, Mustafa, Maurya, and Javed]{GMM1}
Y.~Liu, G.~Mustafa, S.~K. Maurya, and F.~Javed.
\newblock Orbital motion and quasi-periodic oscillations with periastron and
  lense--thirring precession of slowly rotating einstein--\ae ther black hole.
\newblock \emph{Eur. Phys. J. C}, 83:\penalty0 584, 2023.
\newblock \doi{10.1140/epjc/s10052-023-11702-9}.

\bibitem[Mustafa et~al.(2025{\natexlab{a}})Mustafa, Ditta, Naseer, Maurya,
  Channuie, Ibraheem, and Atamurotov]{GMM2}
G.~Mustafa, Allah Ditta, Tayyab Naseer, S.~K. Maurya, Phongpichit Channuie,
  Awad~A. Ibraheem, and Farruh Atamurotov.
\newblock Circular motion, qpos testing, emission energy and thermal
  fluctuations around a non-singular hairy bardeen black hole.
\newblock \emph{Eur. Phys. J. C}, 85:\penalty0 575, 2025{\natexlab{a}}.
\newblock \doi{10.1140/epjc/s10052-025-14235-5}.

\bibitem[Saleem et~al.(2026)Saleem, Majeed, Ali, Ditta, Alimova, Channuie, and
  Atamurotov]{GMM3}
Amna Saleem, Bushra Majeed, Zulfiqar Ali, Allah Ditta, Asalkhon Alimova,
  Phongpichit Channuie, and Farruh Atamurotov.
\newblock Impact of nonlinear electrodynamics on particle motion around a
  charged black hole with matter coupling.
\newblock \emph{Eur. Phys. J. C}, 86:\penalty0 7, 2026.
\newblock \doi{10.1140/epjc/s10052-025-15166-x}.

\bibitem[Mustafa et~al.(2026)Mustafa, Javed, Ghosh, Maurya, and
  Atamurotov]{GMM4}
G.~Mustafa, F.~Javed, S.~G. Ghosh, S.~K. Maurya, and F.~Atamurotov.
\newblock Epicyclic frequencies around charged regular black hole: constraints
  using different quasars data.
\newblock \emph{Eur. Phys. J. C}, 86:\penalty0 6, 2026.
\newblock \doi{10.1140/epjc/s10052-025-15223-5}.

\bibitem[Mustafa et~al.(2025{\natexlab{b}})Mustafa, Alimova, Atamurotov,
  Ibraheem, Channuie, and Bahaddinova]{GMM5}
G.~Mustafa, Asalkhon Alimova, Farruh Atamurotov, Awad~A. Ibraheem, Phongpichit
  Channuie, and Gunel Bahaddinova.
\newblock {Testing regular black holes in the framework of asymptotically safe
  gravity using particle dynamics, QPOs, and shadow constraints}.
\newblock \emph{Eur. Phys. J. C}, 85\penalty0 (741):\penalty0 741,
  2025{\natexlab{b}}.
\newblock \doi{10.1140/epjc/s10052-025-14431-3}.

\bibitem[Mustafa et~al.(2025{\natexlab{c}})Mustafa, Mustafa, Maurya, Naseer,
  Cilli, Güdekli, Abd-Elmonem, and Alhubieshi]{GMM6}
G.~Mustafa, Ghulam Mustafa, S.~K. Maurya, Tayyab Naseer, Arzu Cilli, Ertan
  Güdekli, Assmaa Abd-Elmonem, and Neissrien Alhubieshi.
\newblock {Particle motion and QPOs around Euler-Heisenberg black hole immersed
  in cold dark matter halo}.
\newblock \emph{Nucl. Phys. B}, 1012:\penalty0 116812, 2025{\natexlab{c}}.
\newblock \doi{10.1016/j.nuclphysb.2025.116812}.

\bibitem[Mustafa et~al.(2025{\natexlab{d}})Mustafa, Javed, Maurya, Abd-Elmonem,
  Atamurotov, Alhubieshi, and Channuie]{GMM7}
G.~Mustafa, Faisal Javed, S.~K. Maurya, Assmaa Abd-Elmonem, Farruh Atamurotov,
  Neissrien Alhubieshi, and Phongpichit Channuie.
\newblock {Impact of Barrow’s nonlinear charge on particle motion,
  trajectories, QPOs, and center of mass energy around a black hole}.
\newblock \emph{Phys. Dark Univ.}, 47:\penalty0 101825, 2025{\natexlab{d}}.
\newblock \doi{10.1016/j.dark.2025.101825}.

\bibitem[Maurya et~al.(2025)Maurya, Mustafa, Ditta, Abd-Elmonem, Alhubieshi,
  Caliskan, and Güdekli]{GMM8}
Sunil~Kumar Maurya, G.~Mustafa, Allah Ditta, Assmaa Abd-Elmonem, Neissrien
  Alhubieshi, Aylin Caliskan, and Ertan Güdekli.
\newblock {Circular motion of test particles, trajectories and QPOs around the
  non-rotating black hole in the background of the Symmergent gravity}.
\newblock \emph{Phys. Dark Univ.}, 47:\penalty0 101806, 2025.
\newblock \doi{10.1016/j.dark.2025.101806}.

\bibitem[Mustafa et~al.(2025{\natexlab{e}})Mustafa, Channuie, Javed, Bouzenada,
  Maurya, Cilli, and Güdekli]{GMM9}
G.~Mustafa, Phongpichit Channuie, Faisal Javed, Abdelmalek Bouzenada, S.~K.
  Maurya, Arzu Cilli, and Ertan Güdekli.
\newblock {Orbital motion and epicyclic oscillations around a black hole with
  magnetic charge}.
\newblock \emph{Phys. Dark Univ.}, 47:\penalty0 101765, 2025{\natexlab{e}}.
\newblock \doi{10.1016/j.dark.2024.101765}.

\bibitem[Mustafa et~al.(2025{\natexlab{f}})Mustafa, Maurya, Channuie,
  Bouzenada, Abd-Elmonem, and Alhubieshi]{GMM10}
G.~Mustafa, S.~K. Maurya, Phongpichit Channuie, Abdelmalek Bouzenada, Assmaa
  Abd-Elmonem, and Neissrien Alhubieshi.
\newblock {Epicyclic oscillations around slowly rotating charged black hole in
  Bumblebee gravity}.
\newblock \emph{Phys. Dark Univ.}, 47:\penalty0 101753, 2025{\natexlab{f}}.
\newblock \doi{10.1016/j.dark.2024.101753}.

\bibitem[Mustafa et~al.(2025{\natexlab{g}})Mustafa, Javed, Maurya, Abd-Elmonem,
  Cilli, Atamurotov, and Güdekli]{GMM11}
G.~Mustafa, Faisal Javed, S.~K. Maurya, Assmaa Abd-Elmonem, Arzu Cilli, Farruh
  Atamurotov, and Ertan Güdekli.
\newblock {Particle dynamics and QPOs around a black hole coupled with
  nonlinear electrodynamics and cloud of strings}.
\newblock \emph{Chin. J. Phys.}, 93:\penalty0 1--17, 2025{\natexlab{g}}.
\newblock \doi{10.1016/j.cjph.2024.11.002}.

\bibitem[Kokkotas and Schmidt(1999)]{isz41}
Kostas~D. Kokkotas and Bernd~G. Schmidt.
\newblock {Quasi-normal modes of stars and black holes}.
\newblock \emph{Living Rev. Rel.}, 2:\penalty0 2, 1999.
\newblock \doi{10.12942/lrr-1999-2}.

\bibitem[Konoplya and Zhidenko(2011)]{isz42}
R.~A. Konoplya and Alexander Zhidenko.
\newblock {Quasinormal modes of black holes: From astrophysics to string
  theory}.
\newblock \emph{Rev. Mod. Phys.}, 83:\penalty0 793--836, 2011.
\newblock \doi{10.1103/RevModPhys.83.793}.

\bibitem[Schutz and Will(1985)]{isz43}
Bernard~F. Schutz and Clifford~M. Will.
\newblock {Black hole normal modes: A semianalytic approach}.
\newblock \emph{Astrophys. J. Lett.}, 291:\penalty0 L33--L36, 1985.
\newblock \doi{10.1086/184453}.

\bibitem[Iyer and Will(1987)]{isz44}
Sai Iyer and Clifford~M. Will.
\newblock {Black-hole normal modes: A WKB approach. I. Foundations and
  application of a higher-order WKB analysis of potential-barrier scattering}.
\newblock \emph{Phys. Rev. D}, 35:\penalty0 3621--3631, 1987.
\newblock \doi{10.1103/PhysRevD.35.3621}.

\bibitem[Cardoso et~al.(2009)Cardoso, Miranda, Berti, Witek, and
  Zanchin]{isz45}
Vitor Cardoso, Alex~S. Miranda, Emanuele Berti, Helvi Witek, and Vilson~T.
  Zanchin.
\newblock {Geodesic stability, Lyapunov exponents and quasinormal modes}.
\newblock \emph{Phys. Rev. D}, 79:\penalty0 064016, 2009.
\newblock \doi{10.1103/PhysRevD.79.064016}.

\bibitem[Visser(1999)]{isz46}
Matt Visser.
\newblock {Some general bounds for one-dimensional scattering}.
\newblock \emph{Phys. Rev. A}, 59:\penalty0 427--438, 1999.
\newblock \doi{10.1103/PhysRevA.59.427}.

\bibitem[Harmark et~al.(2010)Harmark, Natario, and Schiappa]{isz47}
Troels Harmark, Jose Natario, and Ricardo Schiappa.
\newblock {Greybody Factors for d-Dimensional Black Holes}.
\newblock \emph{Adv. Theor. Math. Phys.}, 14\penalty0 (3):\penalty0 727--794,
  2010.
\newblock \doi{10.4310/ATMP.2010.v14.n3.a1}.

\bibitem[Aydiner et~al.(2025)Aydiner, Sucu, and Sakall{\i}]{Aydiner:2025eii}
Ekrem Aydiner, Erdem Sucu, and {\.I}zzet Sakall{\i}.
\newblock {Regular magnetically charged black holes from nonlinear
  electrodynamics: Thermodynamics, light deflection, and orbital dynamics}.
\newblock \emph{Phys. Dark Univ.}, 50:\penalty0 102164, 2025.
\newblock \doi{10.1016/j.dark.2025.102164}.

\bibitem[Sucu et~al.(2025)Sucu, Sakall{\i}, Sert, and Sucu]{Sucu:2025fwa}
Erdem Sucu, {\.I}zzet Sakall{\i}, {\"O}zcan Sert, and Yusuf Sucu.
\newblock {Quantum-corrected thermodynamics and plasma lensing in non-minimally
  coupled symmetric teleparallel black holes}.
\newblock \emph{Phys. Dark Univ.}, 50:\penalty0 102063, 2025.
\newblock \doi{10.1016/j.dark.2025.102063}.

\bibitem[Sucu and Sakall{\i}(2025)]{Sucu:2025olo}
Erdem Sucu and {\.I}zzet Sakall{\i}.
\newblock {Quantum tunneling and Aschenbach effect in nonlinear
  Einstein-Power-Yang-Mills AdS black holes}.
\newblock \emph{Chin. Phys.}, 49\penalty0 (10):\penalty0 105101, 2025.
\newblock \doi{10.1088/1674-1137/add8fe}.

\bibitem[G{\"u}rsel et~al.(2025)G{\"u}rsel, Mangut, and Sucu]{Gursel:2025wan}
Huriye G{\"u}rsel, Mert Mangut, and Erdem Sucu.
\newblock {Thermodynamics of Einstein-Euler-Heisenberg Black Holes with Thermal
  Fluctuations and Nonlinear Electromagnetic Fields}.
\newblock \emph{Class. Quant. Grav.}, 42:\penalty0 135015, 2025.
\newblock \doi{10.1088/1361-6382/ade7ea}.

\bibitem[Bronnikov(2001)]{KAB2001}
K.~A. Bronnikov.
\newblock {Regular magnetic black holes and monopoles from nonlinear
  electrodynamics}.
\newblock \emph{Phys. Rev. D}, 63:\penalty0 044005, 2001.
\newblock \doi{10.1103/PhysRevD.63.044005}.

\bibitem[d'Inverno and Vickers(2022)]{RI2022}
Ray d'Inverno and James Vickers.
\newblock \emph{{Introducing Einstein's Relativity: A Deeper Understanding}}.
\newblock Oxford Univ. Pr., Oxford, 2nd edition, 2022.

\bibitem[Zhang et~al.(2021)Zhang, Chen, Ma, He, and Deng]{HXZ2021}
Hong-Xu Zhang, Yuan Chen, Tian-Chi Ma, Peng-Zhang He, and Jian-Bo Deng.
\newblock {Shadow of Schwarzschild black hole surrounded by dark matter}.
\newblock \emph{Chin. Phys. C}, 45\penalty0 (5):\penalty0 055103, 2021.
\newblock \doi{10.1088/1674-1137/abe84c}.

\bibitem[Dymnikova(1992)]{sec2is04}
I.~Dymnikova.
\newblock {Vacuum nonsingular black hole}.
\newblock \emph{Gen. Rel. Grav.}, 24:\penalty0 235--242, 1992.
\newblock \doi{10.1007/BF00760226}.

\bibitem[Lemos and Zanchin(1996)]{sec2is05}
Jose P.~S. Lemos and Vilson~T. Zanchin.
\newblock {Rotating charged black strings in general relativity}.
\newblock \emph{Phys. Rev. D}, 54:\penalty0 3840--3853, 1996.
\newblock \doi{10.1103/PhysRevD.54.3840}.

\bibitem[Li and Yang(2012)]{sec2is07}
Miao-Hui Li and Kuo-Cheng Yang.
\newblock {Galactic dark matter in the phantom field}.
\newblock \emph{Phys. Rev. D}, 86:\penalty0 123015, 2012.
\newblock \doi{10.1103/PhysRevD.86.123015}.

\bibitem[Nascimento et~al.(2024)Nascimento, Bezerra, and Toledo]{FFN2024}
F.~F. Nascimento, V.~B. Bezerra, and J.~M. Toledo.
\newblock Quasinormal modes of black holes with quintessence and string clouds.
\newblock \emph{Ann. Phys. (NY)}, 460:\penalty0 169548, 2024.
\newblock \doi{10.1016/j.aop.2023.169548}.

\bibitem[Sood et~al.(2024)Sood, Ali, Singh, and Ghosh]{AS2024}
Aditya Sood, Md~Sabir Ali, Jitendra~Kumar Singh, and Sushant~G. Ghosh.
\newblock Black holes in perfect fluid dark matter.
\newblock \emph{Chin. Phys. C}, 48\penalty0 (6):\penalty0 065109, 2024.
\newblock \doi{10.1088/1674-1137/ad361f}.

\bibitem[Bambi and Freese(2009)]{sec2is08}
Cosimo Bambi and Katherine Freese.
\newblock {Apparent shape of super-spinning black holes}.
\newblock \emph{Phys. Rev. D}, 79:\penalty0 043002, 2009.
\newblock \doi{10.1103/PhysRevD.79.043002}.

\bibitem[Wald(1984)]{sec2is10}
Robert~M. Wald.
\newblock \emph{{General Relativity}}.
\newblock Chicago Univ. Pr., Chicago, USA, 1984.
\newblock \doi{10.7208/chicago/9780226870373.001.0001}.

\bibitem[Frolov and Zelnikov(2011)]{FrolovZelnikov2011}
Valeri~P. Frolov and Andrei Zelnikov.
\newblock \emph{{Introduction to Black Hole Physics}}.
\newblock Oxford Univ. Pr., Oxford, 2011.
\newblock \doi{10.1093/acprof:oso/9780199692293.001.0001}.

\bibitem[Ahmed et~al.(2025{\natexlab{a}})Ahmed, Al-Badawi, and Sakall{\i}]{gm1}
Faizuddin Ahmed, Ahmad Al-Badawi, and {\.I}zzet Sakall{\i}.
\newblock {Photon spheres, gravitational lensing/mirroring, and greybody
  radiation in deformed AdS-Schwarzschild black holes with phantom global
  monopole}.
\newblock \emph{Phys. Dark Univ.}, 49:\penalty0 101988, 2025{\natexlab{a}}.
\newblock \doi{10.1016/j.dark.2025.101988}.

\bibitem[Al-Badawi et~al.(2025)Al-Badawi, Ahmed, and Sakall{\i}]{gm2}
Ahmad Al-Badawi, Faizuddin Ahmed, and {\.I}zzet Sakall{\i}.
\newblock Dunkl black hole with phantom global monopoles: geodesic analysis,
  thermodynamics and shadow.
\newblock \emph{Eur. Phys. J. C}, 85:\penalty0 660, 2025.
\newblock \doi{10.1140/epjc/s10052-025-14402-8}.

\bibitem[Ahmed et~al.(2025{\natexlab{b}})Ahmed, Al-Badawi, Sakall{\i}, and
  Kanzi]{gm3}
Faizuddin Ahmed, Ahmad Al-Badawi, {\.I}zzet Sakall{\i}, and Sara Kanzi.
\newblock Motions of test particles in gravitational field, perturbations and
  greybody factor of bardeen-like ads black hole with phantom global monopoles.
\newblock \emph{Phys. Dark Univ.}, 48:\penalty0 101907, 2025{\natexlab{b}}.
\newblock \doi{10.1016/j.dark.2025.101907}.

\bibitem[Ahmed et~al.(2025{\natexlab{c}})Ahmed, Al-Badawi, and Sakall{ı}]{gm4}
Faizuddin Ahmed, Ahmad Al-Badawi, and {\.I}zzet Sakall{ı}.
\newblock {Gravitational lensing by black holes in modified gravity theories}.
\newblock \emph{Int. J. Mod. Phys. D}, 34:\penalty0 2550054,
  2025{\natexlab{c}}.
\newblock \doi{10.1142/S0218271825500543}.

\bibitem[Donmez(2025)]{OD2025}
O.~Donmez.
\newblock Accretion dynamics and qpo signatures around quantum-corrected black
  hole: a comparison with kerr spacetime.
\newblock \emph{Eur. Phys. J. C}, 85:\penalty0 1019, 2025.
\newblock \doi{10.1140/epjc/s10052-025-14779-6}.

\bibitem[Stuchlik et~al.(2020)Stuchlik, Kolos, Kovar, Slany, and
  Tursunov]{ZS2020}
Zdenek Stuchlik, Martin Kolos, Jiri Kovar, Petr Slany, and Arman Tursunov.
\newblock {Influence of Cosmic Repulsion and Magnetic Fields on Accretion Disks
  Rotating around Kerr Black Holes}.
\newblock \emph{Universe}, 6\penalty0 (2):\penalty0 26, 2020.
\newblock \doi{10.3390/universe6020026}.

\bibitem[B{\'e}car et~al.(2024)B{\'e}car, Gonz{\'a}lez, Papantonopoulos, and
  V{\'a}squez]{Becar2023}
Ram{\'o}n B{\'e}car, P.~A. Gonz{\'a}lez, Eleftherios Papantonopoulos, and Yerko
  V{\'a}squez.
\newblock {Massive scalar field perturbations of black holes surrounded by dark
  matter}.
\newblock \emph{Eur. Phys. J. C}, 84\penalty0 (3):\penalty0 329, 2024.
\newblock \doi{10.1140/epjc/s10052-024-12553-8}.

\bibitem[Tan et~al.(2025)Tan, Liu, Liang, and Long]{QT2025}
Qun Tan, Dong Liu, Jie Liang, and Zheng-Wen Long.
\newblock {Testing black holes in a perfect fluid dark matter environment using
  quasinormal modes}.
\newblock \emph{Eur. Phys. J. C}, 85:\penalty0 687, 2025.
\newblock \doi{10.1140/epjc/s10052-025-14407-3}.

\bibitem[Liang et~al.(2025)Liang, Cai, Liu, and Long]{YL2023}
Qi-Qi Liang, Ziqiang Cai, Dong Liu, and Zheng-Wen Long.
\newblock {Observational properties and quasinormal Modes of the Hayward black
  Hole surrounded by a cloud of strings}, 11 2025.

\bibitem[Boonserm and Visser(2008)]{gf3}
Petarpa Boonserm and Matt Visser.
\newblock {Bounding the Bogoliubov coefficients}.
\newblock \emph{Annals Phys.}, 323:\penalty0 2779--2798, 2008.
\newblock \doi{10.1016/j.aop.2008.02.002}.

\bibitem[Boonserm(2009)]{gf4}
Petarpa Boonserm.
\newblock \emph{{Rigorous bounds on Transmission, Reflection, and Bogoliubov
  coefficients}}.
\newblock PhD thesis, Victoria University of Wellington, 2009.

\bibitem[Boonserm et~al.(2023)Boonserm, Phalungsongsathit, Sansuk, and
  Wongjun]{gf5}
Petarpa Boonserm, Sattha Phalungsongsathit, Kunlapat Sansuk, and Pitayuth
  Wongjun.
\newblock {Greybody factors for massive scalar field emitted from black holes
  in dRGT massive gravity}.
\newblock \emph{Eur. Phys. J. C}, 83\penalty0 (7):\penalty0 657, 2023.
\newblock \doi{10.1140/epjc/s10052-023-11843-x}.

\bibitem[Ahmed et~al.(2025{\natexlab{d}})Ahmed, Al-Badawi, and Sakall{ı}]{gf6}
Faizuddin Ahmed, Ahmad Al-Badawi, and {\.I}zzet Sakall{ı}.
\newblock {Greybody factors of black holes in modified gravity}.
\newblock \emph{Eur. Phys. J. C}, 85:\penalty0 668, 2025{\natexlab{d}}.
\newblock \doi{10.1140/epjc/s10052-025-14405-5}.

\bibitem[Matyjasek and Opala(2017)]{MatyjasekOpala2017}
Jerzy Matyjasek and Micha{\l} Opala.
\newblock Quasinormal modes of black holes: The improved semianalytic approach.
\newblock \emph{Phys. Rev. D}, 96\penalty0 (2):\penalty0 024011, 2017.
\newblock \doi{10.1103/PhysRevD.96.024011}.

\bibitem[Konoplya(2023)]{RAK2023}
R.~A. Konoplya.
\newblock Further clarification on quasinormal modes/circular null geodesics
  correspondence.
\newblock \emph{Phys. Lett. B}, 838:\penalty0 137674, 2023.
\newblock \doi{10.1016/j.physletb.2023.137674}.

\bibitem[Konoplya and Stuchl{\'\i}k(2017)]{RAK2017}
R.~A. Konoplya and Zden{\v e}k Stuchl{\'\i}k.
\newblock Are eikonal quasinormal modes linked to the unstable circular null
  geodesics?
\newblock \emph{Phys. Lett. B}, 771:\penalty0 597--602, 2017.
\newblock \doi{10.1016/j.physletb.2017.06.021}.

\bibitem[Bolokhov(2024)]{SVB2024}
S.~V. Bolokhov.
\newblock Long-lived quasinormal modes and asymptotic tails of regular black
  holes.
\newblock \emph{Phys. Lett. B}, 856:\penalty0 138879, 2024.
\newblock \doi{10.1016/j.physletb.2024.138879}.

\bibitem[Ghosh(2015)]{isconc2}
Sushant~G. Ghosh.
\newblock {A nonsingular rotating black hole}.
\newblock \emph{Eur. Phys. J. C}, 75\penalty0 (11):\penalty0 532, 2015.
\newblock \doi{10.1140/epjc/s10052-015-3740-y}.

\bibitem[Kumar and Ghosh(2020)]{isconc3}
Rahul Kumar and Sushant~G. Ghosh.
\newblock Rotating black holes in $4d$ einstein--gauss--bonnet gravity and its
  shadow.
\newblock \emph{JCAP}, 07:\penalty0 053, 2020.
\newblock \doi{10.1088/1475-7516/2020/07/053}.

\bibitem[Gundlach et~al.(1994)Gundlach, Price, and Pullin]{isconc4}
Carsten Gundlach, Richard~H. Price, and Jorge Pullin.
\newblock Late time behavior of stellar collapse and explosions: 1. linearized
  perturbations.
\newblock \emph{Phys. Rev. D}, 49:\penalty0 883--889, 1994.
\newblock \doi{10.1103/PhysRevD.49.883}.

\bibitem[Leaver(1985)]{isconc5}
E.~W. Leaver.
\newblock An analytic representation for the quasi normal modes of kerr black
  holes.
\newblock \emph{Proc. Roy. Soc. Lond. A}, 402:\penalty0 285--298, 1985.
\newblock \doi{10.1098/rspa.1985.0119}.

\bibitem[Rodrigues et~al.(2018)Rodrigues, Junior, and de~Sousa~Silva]{isconc8}
Manuel~E. Rodrigues, Ednaldo L.~B. Junior, and Marcos~V. de~Sousa~Silva.
\newblock Using dominant and weak energy conditions to build a new class of
  regular black holes.
\newblock \emph{JCAP}, 02:\penalty0 059, 2018.
\newblock \doi{10.1088/1475-7516/2018/02/059}.

\bibitem[Toledo and Bezerra(2019)]{isconc9}
J.~M. Toledo and V.~B. Bezerra.
\newblock {Kerr-Newman-AdS black hole with quintessence and cloud of strings}.
\newblock \emph{Gen. Rel. Grav.}, 51:\penalty0 41, 2019.
\newblock \doi{10.1007/s10714-019-2528-z}.

\bibitem[Bozza(2002)]{isconc10}
Valerio Bozza.
\newblock {Gravitational lensing in the strong field limit}.
\newblock \emph{Phys. Rev. D}, 66:\penalty0 103001, 2002.
\newblock \doi{10.1103/PhysRevD.66.103001}.

\bibitem[Virbhadra and Ellis(2000)]{isconc11}
K.~S. Virbhadra and George F.~R. Ellis.
\newblock {Schwarzschild black hole lensing}.
\newblock \emph{Phys. Rev. D}, 62:\penalty0 084003, 2000.
\newblock \doi{10.1103/PhysRevD.62.084003}.

\bibitem[Tsukamoto(2017)]{isconc12}
Naoki Tsukamoto.
\newblock {Deflection angle in the strong deflection limit in a general
  asymptotically flat, static, spherically symmetric spacetime}.
\newblock \emph{Phys. Rev. D}, 95\penalty0 (6):\penalty0 064035, 2017.
\newblock \doi{10.1103/PhysRevD.95.064035}.

\bibitem[Toshmatov et~al.(2018)Toshmatov, Stuchlik, and Ahmedov]{isconc13}
Bobir Toshmatov, Zdenek Stuchlik, and Bobomurat Ahmedov.
\newblock {Electromagnetic perturbations of black holes in general relativity
  coupled to nonlinear electrodynamics}.
\newblock \emph{Phys. Rev. D}, 97\penalty0 (8):\penalty0 084058, 2018.
\newblock \doi{10.1103/PhysRevD.97.084058}.

\bibitem[Moreno and Sarbach(2003)]{isconc14}
Claudia Moreno and Olivier Sarbach.
\newblock {Stability properties of black holes in selfgravitating nonlinear
  electrodynamics}.
\newblock \emph{Phys. Rev. D}, 67:\penalty0 024028, 2003.
\newblock \doi{10.1103/PhysRevD.67.024028}.

\bibitem[Psaltis et~al.(2020)]{isconc15}
Dimitrios Psaltis et~al.
\newblock {Gravitational Test beyond the First Post-Newtonian Order with the
  Shadow of the M87 Black Hole}.
\newblock \emph{Phys. Rev. Lett.}, 125\penalty0 (14):\penalty0 141104, 2020.
\newblock \doi{10.1103/PhysRevLett.125.141104}.

\bibitem[Vagnozzi et~al.(2023)]{isconc16}
Sunny Vagnozzi et~al.
\newblock {Horizon-scale tests of gravity theories and fundamental physics from
  the Event Horizon Telescope image of Sagittarius A*}.
\newblock \emph{Class. Quant. Grav.}, 40\penalty0 (16):\penalty0 165007, 2023.
\newblock \doi{10.1088/1361-6382/acd97b}.

\bibitem[Abbott et~al.(2016{\natexlab{b}})]{isconc17}
B.~P. Abbott et~al.
\newblock {Tests of General Relativity with GW150914}.
\newblock \emph{Phys. Rev. Lett.}, 116\penalty0 (22):\penalty0 221101,
  2016{\natexlab{b}}.
\newblock \doi{10.1103/PhysRevLett.116.221101}.

\bibitem[Cardoso and Pani(2019)]{isconc18}
Vitor Cardoso and Paolo Pani.
\newblock {Testing the nature of dark compact objects: a status report}.
\newblock \emph{Living Rev. Rel.}, 22:\penalty0 4, 2019.
\newblock \doi{10.1007/s41114-019-0020-4}.

\end{thebibliography}
\bibliographystyle{apsrev4-2}

\end{document}